\documentclass[preprint]{aastex}

\slugcomment{Accepted for publication in Astronomical Journal}

\shorttitle{$\gamma$~Doradus Survey}
\shortauthors{Henry, Fekel, \& Henry}

\newcommand{\kms}{km~s$^{-1}$}

\begin{document}

\title{A VOLUME-LIMITED PHOTOMETRIC SURVEY OF 114 $\gamma$~DORADUS CANDIDATES}

\author{Gregory W. Henry, Francis C. Fekel\altaffilmark{1}}
\affil{Center of Excellence in Information Systems, Tennessee State University,
    \\ 3500 John A. Merritt Blvd., Box 9501, Nashville, TN 37209}
\altaffiltext{1}{Visiting Astronomer, Kitt Peak National Observatory, National
Optical Astronomy Observatory, operated by the Association of Universities
for Research in Astronomy, Inc., under cooperative agreement with the
National Science Foundation.}

\author{Stephen M. Henry}
\affil{Department of Mathematical Sciences, Clemson University,\\
O-110 Martin Hall, Box 340975, Clemson, SC 29634}

\email{gregory.w.henry@gmail.com}
\email{fekel@evans.tsuniv.edu}
\email{smhenry@clemson.edu}

\begin{abstract}

We have carried out a photometric survey of a complete, volume-limited 
sample of $\gamma$~Doradus candidates.  The sample was extracted from the 
{\it Hipparcos} catalog and consists of 114 stars with colors and absolute 
magnitudes within the range of known $\gamma$~Doradus stars and that also 
lie within a specified volume of 266,600 pc$^3$.  We devoted one year of 
observing time with our T12 0.8~m automatic photometric telescope to acquire 
nightly observations of the complete sample of stars.  From these survey 
observations, we identify 37 stars with intrinsic variability of 0.002 mag 
or more.  Of these 37 variables, eight have already been confirmed as 
$\gamma$~Doradus stars in our earlier papers; we scheduled the remaining 
29 variables on our T3 0.4~m automatic telescope to acquire more intensive 
observations over the next two years.  As promising new $\gamma$~Doradus 
candidates were identified from the photometry, we obtained complementary 
spectroscopic observations of each candidate with the Kitt Peak coud\'e feed 
telescope.  Analysis of our new photometric and spectroscopic data reveals 
15 new $\gamma$~Doradus variables (and confirms two others), eight new 
$\delta$~Scuti variables (and confirms one other), and three new variables 
with unresolved periodicity.  Therefore, of the 114 $\gamma$~Doradus 
candidates in our volume-limited sample, we find 25 stars that are new or 
previously-known $\gamma$~Doradus variables.  This results in an incidence 
of 22\% for $\gamma$~Doradus variability among candidate field stars for 
this volume of the solar neighborhood.  The corresponding space density 
of $\gamma$~Doradus stars in this volume of space is 0.094 stars per 
$10^3~pc^3$ or 94 stars per $10^6~pc^3$.  We provide an updated list of 
86 bright, confirmed, $\gamma$~Doradus field stars.

\end{abstract}


\section{INTRODUCTION}

Two decades ago, \citet{k93} and others began finding a small number of 
early-F stars that seemed to be ``variables without a cause.''  The first 
recognized examples of these low-amplitude variable stars included 9~Aurigae 
\citep{ketal93}, $\gamma$~Doradus \citep{betal94,bkc94}, HD~111828 
\citep{mpa91}, HD~224638 and HD~224945 \citep{mp91,mpz94}.  All five 
stars were found to vary with at least two photometric periods of order one 
day and to exhibit spectroscopic line-profile variations with the same 
timescale.  All five were also found to lie on or near the main 
sequence in the Hertzsprung-Russell (H-R) diagram and to cluster around 
the cool edge of the $\delta$~Scuti instability strip \citep{kh95}, despite 
the fact that their periods are too long to be $p$-mode $\delta$~Scuti 
pulsations.  Results from a multi-site photometric and spectroscopic 
observing campaign on $\gamma$~Doradus ($MUSICOS-94$) led \citet{betal96} 
to the conclusion that ``non-radial pulsation is the only viable explanation''
for their variability. 

Given that $\gamma$~Doradus is the brightest of these early-F pulsators 
($V=4.26$) and was the first to be detected as a variable star \citep{cw63}, 
it serves as the prototype of this new, slowly-growing group of variable
stars \citep{mpz94,kh95,betal96,ketal99a}.  By 1999, the group of 
$\gamma$~Doradus variables had expanded to 13 stars, and \citet{ketal99a} 
described their primary observational characteristics to be (1) spectral 
type A7--F5, (2) luminosity class IV, IV-V, or V, (3) low-amplitude 
photometric variability with one or more periods in the range 0.4--3 days, 
and (4) spectroscopic line-profile variations accompanied by low-amplitude, 
radial velocity variability.  Kaye et al. plotted the 13 known 
$\gamma$~Doradus stars in the H-R diagram and found that their larger 
sample still clustered around the lower right corner of the $\delta$~Scuti 
instability strip.

The most recent list of $\gamma$~Doradus stars is given by \citet{hfh07} 
and contains 66 members (their Table~6).  Over half of these stars were
identified as $\gamma$~Doradus candidates by \citet{han99} from his
analysis of {\it Hipparcos} photometry and then confirmed with additional
photometry and spectroscopy from our automatic telescopes at Fairborn 
Observatory.  The $\gamma$~Doradus stars in this larger sample usually 
have two or more photometric periods in the range 0.3 to 2.6 days and 
sinusoidal light curves with amplitudes between 0.002 and 0.10 mag 
\citep[see][and references therein]{hfh07}.  Radial velocity variations of 
2--4 \kms\ and changing spectroscopic line profiles are also seen in many 
members of this larger group 
\citep[e.g.,][]{ketal95,betal96,h98,ketal99b,ketal99c,fh03,metal04,dcetal06}.  
Many observing campaigns on individual $\gamma$~Doradus stars have confirmed 
that their photometric and spectroscopic variations arise from high-order, 
non-radial, $g$-mode pulsations 
\citep[see, e.g.,][]{betal96,h98,betal97,ak01,aetal04,retal06b}.  

Significant theoretical work has been accomplished in the past decade, 
including determining the pulsational driving mechanisms 
\citep[e.g.,][]{getal00,getal05,detal05b,g10}, understanding the limits of 
$\gamma$~Doradus pulsation in the H-R diagram \citep[e.g.,][]{wkg03,detal04,
getal04}, mode identification \citep[e.g.,][]{metal05,setal05,detal05a,
mmne08}, and asteroseismic modeling \citep[e.g.,][]{rak06,metal08,petal08}.  
We refer the reader to \citet{detal07} and \citet{p09} for recent reviews 
of the observational and theoretical status of the $\gamma$~Doradus stars.

In this paper, we describe our photometric survey of a complete, 
volume-limited sample of 114 late-A to early-F dwarfs and subgiants that 
lie within a nearby, well-defined region of the solar neighborhood with 
a volume of 266,600 $pc^3$.  The stars in this sample all lie within the 
observed $\gamma$~Doradus instability strip.  We use our results to determine 
the incidence of $\gamma$~Doradus variables within the instability strip and 
to compute the space density of $\gamma$~Doradus stars at our location in 
the Orion Spur of the Milky Way galaxy.  In the process, we have increased 
the number of bright $\gamma$~Doradus stars by 23\%.  Finally, we provide 
a definitive list of 86 bright, nearby $\gamma$~Doradus field stars that
will be excellent targets for future multi-site, multi-technique, and/or
space-based observing campaigns.

\section{PHOTOMETRIC SURVEY OF 114 $\gamma$~DORADUS CANDIDATES}

\subsection{Creating the Volume-Limited Sample}

Our first step was to construct a volume-limited sample of approximately 
100 stars that would be well suited to the capabilities of our T12 0.8~m 
APT and its location at Fairborn Observatory in southern Arizona.  The 
stars would need to be within a $V$ magnitude range of approximately 5.5 to 
8.0 to assure acceptably low coincidence-count corrections for the brighter 
stars and count rates that are still scintillation limited rather than 
photon limited for the fainter stars.  The stars should lie within a 
Declination range of $-$10\arcdeg\ to $+$65\arcdeg\ so they can be observed 
at airmass values less than 1.5 from the latitude of Fairborn Observatory.  
Based on H-R diagrams of the small samples of $\gamma$~Doradus stars in 
\citet{aetal98}, \citet{han99}, and \citet{ketal99a}, we set $B-V$ limits 
of 0.25--0.38 mag and absolute magnitude $M_V$ limits of 1.6--3.7 for the 
sample.

The {\it Hipparcos} catalog \citep{petal97} provides an excellent source 
from which to draw stars for our volume-limited survey because it is complete 
for $V \le 7.5$ and ``largely complete'' down to $V=9$.  We extracted
various samples from the {\it Hipparcos} catalog, varying the parallax 
limits each time, to find a spherical shell of space centered on the Sun 
that contained roughly the desired number of stars that satisfied our 
chosen limits for $V$, $B-V$, $M_V$, and declination.  We converged to a 
final parallax range of 15 to 20 mas inclusive, which corresponds to a 
distance of 50.0--66.7 pc.  The spherical shell, truncated at declinations 
of $-$10\arcdeg and $+$65\arcdeg , has a volume of 266,600 pc$^3$ and 
contains 126 stars that meet our search criteria.  From this 126-star 
sample, we eliminated 11 close visual doubles and one eclipsing binary 
for which the {\it Hipparcos} magnitudes and colors represented the combined 
value of the components and not the individual components themselves.  The 
final sample contains 114 stars, only one of which is slightly fainter than 
$V=7.5$ (HD~62196; $V=7.67$).  Therefore, given the completeness limits of 
the {\it Hipparcos} catalog cited above, we can be certain that our sample of 
114 stars constitutes a statistically-complete, volume-limited sample of 
$\gamma$~Doradus candidates.

The selection criteria for the survey stars are summarized in Table~1, and 
the final sample of 114 stars is given in Table~2.  Columns (4), (5), and (6) 
list the {\it Hipparcos} $V$ magnitudes, $B-V$ color indices, and parallax 
values for all stars; column (7) gives the absolute magnitudes computed from 
the $V$ magnitudes and parallaxes (ignoring interstellar extinction).  
Column (8) lists the {\it Hipparcos} variability type; we note that only 10 
of the 114 stars are flagged as possible variable stars with only three as 
periodic variables.

\subsection{Photometric Observations of the 114 Star Sample}

The initial photometric survey of the sample of 114 stars was carried 
out between 2001 April and 2002 July with our T12 0.8~m APT at Fairborn
Observatory.  This APT and its two-channel precision photometer are 
functionally identical to our T8 APT described in \citet{h99}.  The
photometer uses two temperature-stabilized EMI 9124QB photomultiplier tubes 
mounted behind a dichroic beam splitter to measure photon count rates 
simultaneously through Str\"omgren $b$ and $y$ filters.  We programmed the 
telescope to make nightly observations of all target stars that were observable 
on a given night.  Three comparison stars in the vicinity of each target
star, designated A, B, and C, were measured along with the target star, 
D, in the following sequence, termed a group observation:  DARK, A, B, C, D, 
A, SKY$_{\rm A}$, B, SKY$_{\rm B}$, C, SKY$_{\rm C}$, D, SKY$_{\rm D}$, 
A, B, C, D.  We used a diaphragm $45\arcsec$ in diameter and an integration 
time of 20 seconds for all measurements.

Each completed group observation was reduced to form three independent measures 
of each of the six differential magnitudes D$-$A, D$-$B, D$-$C, C$-$A, C$-$B, 
and B$-$A.  These differential magnitudes were corrected for differential 
extinction with nightly extinction coefficients and transformed to the 
standard Str\"omgren system with yearly mean transformation coefficients.  
The three independent measures of each differential magnitude were combined, 
giving one mean data point per complete group sequence for each of the six 
differential magnitudes.  To increase the precision of the observations, 
we combined the Str\"omgren $b$ and $y$ differential magnitudes 
into a single $(b+y)/2$ passband.  The typical precision of a group mean 
acquired in good conditions is 0.0015--0.0020 mag.  Data taken in 
non-photometric conditions were eliminated by discarding all group sequences 
in which one or more of the mean differential magnitudes had a standard 
deviation of 0.01 mag or greater.  For additional information on the 
telescope, photometer, observing procedures, data reduction techniques, and 
photometric precision, see \citet{h99} and \citet{ehf03}.

\subsection{Photometric Results from the 114 Star Sample}

Columns (9), (10), and (11) of Table~2 give the results of the 2001--2002 
photometric survey with the T12 APT.  The APT typically acquired several tens 
of group observations over the course of each star's observing season, as 
shown in Column~(9), where {$N_{obs}$} is the number of good group observations 
that survived the 0.01-mag ``cloud filter'' described above and also passed
a visual inspection to eliminate any remaining outliers.  A few of the stars 
were reobserved in 2006 for a couple of nights of monitoring observations 
and therefore have up to several hundred observations.  Column~(10) gives 
the Julian date range for each star over which the survey observations 
were acquired.  Column~(11) gives an estimate of each star's intrinsic 
brightness variability, $\sigma_{star}$, in the combined $(b+y)/2$ passband.
This variability estimate is computed by subtracting the total 
variance of the most constant set of comparison star differential magnitudes 
($\sigma^2_{C-A}$, $\sigma^2_{C-B}$, or $\sigma^2_{B-A}$) from the total 
variance of the program star differential magnitudes computed against 
the mean brightness of the two best comparison stars ($\sigma^2_{D-CA}$, 
$\sigma^2_{D-CB}$, or $\sigma^2_{D-BA}$).  Therefore, our brightness 
variability estimate, given as the standard deviation $\sigma_{star}$ in 
column~(11) of Table~2, represents an approximation of the intrinsic stellar 
variability in each target star, corrected for measurement errors and 
possible low-level variability in the comparison stars.

In cases where the program star and the two best comparison stars 
are all constant or nearly so, then $\sigma^2_{star}$, the difference of
two variances, can have either a small positive {\it or} a small negative 
value due to random measurement uncertainties.  If $\sigma^2_{star}$ 
is negative, then $\sigma_{star}$ is imaginary so we set its value to 
zero in Table~2 since, in those cases, we clearly have not resolved any 
variability in the program star above the expected uncertainties in the 
measurements.  Small positive values of $\sigma_{star}$ are recorded as 
such in Table~2, but long experience with our APTs \citep[e.g.,][]{hfh95,h99,
hetal00,letal07} has shown that only $\sigma_{star}$ values of 0.002 mag 
or higher reliably indicate variability for stars brighter than $V=8.0$.
Therefore, we classify the 77 stars in Table~2 with $\sigma_{star} < 0.002$ 
mag as constant stars (column~(12)).  The remaining 37 stars have 
$\sigma_{star} \geq 0.002$ mag and so are excellent $\gamma$~Doradus 
candidates.  

Frequency analyses of the 114 survey stars confirmed the presence of 
short-period photometric variability in all 37 stars with 
$\sigma_{star} \geq 0.002$ mag.  All 114 stars are plotted in the H-R 
diagram in Figure~1, where the 37 variables are plotted as filled circles.  
As expected, the frequency spectra exhibit severe aliasing because the 
stars were generally observed only once per night.  Based on these 
frequency spectra alone, we cannot distinguish between $\gamma$~Doradus 
stars with periods in the range 0.3--3.0 days and the more rapidly 
pulsating $\delta$~Scuti stars with periods between 0.02 and 0.25 days.  
We designed this initial survey only to {\it find} all low-amplitude 
variables in the survey sample; we then made additional photometric 
and spectroscopic observations of the variable stars in the survey to improve 
the frequency analyses, to detect any binary stars among the variables, and to 
enable reliable variable star classifications and confirmation of new 
$\gamma$~Doradus stars.

\section{FOLLOW-UP PHOTOMETRY OF THE $\gamma$~DORADUS CANDIDATES}

Eight of the 37 $\gamma$~Doradus candidates found in the 114-star survey 
are known $\gamma$~Doradus stars from our earlier work; we kept them in 
the T12 survey for completeness and as a quality control measure to confirm 
they would be ``discovered'' in the survey.  These eight stars include 
HD~69715, HD~99329, HD~124248, HD~165645, HD~167858, HD~207233, and HD~213617, 
confirmed as $\gamma$~Doradus stars by \citet{hfh07}, and the hybrid 
$\gamma$~Doradus / $\delta$~Scuti pulsator HD~8801 found by \citet{hf05}.  
All eight of these known $\gamma$~Doradus stars were detected as variable 
stars in our survey observations.  Since detailed analyses of these eight 
stars have already been published, we did not make follow-up observations 
of them.  We limited our follow-up photometry and spectroscopy to the 
remaining 29 variable stars.

\subsection{Observations}

We used our T3 0.4~m APT at Fairborn Observatory to acquire the new 
higher-cadence Johnson $BV$ photometric follow-up observations of the 
remaining 29 variables identified in the T12 survey.  The observations were 
acquired between 2004 September and 2006 July; each star was observed over 
a single observing season.  T3 is the same telescope used in most of our 
previous work on $\gamma$~Doradus stars.  We shortened the group observing 
sequence used for the survey observations by retaining only the two best 
comparison stars from the survey and scheduling them as the check and 
comparison stars in the following sequence:  K, S, C, V, C, V, C, V, C, S, K, 
where K is a check star, C is the primary comparison star, V is the program 
star, and S is a sky reading.  

Up to five group observations of each program star were acquired every clear 
night at intervals of two to three hours throughout each star's observing 
season.  Each star was also observed continuously for several hours on one 
night near its opposition.  This observing strategy helps to weaken the 
one-day aliases in our period analyses and allows us to discriminate between 
$\gamma$~Doradus variability (0.3--3.0 days) and $\delta$~Scuti variability 
(0.02--0.25 days).  We refer the reader to \cite{hfh07} for a more complete 
description of the instrument, observing techniques, and data reduction 
procedures. 

\subsection{Analyses}

We performed period-search analyses of the new follow-up photometry on 
the remaining 29 variables from the survey.  Our period-search technique, 
based on least-squares fitting of sinusoids, is described in our earlier 
papers \citep[e.g.,][]{hetal01} and is well suited for finding multiple 
signals in our low-amplitude light curves.  Briefly, we search for 
periodicity in the program star minus comparison star ($V-C$) differential 
magnitudes over the frequency range 0.01--30.0 day$^{-1}$, which corresponds 
to the period range 0.033--100 days.  We fix the first detected frequency 
but not its associated amplitude, phase, or mean brightness level and then 
introduce that frequency as a fixed parameter into a new search for additional 
frequencies.  In an iterative process, each new search is carried out while 
simultaneously fitting a single new mean brightness level along with the 
amplitudes and phases of all frequencies introduced as fixed parameters.  In 
the resulting least-squares frequency spectra, we plot the fractional 
reduction of the total variance (the reduction factor) versus the trial 
frequency.

We found 17 of the 29 stars to exhibit periodic variability within the 
period range of $\gamma$~Doradus stars.  Two of those stars (HD~65526 and 
HD~224945) were previously confirmed to be $\gamma$~Doradus variables 
by \citet{hs02} and \citet{mpz94}, respectively; the other 15 stars are 
new variables.  In addition, we found nine stars with variability 
on $\delta$~Scuti timescales, only one of which was a known $\delta$~Scuti 
star \citep{petal98}.  Also, there were three stars among the 29 variables 
for which we could not determine definitive periods.

The analyses and discussion in the following sections are concerned only 
with the 15 new $\gamma$~Doradus candidates plus HD~65526 and HD~224945 
(mentioned above) since those two stars lack extensive APT data sets.  
The parameters of these 17 stars are printed in bold font in Table~2.  We
will refer to these 17 stars as our ``$\gamma$~Doradus candidates.''  The 
analyses of the nine $\delta$~Scuti variables and the three unresolved 
variables may appear in a future paper(s).

Table~3 lists the comparison and check stars used for each of the 17
$\gamma$~Doradus candidates as well as the standard deviations of the
$V-C$ and $K-C$ observations.  The $\sigma_{(K-C)}$ values demonstrate 
that all comparison and check stars are constant to a few millimagnitudes, 
which is approximately the limit of precision for the T3 APT.  The 
individual photometric observations of the 17 $\gamma$~Doradus candidates 
are given in electronic format in Table~4. 

The results of our period analyses of the 17 $\gamma$~Doradus candidates are 
summarized in Table~5.  The frequencies and corresponding periods in both 
$B$ and $V$ are given in columns 5 and 6; the uncertainty in each frequency 
measurement is estimated from the width of its peak in the frequency 
spectrum.  The majority of the B and V frequency pairs agree within 
$\pm~1~\sigma$, the rest within $\pm~2~\sigma$ or a bit more, indicating
that our estimated uncertainties are realistic.  We also note that 
\citet{cetal09} used the 1.2~m Mercator telescope on La Palma over a period 
of four years to monitor a sample of 21 gamma Doradus stars, 19 of which 
we have confirmed as $\gamma$~Doradus variables in our earlier papers.  The 
vast majority of our frequencies in these 19 stars were confirmed in the 
Cuypers paper. 

In several cases, we find very close frequencies in the same star that are 
separated by no more than 0.01 day$^{-1}$ or so.  A conservative criterion 
for frequency resolution is given by \citet{ld78} to be $1.5/T$ where $T$ 
is the total range of the observations in days.  Since most of our data sets 
span $\sim200$ days, the limit of our frequency resolution is typically around 
0.0075 day$^{-1}$, ensuring that we can resolve such close frequencies.  The 
peak-to-peak amplitudes reported in column~(7) of Table~5 are determined for 
each frequency {\it without} prewhitening for the other frequencies.  The $B$ 
amplitudes range from 54 mmag down to 5 mmag and average 1.3 times larger 
than those in $V$.  The individual $B/V$ amplitude ratios and their 
uncertainties are listed for each frequency in column~(8).  Finally, times of 
minimum light for each frequency are given in column~(9); in each case, the 
times of minimum in the two passbands agree within their uncertaintes, 
so there are no detectable phase shifts in our two-color photometry.

In the same way, the ($K-C$) differential magnitudes were also analyzed to
search for periodicities that might exist in the comparison and check stars.
None of the 17 ($K-C$) time series showed any evidence for periodicity,  
so we are assured that all periods reported in Table~5 are intrinsic 
to the program stars.

Least-squares frequency spectra and phase diagrams of the $B$ observations 
for all of the 17 $\gamma$~Doradus candidates are given in \S6 below.  
Although all analyses were done over the frequency range of 
0.01--30.0 day$^{-1}$, the least-squares frequency spectra are plotted over 
more restricted ranges since none of the stars exhibited variability above 
5 day$^{-1}$.  In particular, no higher frequencies that could be 
attributed to $\delta$~Scuti-type variability were found in any of the
17 stars, so we have not identified any new hybrid variables.  The plots 
of the least-squares frequency spectra show the results of successively 
fixing each detected frequency until no further frequencies could be found 
in both passbands.  For illustrative purposes only, the phase diagrams are 
plotted for each frequency after the data sets have been prewhitened to 
remove the other detected frequencies.

\section{FOLLOW-UP SPECTROSCOPY OF THE $\gamma$~DORADUS CANDIDATES}

\subsection{Observations}

We obtained our high-dispersion spectroscopic observations of the  
$\gamma$~Doradus candidates at the Kitt Peak National Observatory with 
the coud\'e feed telescope, coud\'e spectrograph, and a CCD detector.  
Except for two spectrograms of HD~99267 acquired in 1995 April, our 
observations were collected from 2003 October to 2008 September.  We 
have three to 13 spectra of each star.  The vast majority of the 
spectrograms were obtained with a TI CCD, are centered at 6430~\AA, cover 
a wavelength range of about 80~\AA, and have a two-pixel resolution of 
0.21~\AA\ or a resolving power of just over 30,000.  The TI CCD was 
unavailable in 2008 September, and so a Tektronics CCD, designated T1KA, 
was used instead.  With that CCD the spectrum was centered at 6400~\AA.  
Although the wavelength range covered by the T1KA increased to 172~\AA, 
the two-pixel resolution was 0.34~\AA, reducing the resolving power to 
19,000.  Typical signal-to-noise ratios of our spectra range between 
150 and 250.

\subsection{Analyses}

The reduction and analysis of the spectroscopic data  are extensively 
described in \citet{fwk03}.  Here, we provide a summary of our analysis 
procedures but also include additional information that is specific to some 
of the program stars in the current work.  The basic properties of the 
17 $\gamma$~Doradus candidates, determined from our spectroscopic 
observations, are summarized in Table~6.

The spectra of our program stars were compared with the spectra of various 
reference stars with well-determined spectral types.  Because the lines of 
late A- and F-type stars in the 6430~\AA\ region have little sensitivity 
to luminosity, we can only determine the spectral classes of the program 
stars from our spectra, which we list in column~(6) of Table~6.  To determine 
the luminosity classes given in column~(7), we computed absolute magnitudes 
from the {\it Hipparcos} magnitudes and parallaxes \citep{petal97}, which 
were then compared with canonical values of \citet{g92}. Because we  
do not derive our luminosity classes from an examination of the spectrum
itself, we do not use Roman numerals to represent those classes
but instead call them dwarfs. 

The strongest features in the 6430~\AA\ region are primarily iron and 
calcium lines, enabling us to estimate the iron abundances of our 
program stars because our reference stars have spectroscopically 
determined [Fe/H] values in the literature.  Unless otherwise noted in 
the discussions of the individual systems, the program stars have [Fe/H] 
$\sim$0.0, indicating a metallicity close to the solar value.  Because 
the 6430~\AA\ region has both iron and calcium lines, we are able 
to identify the abundance peculiarities associated with Am stars. 

We also have determined the projected rotational velocities of the 17
$\gamma$~Doradus candidates.  For {\it v}~sin~{\it i} values 
$\leq$ 60 km~s$^{-1}$, we followed the procedure of \citet{f97}.  However,
several of the stars have larger values.  In such cases, as well as for 
the stars having composite spectra, we determined the {\it v}~sin~{\it i} 
values from spectrum addition fits \citep{fwk03}.   
\citet{fwk03} provide the estimated uncertainties of these rotational 
velocities, which depend on the individual {\it v}~sin~{\it i} values.

A colon after a derived value in Table~6 indicates greater than usual 
uncertainty. For the spectral classes and projected rotational velocities, 
this is typically because a star has very broad and shallow lines or because 
the components of a binary have a large magnitude difference, making 
the lines of the secondary component very weak.  Weak, broad lines also
reduce the precision of the measured radial velocities. 

Individual radial velocities, determined by cross correlation, 
are listed in Table~7 along with comments about the spectra.  The
velocities of our standard stars are adopted from \citet{setal90}. 
The precision of our velocities depends primarily on the projected
rotational velocities of the stars.  For narrow lined stars with
$v$~sin~$i$ values that are less than 40 km~s$^{-1}$, we estimate
radial velocity uncertainties of 0.2--0.3 km~s$^{-1}$.  For
faster rotating stars with $v$~sin~$i$ values between 40 and 80 
km~s$^{-1}$ we estimate velocity uncertainties of 0.5 km~s$^{-1}$. 
For the most rapidly rotating stars in our sample, those with $v$~sin~$i$ 
values of 90 km~s$^{-1}$ or larger, we estimate uncertainties of 1--2 
km~s$^{-1}$.
Our new velocities, supplemented by older and generally lower precision 
velocities, enable us to establish in most cases whether a star is 
single or binary.  However, the $\gamma$~Doradus variables also have 
line-profile variations resulting from pulsation, which usually cause 
velocity variability of at least several km~s$^{-1}$ 
\citep[e.g.,][]{metal04}.  In the sections on the individual $\gamma$~Doradus 
variables we discuss a few cases where the determination of whether a star is 
single or a single-lined spectroscopic binary is problematic.
In column~(9) of Table~6, stars for which we believe orbital motion 
is detected are indicated as ``Binary'', while the mean radial velocities
and their standard deviations are listed for the stars that we conclude 
are single. 

For single stars and single-lined spectroscopic binaries, the determination of 
basic properties is relatively straightforward.  However, the spectra of four 
of the 17 stars in our sample (HD~38309, HD~114447, HD~145005, HD~220091) 
show two sets of lines (see Figures 2, 3, 4, and 5), making the determination 
of the magnitudes and colors of the individual components more difficult 
than for a single star.  To obtain individual magnitudes and colors for 
the components of a binary, a $V$ magnitude difference is needed.  The 
spectrum addition method described in \citet{fwk03} produces a continuum 
luminosity ratio, which results in a magnitude difference that is a 
minimum value if the secondary has a later spectral type than the primary.  
If the spectral type difference is small, the continuum luminosity ratio 
can be adopted as the luminosity ratio from which the magnitude difference 
can be determined.  Otherwise the continuum luminosity ratio must be 
adjusted for the difference in spectral types before estimating a magnitude 
difference.  Results for the individual components of the four stars
mentioned above are included in Table~6.

\section{CRITERIA FOR CONFIRMING $\gamma$~DORADUS VARIABILITY}

Throughout our series of papers on $\gamma$~Doradus stars 
\citep[see][and references therein]{hfh07}, we have consistently used 
the following criteria for confirming stars as $\gamma$~Doradus variables: 
(1) late-A or early-F spectral class, (2) luminosity class IV or V, and 
(3) periodic photometric variability in the $\gamma$~Doradus period range 
that is attributable to pulsation.  Our spectroscopic observations establish 
the spectral types of each candidate.  Our photometric observations are 
numerous and extensive enough (typically several hundred observations over 
a full observing season) to minimize the effects of one cycle per day 
aliasing and provide the correct period identifications.  This is important 
since the cadence of the {\it Hipparcos} observations, from which Handler's 
candidates were identified, can result in spurious photometric periods, 
especially for multiperiodic stars \citep[e.g.,][]{eg00}.

Our photometric and spectroscopic observations are also used to confirm 
the variability mechanism(s), especially for stars with only one photometric 
period.  These stars could be ellipsoidal variables in close binary systems 
or rapidly-rotating starspot variables rather than pulsating stars.  Multiple 
spectroscopic observations that do not exhibit large-amplitude, short-period 
radial velocity variations argue strongly against the ellipticity effect.  
The early-F spectral types of the candidates and the high level of coherence 
in the light curves over hundreds of cycles both argue strongly against 
starspot variability.  Furthermore, our observed photometric $B/V$ amplitude 
ratios are indicative of pulsations in these stars.  \citet{hetal00} 
demonstrated that ellipsoidal variables have $B/V$ amplitude ratios close to 
1.00 while starspot variables have typical $B/V$ ratios around 1.12--1.14.  
The 17 stars analyzed in this paper have a weighted mean $B/V$ amplitude 
ratio of $1.33 \pm 0.03$, in agreement with theoretical models of 
$\gamma$~Doradus stars with low spherical degree ($\ell = 1,2$) non-radial 
pulsations \citep[e.g.,][]{g00}.

Our Johnson $BV$ photometry and limited spectroscopic observations are
{\it not} sufficient to allow us to identify uniquely the spherical degree
($\ell$) or the azimuthal order ($m$) of the pulsations.  \citet{sw81}
demonstrated for non-radial pulsations that the wavelength dependence of the
photometric amplitude and the phase shift between various photometric bands
is a function of $\ell$ but not $m$ (with an additional dependence of the
amplitude on the inclination of the pulsating star's rotation axis).  In
practice, the identification of the spherical degree from photometric
observations has many subtle difficulties \citep[e.g.,][]{g00,s02}.  We can 
only note that our observed $B/V$ amplitude ratios and lack of detectable 
phase shifts between the two photometric bands (Table~5) are consistent with 
spherical degree $\ell = 1$ or $\ell = 2$ and probably inconsistent with 
$\ell = 3$ \citep{g00}.  Our spectroscopic observations, obtained primarily 
to determine spectral class, {\it v}~sin~{\it i}, and to search for evidence 
of duplicity, are not nearly numerous enough for line-profile variability 
and mode-identification studies.  Such studies require much more extensive
multi-site, multi-technique observing campaigns
\citep[e.g.,][]{hetal02,aetal04,metal04}.

\section{CONFIRMATION OF 17 NEW $\gamma$~DORADUS VARIABLES}

In this section, we examine in detail our photometric and spectroscopic 
observations of the 17 remaining $\gamma$~Doradus candidates that, along 
with previous results from the literature, allow us to determine the 
properties of these 17 stars and, thereby, confirm them as new $\gamma$~Doradus 
variables.

\subsection{HD 6568 = HIP 5209}

We have obtained 13 red-wavelength spectrograms of HD~6568, which show
only a single component with a moderate projected rotational velocity of
55 km~s$^{-1}$.  We determine that the star is an F1 dwarf.  Our radial
velocities from eight different epochs have a range of 23 km~s$^{-1}$,
indicating that the star is a spectroscopic binary with an, as yet, unseen
lower-mass companion.  The velocities that have been obtained to date
suggest an orbital period of 126.5 days if the orbit is circular or
nearly circular.  Observations are continuing to determine orbital
elements.

The {\it Hipparcos} catalog does not flag HD 6568 as a variable star 
\citep{petal97}, but our T12 survey observations (Table~2) reveal definite 
variability with a standard deviation of 0.0066 mag.  Least-squares 
frequency spectra of our T3 follow-up observations in the $B$ are plotted 
in Figure~6, and the results of our period analyses are given in Table~5.  
We find two closely spaced periods of 0.73421 and 0.74538 days with 
amplitudes of 18 and 15 mmag, respectively; we detect the same two periods 
in the $V$ band observations.  The $B$ observations are phased with these 
periods and times of minimum given in Table~5 and plotted in Figure~7, 
which shows clear sinusoidal variations at both periods.  The weighted 
mean of the ratio of the photometric amplitude in $B$ to the amplitude in 
$V$ is $1.36~\pm~0.13$ for the two periods, consistent with other 
$\gamma$~Doradus variables \citep[e.g.,][]{hfh07} and inconsistent with 
the ellipticity effect or starspots \cite[][their Table~8]{hetal00}.  
Given the star's F1 dwarf classification, the two independent periods in 
the $\gamma$~Doradus period range, and the $B/V$ amplitude ratio, we 
confirm the primary component of HD~6568 as a new $\gamma$~Doradus variable.

\subsection{HD 17163 = HR 816}

The only previous MK spectral classification of HD~17163 is by \citet{am95}, 
who called the star an A9~III.  We estimate an F1: spectral class but find 
that it has a dwarf luminosity.  The broad-lined nature of HD~17163, as 
discussed below, makes our classification more difficult than usual.  The 
star may be somewhat metal rich compared to the Sun.

\citet{df72} reported {\it v}~sin~{\it i} = 90 km~s$^{-1}$, while
\citet{am95} found 108~km~s$^{-1}$, which \citet{retal02a} rescaled to
120 km~s$^{-1}$.  Our projected rotational velocity value of 105 
km~s$^{-1}$ is in reasonable agreement with the past determinations.

HD~17163 was a part of several extensive radial velocity surveys.  Based on 
four spectra, \citet{sa32} reported an average velocity of 19.9 $\pm$ 2.5 
km~s$^{-1}$.  From three observations \citet{h37} gave it a mean velocity 
of 16.0 $\pm$ 4.9 km~s$^{-1}$ and commented that its spectrum had 
numerous fuzzy lines.  \citet{wj50} presented results from three 
additional spectrograms that have dispersions of 36 \AA~mm$^{-1}$ 
\citep{a70}.  Those Mount Wilson observations produce an average velocity 
of 22.0 $\pm$ 7.4 km~s$^{-1}$ and a range of 25 km~s$^{-1}$.  Despite 
noting the broad lines of the star and the reasonable agreement of their
average velocity with previous means, \citet{wj50} concluded that the
velocity of HD~17163 is variable.  Measurement of our three KPNO
spectrograms results in an average velocity of 26.2 $\pm$ 0.3 km~s$^{-1}$,  
so there is a 10 km~s$^{-1}$ range in the average velocities found at 
the four different observatories.  Even so, given the broad, shallow 
lines of this star and the pulsation that we have discovered, there is 
no strong evidence of velocity variability resulting from orbital motion, 
and so we conclude that the star is single.
 
The {\it Hipparcos} satellite observed HD~17163 58 times during the course 
of its 3.5 year mission, resulting in its classification as a constant star 
\citep{petal97}.  Nonetheless, our T12 survey observed it 53 times and found 
it to be slightly variable with a standard deviation of 0.0030 mag (Table~2).

We find a single period of 0.42351 days in the T3 APT follow-up photometry
with a $B$ amplitude of 8 mmag (Figs. 8 and 9; Table~5).  The light
curve approximates a sinusoid when phased with this period. The $B/V$ 
amplitude ratio is $1.16~\pm~0.20$, consistent with pulsation but also 
consistent with ellipsoidal variation and the rotational modulation of 
starspots due to the large uncertainty of the amplitude ratio.  Ellipsoidal 
variation in a close binary system is precluded by the fact that our 
spectroscopy does not reveal large radial velocity changes.  Likewise, 
starspot activity (implying a stellar rotation period of 0.42351 days) is 
unlikely because of the star's early spectral type and the coherency of 
the light curve over hundreds of cycles.  As a result, we conclude that 
the photometric variability is due to pulsation and confirm this F1 dwarf 
as a new $\gamma$~Doradus star.

\subsection{HD 25906 = HIP 19408}

To compile basic data for stars that were included in the {\it Hipparcos} 
mission, \citet{detal95} acquired six objective-prism spectra with a 
dispersion of 80~\AA~mm$^{-1}$.  From those plates they measured a mean 
radial velocity of 37 km~s$^{-1}$ with an rms of 4.2 km~s$^{-1}$.

From our spectrograms we have determined an F1 spectral class for HD~25906, 
and its {\it Hipparcos} parallax \citep{petal97} results in a dwarf 
luminosity class.  The star has a moderate projected rotational velocity of 
64 km~s$^{-1}$.  Our seven observations show modest night-to-night radial 
velocity variations, and their mean velocity is 6.5 $\pm$ 1.8 km~s$^{-1}$.  
The lines of the two spectra with the largest velocity differences have 
significant asymmetries, which presumably result from pulsation.  However, 
our average velocity, with observations from five different epochs, differs 
by 30 km~s$^{-1}$ from the mean of \citet{detal95}.  Those earlier 
velocities are from low dispersion spectra, and so high precision is 
not expected.  Indeed, the individual velocities, which span a three yr 
period, range from 25 to 53 km~s$^{-1}$ and differ on consecutive nights by 
16 km~s$^{-1}$ \citep{detal95}.  Their velocity range does not overlap ours 
(Table~7), and so we conclude that the star is probably a spectroscopic 
binary but note that additional observations will be needed to confirm 
its status.  In Table~6 we list HD~25906 as having a variable orbital 
velocity, but append a question mark to indicate our uncertainty. 

The {\it Hipparcos} mission obtained 119 photometric measurements of
HD~25906, but, based on those observations, the star could not be 
classified as either constant or variable \citep{petal97}.  Our 249 T12 survey 
observations indicate a brightness variability of 0.0050 mag (Table~2).  
Analysis of the T3 follow-up observations found three independent periods
of 0.79164, 0.81646, and 1.34192 days with Johnson $B$ amplitudes of
9, 6, and 6 mmag, respectively (Figs. 10 and 11; Table~5).  The light
curves at all three periods are sinusoidal, and the weighted mean of
the three $B/V$ amplitude ratios from Table~7 is $1.26~\pm~0.15$.  
Therefore, based on the star's F1 dwarf spectral type, asymmetric 
spectroscopic line profiles, multiple photometric periods in the 
$\gamma$~Doradus period range, and its $B/V$ amplitude ratio, we confirm 
HD~25906 as a new $\gamma$~Doradus variable.

\subsection{HD 31550 = HIP 23117}

We obtained five spectra of HD~31550 and find it to be a F0 dwarf.  The 
line strengths of HD~31550, compared with those of our reference stars, 
suggest that HD~31550 may be modestly metal poor compared to the Sun.  
Our mean projected rotational velocity is 32 km~s$^{-1}$.  Our five 
velocities have an average of 17.6 $\pm$ 0.5 km~s$^{-1}$ and show no 
evidence of orbital motion.  Earlier, \citet{metal01b} obtained a lone
speckle observation of HD~31550 and found no duplicity.  We conclude that 
the star is single.

The {\it Hipparcos} catalog has 60 photometric observations but does not
claim intrinsic variability \citep{petal97}.  Our T12 survey obtained 247
brightness measurements that scattered about their mean with a standard 
deviation of only 0.0020 mag (Table~2), equal to our defined lower limit 
of detectable variability.  Even so, we find a variability period of 
1.49566 days in the T3 photometry with a $B$ amplitude of 7 mmag 
(Figs. 12 and 13; Table~5).  The light curve is sinusoidal when phased 
with this period, although it is badly binned in phase because the period 
is almost exactly 1.5 days.  The $B/V$ amplitude ratio is $1.13~\pm~0.21$, 
which is consistent with both ellipsoidal and spot variation as well 
as pulsation.  However, ellipsoidal variation in a close binary system 
can be ruled out by the fact that our spectroscopy, with observations
taken both one and two days apart, does not reveal large radial velocity 
changes.  Likewise, starspot activity is unlikely because of the early 
spectral type of the star and the coherency of the light curve over hundreds 
of cycles.  We conclude that the photometric variability in HD~31550 is 
due to pulsation and confirm this F1 dwarf as a new $\gamma$~Doradus star.

\subsection{HD 38309 = HR 1978 = ADS 4333A = HIP 27118}

HD 38309 is component A of a visual triple system with components
B and C separated from A by 17\arcsec.  According to the
notes of The Bright Star Catalogue \citep{h82}, these two additional 
components are optical companions of component A.  \citet{am95} classified 
HD~38309 as F0~IV.  \citet{df72} found a {\it v}~sin~{\it i} value 
of 100 km~s$^{-1}$, which \citet{ws97} revised to 86 km~s$^{-1}$. 
\citet{am95} estimated a projected rotational velocity of 75 km~s$^{-1}$, 
which \citet{retal02a} rescaled to 86 km~s$^{-1}$.  \citet{a70} listed 
a single Mount Wilson Observatory radial velocity of 9.0 km~s$^{-1}$, 
while \citet{sa32} found a similar mean velocity of 7.0 $\pm$ 2.1 km~s$^{-1}$ 
from four plates.

HD 38309 is listed in Part G, the Acceleration Solution section, of the 
Double and Multiple Star Appendix of the {\it Hipparcos} catalog 
\citep{petal97}. Unlike most of the stars observed by {\it Hipparcos},
the successive positions of HD~38309 during the three yr mission could 
not be satisfactorily modeled with just five parameters (position, 
parallax, and proper motion).  Instead, a seven parameter model that 
included two additional terms for acceleration, which is the 
time derivative of the proper motion, was adopted to account for a 
curvature in the stellar trajectory. A star with such a solution 
often turns out to have a companion with an orbital period that is 
too long to be derived from only the {\it Hipparcos} data.

Our eight red-wavelength spectra show a composite spectrum as does a
single spectrum in the ELODIE archive \citep{metal04b}.  Each line 
of HD~38309 consists of a broad component with a weak, narrow 
component near its center (Figure~2).  As suggested by its listing 
in Part G of the {\it Hipparcos} catalog, HD~38309 is indeed a binary.  
The broad-lined star, component Aa, we classify as F1, while the spectral 
class of the narrow-lined component, Ab, is G1:, and both stars are  
dwarfs.  Guided by our spectrum addition fits, we adopt a magnitude
difference, $\Delta$$V$, of 2.1.  We also measure projected rotational 
velocities of 95: and 5: km~s$^{-1}$, for components Aa and Ab, 
respectively, so our {\it v}~sin~{\it i} value of component Aa is 
similar to those previously determined.

To date, we have obtained nine radial velocity observations of HD~38309 at
eight different epochs that span four yr.  Radial velocities of component Ab,
the narrow-lined star, show a systematic velocity decrease of nearly
5 km~s$^{-1}$ (Table~7) during that period, indicating that it is a
spectroscopic binary.  The velocity of component Aa also shows variability,
but because it has an early-F spectral type, its velocity changes are a
combination of pulsation plus orbital motion. We assume that Aa is the
binary companion of Ab.

The {\it Hipparcos} catalog lists 66 brightness measurements for HD~38309 
but is unable to classify it as constant or variable.  The 214 brightness 
measurements from our T12 survey (Table~2) have a standard deviation of 
0.0033 mag and thereby show that HD~38309 is, indeed, a low-amplitude 
variable.  Analysis of the follow-up observations from the T3 APT revealed 
three independent periods of 0.37703, 0.35993, and 0.34713 days with $B$ 
amplitudes of 9, 7, and 5 mmag, respectively (Figs. 14 and 15; Table~5).  
The light curves at all three periods are sinusoidal, and the weighted 
mean of the three $B/V$ amplitude ratios from Table~5 is $1.07~\pm~0.14$.  
This is somewhat low for a $\gamma$~Doradus variable, but we note that 
our photometric measurements (made with a $55\arcsec$ diaphragm) were 
diluted by the G1 dwarf spectroscopic companion as well as by the two 
visual components B and C.  Consequently, the $B/V$ amplitude ratio is 
somewhat biased.  Based on the star's F1 dwarf spectral type, asymmetric 
spectroscopic line profiles, and multiple photometric periods in the 
$\gamma$~Doradus period range, we confirm component Aa of HD~38309 as a 
new $\gamma$~Doradus variable.

\subsection{HD 45638 = HR 2351}

\citet{cf74} called HD~45638 an A9~IV star, while \citet{am95} similarly 
classified the star as F0~IV.  \citet{gg89} designated it a mild
Am star with spectral classes of A9/F0/F2 for the Ca K line, hydrogen lines, 
and metal lines, respectively.  Our spectrum of the metal lines in the 
6430~\AA\ region is best fitted by an A9 reference star, and the Ca~I lines
in our spectrum are consistent with that classification. 
We find that the star is a dwarf, making it less luminous than previous
classifications.

\citet{am95} found a {\it v}~sin~{\it i} value of 35 km~s$^{-1}$, which
was revised by \citet{retal02a} to 44 km~s$^{-1}$.  Our average projected
rotational velocity is 38 km~s$^{-1}$ and in reasonable agreement 
with both results.

The only previous velocities of HD~45638 are those of \citet{y45}, who
included it in a radial velocity survey of 681 stars at the David Dunlap 
Observatory. He determined a mean velocity of 40.6 $\pm$ 1.3 km~s$^{-1}$ 
from four plates.  Our four velocities are in excellent agreement,
having an average of 42.4 $\pm$ 0.4 km~s$^{-1}$; the star appears to 
be single.

The results of our photometric survey with the T12 APT are given in Table~2.  
We derived a standard deviation of 0.0039 mag from our 160 brightness 
measurements, indicating that HD~45638 is a low-amplitude variable.  The
{\it Hipparcos} catalog contains only 43 brightness measurements and does
not rule on the constancy or variability of the star.  Analysis of our 
follow-up observations from T3 gives a single period of 0.86046 day with a $B$
amplitude of 11 mmag (Figs. 16 and 17; Table~5).  The sinusoidal light
variations have a $B/V$ amplitude ratio of $1.41\pm0.15$ mag, which is 
consistent with pulsation.  Ellipticity as the cause of the light variability 
is ruled out by the observed lack of radial velocity variations as well as 
by the amplitude ratio.  We confirm HD~45638 as a new $\gamma$~Doradus
variable.

\subsection{HD 62196 = HIP 37802}

From its Str\"omgren photometric indices, \citet{o80} concluded that 
HD~62196 is a metal-weak F star, and, in a major survey of F and
G dwarfs in the solar neighborhood, \citet{netal04} computed 
[Fe/H] = $-$0.64 from those observations.  \citet{jetal89} included it 
in a spectroscopic survey of 350 F-type stars.  Using moderate dispersion 
spectra, they found it to be metal weak, classifying its hydrogen lines 
as F3 but giving its metal lines a spectral class of A9.  We determine 
that the star is an F2 dwarf and estimate [Fe/H] $\sim$$-$0.5 from the 
reference star fits to our spectrum in the 6430~\AA\ region. Thus, we 
concur that the star is metal weak.  

\citet{fetal87a} obtained four objective prism spectra of HD~62196 and found
an average velocity of $-$3 km~s$^{-1}$.  Over a timespan of 11 years 
\citet{netal04} obtained eight radial velocities that have a mean  
of $-$9.0 km~s$^{-1}$ but a large standard deviation of 5.1 km~s$^{-1}$, 
indicating that this star is a spectroscopic binary.  They also determined 
a {\it v}~sin~{\it i} value of 7 km~s$^{-1}$ in agreement with an 
average {\it v}~sin~{\it i} of 6 km~s$^{-1}$ determined from our spectra.  
Our eight radial velocities, acquired at five epochs, support the 
conclusion that HD 62196 is a spectroscopic binary and suggest a period
of $\sim$3.3 yr.  We are obtaining additional observations to determine an 
orbit.

The {\it Hipparcos} catalog has only 37 brightness measurements of HD~62196 
and does not classify the star as either constant or variable.  Our T12 
survey, however, reveals brightness variability of 0.0049 mag (Table~2). 
Analysis of our T3 follow-up photometry reveals two closely-spaced periods 
of 1.00341 and 0.99236 days with $B$ amplitudes of 14 and 13 mmag, 
respectively (Figs. 18 and 19; Table~5).  The prewhitened light curves in 
Figure~19 both exhibit sinusoidal light variations of nearly identical 
amplitudes.  Ellipticity as the cause of the light variability is ruled out 
by the observed lack of short-term radial velocity variations as well as 
by the presence of two independent periods.  Brightness modulation by 
starspots is ruled out by the large weighted mean $B/V$ amplitude ratio of 
$1.78\pm0.10$ mag.  We are puzzled by this large amplitude ratio, which is 
much larger than the typical value of $\sim1.3$ for $\gamma$~Dor stars.  A 
cooler companion (implied by the radial velocity changes) would dilute the 
$V$ amplitudes more than the $B$, but this dilution would be very subtle 
since the companion's light is not seen in our spectra.  As noted above, 
HD~62196 is metal poor and is also the faintest variable star among the 
$\gamma$~Doradus candidates in Figure~1.  We confirm HD~62196 as a new 
$\gamma$~Doradus variable.

\subsection{HD 63436 = DD CMi}

\citet{metal04} included HD~63436 in a spectroscopic survey of $\gamma$
Doradus candidates. They determined {\it v}~sin~{\it i} = 66 km~s$^{-1}$
and noted line-profile variations in the blue wings of the lines.
Our projected rotational velocity of 70~km~s$^{-1}$ is similar to theirs.
From seven observations \citet{metal04} found the velocity to range 
from $-$15 to $-$7 km~s$^{-1}$.  Our five velocities, obtained at four 
different epochs, cover approximately the same velocity range and have 
a mean velocity of $-$13.7 $\pm$ 1.3  km~s$^{-1}$.  The velocity variations
are probably due to pulsation.  Using a probability neural network
technique, \citet{mahdi08} classified HD 63436 as A9~V.  We determine a 
spectral class of F2 and conclude that the star is a dwarf.  

The {\it Hipparcos} catalog lists 51 brightness measurements for HD~63436 and
identifies it as an ``Unsolved Variable'', recognizing that it has significant
brightness variability, but the authors were unable to classify its 
variability type.  Analyzing these observations, \citet{han99} suggested 
HD~63436 as a possible $\gamma$ Doradus variable.  He found several plausible 
periods near 0.7 days but noted that the data produced a weak signal.  
\citet{metal03} obtained five nights of photometry of HD~63436.  Their 
analysis of that data and also a reanalysis of the {\it Hipparcos} photometry 
led to two periods near 0.7 days, so they identified it as a $\gamma$ Doradus 
variable.  Based on those results, \citet{ketal06} gave it the variable star 
name DD~CMi and assigned it to the $\gamma$ Doradus class of variables.

The 38 brightness measurements from the T12 survey (Table~2) have a 
standard deviation of 0.0271 mag, the largest of all the 114 survey 
stars.  Analysis of the follow-up observations from the T3 APT reveals 
four independent periods of 0.68695, 0.71327, 0.69238, and 0.54431 days 
with Johnson $B$ amplitudes of 54, 35, 39, and 24 mmag, respectively 
(Figs. 20 and 21; Table~5).  Two additional, periods of 0.72238 and 
0.60310 days are suspected in the $B$ data but could not be confirmed in 
the $V$ observations; therefore, these two periods are not included in 
Table~5.  The light curves of all four confirmed periods are sinusoidal, 
and the weighted mean of the four $B/V$ amplitude ratios from Table~5 is 
$1.33~\pm~0.12$, very typical for $\gamma$~Dor stars. Therefore, based on 
the star's F1 dwarf spectral type, variable spectroscopic line profiles, 
multiple photometric periods in the $\gamma$~Doradus period range, and 
the $B/V$ ratio of 1.33, we confirm the earlier classification of HD~63436 
as a $\gamma$~Doradus variable.

\subsection{HD 65526 = V769 Mon}

HD 65526 was observed as part of two extensive spectroscopic surveys of 
candidate and confirmed $\gamma$ Doradus stars with the goal of 
characterizing the line-profile variations that result from pulsations.
\citet{metal04} obtained a single spectrum of HD~65526 and did not detect
line-profile asymmetries.  However, the cross correlation functions 
of two spectra obtained by \citet{dcetal06} show clear profile variations. 
\citet{metal04} determined a {\it v}~sin~{\it i} value of 59 km~s$^{-1}$,
while \citet{dcetal06} found a mean value of 56 km~s$^{-1}$. 

We have determined that the star is an A9 dwarf that may be slightly metal
poor compared to the Sun.  Our three spectroscopic observations produce a 
constant velocity of $-$6.0 $\pm$ 0.5 km~s$^{-1}$, and we assume that the 
star is single.  Our projected rotational velocity of 59 km~s$^{-1}$ is in 
excellent agreement with the two previous determinations.

Periodic light variations were found by the {\it Hipparcos} team, who
determined a period of 1.288 day \citep{petal97}.  As a result, the star 
was given the designation V769~Mon by \citet{kaz99}, who assigned it,
somewhat uncertainly, to the $\beta$ Lyrae class of eclipsing binaries. 
\citet{han99} honed in on the correct variability class and 
photometric period when he identified HD~65526 as a prime $\gamma$ Doradus 
candidate and listed two periods of 0.644 and 0.598 days, determined from 
the {\it Hipparcos} photometry.  \citet{hs02} obtained two nights of 
follow-up photometry of HD~65526 that demonstrated pulsation as the cause 
of its multiperiodic light variations.  They concluded that HD~65526 is 
indeed a $\gamma$ Doradus variable. \citet{metal03} obtained five nights 
of photometry and determined a period of 0.6427 days. 

The 47 brightness measurements from our T12 survey (Table~2) have a 
standard deviation of 0.0233 mag, showing that HD~65526 is among the most
variable stars in the survey.  Analysis of the follow-up observations 
from the T3 APT reveals three independent periods of 0.64404, 0.59755, 
and 0.58476 days with Johnson $B$ amplitudes of 52, 21, and 23 mmag, 
respectively (Figs. 22 and 23; Table~5).  These periods are in excellent 
agreement with the various determinations of \citet{han99} and 
\citet{metal03}.  The light curves are sinusoidal at all three periods; 
the weighted mean of the three $B/V$ amplitude ratios from Table~5 is 
$1.23~\pm~0.09$.  Based on the star's A9 dwarf spectral classification, 
the observed line-profile variations, the photometric periods, and the 
$B/V$ amplitude ratios, we confirm previous claims that HD~65526 is a 
$\gamma$~Doradus star.

\subsection{HD 69682 = HR 3258}

\citet{cc65} classified HD~69682 as F0~IV as did \citet{am95}, while 
\citet{cf74} called it F0~V.  Our result of A9 dwarf is in good agreement
with those three classifications.  When its spectrum is compared to that of 
HR~1613, an A9~V star \citep{am95} with solar abundances \citep{betal02}, 
we conclude that HD~69682 is metal rich.  \citet{df72} found a projected 
rotational velocity of 35 km~s$^{-1}$, which \citet{ws97} rescaled to 
26 km~s$^{-1}$.  \citet{am95} determined a value of 25 km~s$^{-1}$, and
\citet{retal02a} measured a value of 28 km~s$^{-1}$.  Our average 
$v$~sin~$i$ value of 30 km~s$^{-1}$ is in agreement with those results.

Despite its listing in The Bright Star Catalog \citep{h82}, there
are only two previous determinations of the mean radial velocity
of HD 69682.  \citet{y45} reported an average velocity of 10.0 $\pm$
1.0 km~s$^{-1}$ from the measurement of four photographic plates.
From three slit spectra with a dispersion of 80~\AA~mm$^{-1}$
\citet{fetal87b} determined a mean velocity for HD~69682,
which \citet{getal99a} revised to 7.9 $\pm$ 3.1 km~s$^{-1}$.
Our three spectra have an average velocity of 11.4 $\pm 0.3$
km~s$^{-1}$.  The concordance of the results argues that HD~69682 
has no significant velocity variability other than that expected for 
a $\gamma$~Doradus variable, and so we assume that the star is single.

The {\it Hipparcos} catalog lists 89 brightness measurements of HD~69682 
\citep{petal97} but is unable to classify the star as constant or slightly 
variable.  Our 54 T12 survey observations (Table~2) reveal definite 
variability with a standard deviation of 0.0057 mag.  We have over four
hundred follow-up observations with the T3 APT.  We find two closely-spaced
periods of 0.53189 and 0.47703 days (Figs. 24 and 25; Table~5) with
Johnson $B$ amplitudes of 7 and 5 mmag, respectively; the same two 
periods are found in the $V$ band.  Despite the low amplitudes, clear 
sinusoidal variations are seen in Figure~25 at both periods.  The weighted 
mean $B/V$ amplitude ratio for the two periods is $1.28~\pm~0.18$, consistent 
with other $\gamma$~Doradus variables.  Given the star's A9 dwarf 
classification, low-amplitude radial velocity variations, two periods in the 
range of other $\gamma$~Doradus variables, and a $B/V$ amplitude ratio of 
1.28, we confirm HD~69682 as a new $\gamma$~Doradus variable.

\subsection{HD 99267 = HIP 55766}

\citet{y39} included HD~99267 in a radial velocity survey of 500 stars
made at David Dunlap Observatory in the 1930s.  The five observations
have a mean velocity of $-$4.8 km~s$^{-1}$ and show a velocity range of 
30 km~s$^{-1}$.  Young concluded that the velocity was probably variable
but also remarked that the spectrum was difficult to measure; he gave a 
spectral class of A8.  \citet{hfh95} classified the star as F1~V
and from two velocities obtained on the same night argued that the star
was not a binary with a very short-period.  From a single spectrum 
\citet{f97} estimated a projected rotational velocity of 89 km~s$^{-1}$.

For HD~99267 we have a total of five spectra, including the two mentioned 
by \citet{hfh95}.  Our velocities have an average of 2.9 $\pm$ 0.4 
km~s$^{-1}$ and a velocity range of only 2.5 km~s$^{-1}$.  Despite the 
conclusion of \citet{y39} that the velocity is probably variable, our 
higher resolution observations suggest that the star is single, with its 
very modest velocity variations consistent with pulsation.   We determine 
a projected rotational velocity of 95 km~s$^{-1}$ in reasonable agreement 
with that of \citet{f97}.  A reexamination of our spectra of HD~99267 
with an expanded set of reference stars and the {\it Hipparcos} parallax 
\citep{petal97} produce the same spectral classification that 
was found by \citet{hfh95}, an F1 dwarf.  Our comparison also suggests 
that HD~99267 is somewhat metal-rich relative to the Sun. 

Light variability in HD~99267 was first noted by \citet{hfh95}. They 
found its brightness to change by 0.03 mag with a preliminary period of 
$0.575\pm0.001$ days but did not know the cause of variability.  At that 
time, the $\gamma$~Dor stars were not yet recognized as a new variability 
class, although a few early-F, main-sequence stars with short-period, 
low-amplitude variability were known to have similar properties 
\citep[e.g.,][]{ketal93,mpz94}.

The {\it Hipparcos} satellite confirmed the low-amplitude brightness 
variability from its 117 observations, but the analysis team was unable 
to determine a period.  The 53 brightness measurements from our T12 survey 
(Table~2) have a standard deviation of 0.0166 mag, providing another 
confirmation of HD~99267's variability.  Our analysis of the T3 measurements 
resolves a cluster of five independent periods:  0.57465, 0.56529, 0.56173, 
0.58820, and 0.47553 days with Johnson $B$ amplitudes of 25, 19, 20, 20, and 
5 mmag, respectively (Figs. 26 and 27; Table~5).  The first and 
highest-amplitude periodicity matches the 0.575-day period found by 
\citet{hfh95}.  The light curves are sinusoidal at all five periods; the 
weighted mean of the five $B/V$ amplitude ratios from Table~5 is 
$1.23~\pm~0.11$.  Given the star's F1~V spectral classification, its 
low-amplitude velocity variations, the five photometric periods, and the 
$B/V$ amplitude ratios, we confirm that HD~99267 is a $\gamma$~Doradus star.

\subsection{HD 114447 = 17 CVn = HR 4971 = ADS 8805A}

The spectral classification of this moderately bright star has been 
determined a number of times.  \citet{a67} found A9~III-IV, 
\citet{a81} classified it as A9~IV, and \citet{hetal76}, \citet{c76},
and \citet{am95} called it an F0~V, while \citet{gg89} similarly 
determined F0~IV-V.  In addition to their spectral classification of 
HD 114447, \citet{am95} found a {\it v}~sin~{\it i} value of 71 km~s$^{-1}$, 
which \citet{retal02a} revised to 82 km~s$^{-1}$.

In a search for possible duplicity, HD~114447 was one of 85 stars 
observed interferometrically by \citet{m22} with the 2.5 m Hooker 
telescope at Mount Wilson Observatory.  Like the vast majority of 
program stars that he observed, Merrill concluded that HD~114447 was 
apparently single.  Four radial velocities, obtained at Yerkes 
Observatory by \citet{fetal29} as part of a survey of 500 A stars, 
resulted in an average velocity of 0.9 km~s$^{-1}$ and an observed 
range of 22 km~s$^{-1}$, leading \citet{fetal29} to conclude that 
this star is ``perhaps a spectroscopic binary.'' \citet{hetal76} 
included HD~114447 in their study of A and F stars in the region of  
the north galactic pole.  They obtained 17 spectrograms, the vast 
majority of which had a dispersion of 80 \AA~mm$^{-1}$.  They determined 
a mean radial velocity of $-$20.2 $\pm$ 2.9 km~s$^{-1}$, but also stated 
that the star is a possible double-lined spectroscopic binary.  

We have acquired six spectra at three different epochs.  The first
two spectra, obtained within two days of each other, show  
the lines of two components that have very different radial 
velocities (Figure~3).  However, because of the significant line 
broadening of both components, we estimate {\it v}~sin~{\it i}
values of 50: and 90: km~s$^{-1}$ for Aa and Ab, respectively,
the line pairs are partially blended with each other.  In our 
last four spectra the lines of Aa and Ab are completely blended with each 
other.  Thus, our observations confirm the conclusion of \citet{hetal76} 
that HD~114447 is a double-lined binary.  We analyzed one of 
the first two spectra and determined spectral classes of F0: for 
both stars.  The {\it Hipparcos} parallax \citep{petal97} leads to 
a dwarf luminosity class for both components.  Therefore, our results
are in excellent agreement with most other classifications.   
Although the two components have very different projected rotational 
velocities, the magnitude difference between Aa and Ab is small.  
From our spectrum addition fits, $\Delta$$V$ is 0.2 with an estimated 
uncertainty of 0.1 mag.  

\citet{p78} searched for brightness variations in HD~114447 in 10.4 hours 
of Johnson $V$ photometry acquired on four nights and concluded that any 
photometric variations were less than 0.01 mag.  Similarly, no definitive 
variability was detected in the 114 {\it Hipparcos} observations 
\citep{petal97}.  We collected 81 brightness observations during our T12 APT
survey and detected clear variability with a standard deviation of 
0.0074 mag (Table~2).  We find three distinct and independent periods in
the follow-up observations with the T3 APT:  0.88621, 0.74705, and
0.68970 days.  The Johnson $B$ amplitudes are 16, 16, and 11 mmag,
respectively (Figs. 28 and 29; Table~5).  The light curve is sinusoidal
at all three periods, and the weighted mean of the three $B/V$ amplitude 
ratios from Table~5 is $1.23~\pm~0.11$.  Therefore, we can confirm that 
at least one of the F0:~V components, Aa and Ab of HD~114447 A, is a
$\gamma$~Doradus variable. 

\subsection{HD 138936 = HR 5791}

\citet{am95} classified HD~138936 as an A9~V star, while \citet{gg89}
found A9~IV-Vs.  We also determine that the star is an A9 dwarf, which is 
in excellent agreement with the previous results.  \citet{df72} measured 
a {\it v}~sin~{\it i} value of 81 km~s$^{-1}$ that \citet{ws97} rescaled 
to 72 km~s$^{-1}$.  \citet{am95} also determined a value of 81 km~s$^{-1}$ 
that was revised by \citet{retal02a} to 92 km~s$^{-1}$.  More recently, 
\citet{jetal06} determined an even larger value of 100 km~s$^{-1}$ from a 
spectrum synthesis fit to their spectrum of the lithium region at 6700~\AA.  
Our value of 65 km~s$^{-1}$ is less than that of the two revised values 
and substantially less than that of \citet{jetal06}.   

HD~138936 was part of a large survey of bright stars carried out at the 
David Dunlap Observatory \citep{y45}. Young reported, based on five 
spectrograms, that it had an average velocity of 
$-$19.5 $\pm$ 3.0 km~s$^{-1}$.  Our three velocities have an average 
of $-$24.5 $\pm$ 1.0 km~s$^{-1}$, in accord with the result of Young.  
Obtaining a single speckle observation, \citet{metal01b} found no evidence of 
duplicity.  Despite the range of {\it v}~sin~{\it i} values, which might 
result from a double-lined binary with its lines seen blended at various 
phases, we assume that HD~138936 is a single star.  Additional spectra will 
be needed to support this conclusion.

The {\it Hipparcos} satellite made 77 brightness measurements of
HD~138936, and \citet{petal97} classified it as a constant star.  During
our initial survey, the T12 APT acquired 45 measurements with a standard 
deviation of only 0.0029 mag, just above our 0.0020 mag cutoff for claiming
variability (\S2.3).  T3 followed up with over 200 brightness measurements.  
Our analysis of the T3 data reveals three distinct periods of 0.41920, 
0.41635, and 0.45790 days with $B$ amplitudes of 8, 5, and 5 mmag, 
respectively (Figs. 30 and 31; Table~5).  The light curve appears to be 
slightly asymmetric when phased with the first of the three periods.  The mean 
$B/V$ amplitude ratio is $1.33\pm0.16$.  Given these properties, we confirm 
that HD~138936 is a $\gamma$~Doradus star.

\subsection{HD 139478 = HR 5817}

With its $V$ magnitude of 6.70 \citep{petal97}, HD 139478 is a 
star that is fainter than the nominal 6.5 mag limit for inclusion in The 
Bright Star Catalogue \citep{h82}. As a result of its listing
in that prominent catalog, a number of its basic properties have been 
previously determined.  \citet{c76} classified HD~139478 as F2 pec and 
stated that its metal line ratios are peculiar with strontium being 
enhanced.  In contrast, later classifications of F4~III, F0~V, and F1~IV by 
\citet{cb79}, \citet{am95}, and \citet{getal01}, respectively, indicate 
a normal spectrum.  We conclude from our work that the star is a F1
dwarf. 

HD~139478 was part of a sample of 917 mostly A-type stars observed at 
the Dominion Astrophysical Observatory (DAO).  As a result, \citet{h37} 
reported a mean velocity of $-$16.9 km~s$^{-1}$ from three plates and 
noted that ``the numerous narrow lines matched against Procyon standard 
are reliable.''  Two additional velocities, obtained by \citet{netal04}, 
have an average of $-$16.2 km~s$^{-1}$ and so are in excellent agreement 
with the DAO mean velocity.  Our five velocities, obtained at four different 
epochs, produce an average velocity of $-$15.7 $\pm$ 0.1 km~s$^{-1}$, 
enhancing the consensus that this star is single.  \citet{metal87} examined 
the binary star frequency of 672 relatively bright stars, using speckle 
interferometry.   They found no evidence that HD~139478 is a double star.

\citet{am95} measured a {\it v}~sin~{\it i} value of 61 km~s$^{-1}$, which
\citet{retal02a} rescaled to 71 km~s$^{-1}$.  The projected 
rotational velocity determined by \citet{netal04} is substantially smaller, 
19 km~s$^{-1}$, and in excellent agreement with our result of 20.5 
km~s$^{-1}$.  At our request H. Abt (2007, private communication) examined 
his classification dispersion spectrum of HD~139478 and reported that 
the spectrum shows sharp lines that appear inconsistent with the 
{\it v}~sin~{\it i} value given in \citet{am95}, which was based on the
measurement of a different spectrum.  This very substantial 
difference between the value of \citet{am95} and other results, including 
the comments of \citet{h37}, begs the question of whether HD~139478 is a 
double-lined binary that, except for one observation, has been observed 
at phases that are close to its center of mass velocity, producing a
single narrow set of lines.  If it is a binary, it apparently has 
a rather eccentric orbit, since all radial velocities and all but one 
{\it v}~sin~{\it i} value are in excellent accord.  Given the weight of 
the evidence, there is a possibility that the star is a double-lined 
spectroscopic binary, but we currently assume that HD~139478 is a single 
star.  As with HD~138936, additional observations will be needed to confirm 
this conclusion.

HD~139478 has been included in several photometric surveys.  \citet{b69} 
observed it in a search for $\delta$~Scuti-type pulsating stars.  He 
acquired photometry for 3.7 hours but found it constant to the precision 
of his measurements.  The {\it Hipparcos} satellite acquired 107 brightness 
measurements over the duration of the mission, but the {\it Hipparcos} 
team \citep{petal97} left its variability undecided.  \citet{hp99} 
included HD~139478 in a survey for rapidly oscillating Ap stars.  They 
monitored the star for 1.34 hours on one night and found no obvious 
periodicity.  When \citet{ke02} reanalyzed the {\it Hipparcos} 
photometry in \citet{petal97}, they found a possible low-amplitude 
variability (9 mmag) with a period of 0.6578 days.

Our initial T12 survey easily detected photometric variability, gathering 
50 measurements with a standard deviation of 0.0103 mag.  T3 followed up 
with over 300 observations that revealed three distinct, closely-spaced 
periods of 0.68818, 0.71058, and 0.66419 days with $B$ amplitudes of 24, 18, 
and 18 mmag, respectively (Figs. 32 and 33; Table~5).  The third and 
shortest of the three periods, 0.66419 days, is only 0.00639 days longer 
than the period cited by \citet{ke02}.  The T3 light curve is sinusoidal 
when phased with each of the three periods.  The mean $B/V$ amplitude ratio 
is $1.33\pm0.11$.  Thus, we confirm that HD~139478 is a $\gamma$~Doradus 
variable.

\subsection{HD 145005 = HIP 79122}

\citet{getal99b} determined a mean radial velocity of $-$10.9 km~s$^{-1}$ 
from two spectrograms and called the velocity variable based on the 
4.6 km~s$^{-1}$ difference between the two observations.  They also 
measured a {\it v}~sin~{\it i} value of 50 km~s$^{-1}$ as did 
\citet{retal02b}, who analyzed the same spectra with a different 
technique.  \citet{mk05} identified HD~145005 as one of 1929 stars 
with {\it Hipparcos} and Tycho proper motions that differ by more than 
3.5 $\sigma$ in at least one coordinate.  They designated such stars 
as $\Delta$$\mu$ binaries. Our spectrograms show that, like HD~38309, 
HD~145005 is a star that has a composite spectrum with a weak, 
narrow line superposed near the center of each stronger, broad line 
(Figure~4), so we can confirm that HD~145005 is a binary.  Our spectral 
classes for the broad-lined and narrow-lined stars, components A and B, 
are F0 and G4:, respectively.  Their positions in an H-R diagram indicate 
that both stars are dwarfs.  Our spectrum-addition fits provide an estimated 
$\Delta$$V$ = 2.8 mag.  We measure projected rotational velocities of 57 
and 5: for the primary and secondary, respectively.  During the 2.9 
years covered by our observations, we have obtained seven spectra at 
six different epochs.  Over that time span, the velocities of 
the two stars have remained nearly constant, $-$9.3 $\pm$ 0.4 and 
$-$11.0 $\pm$ 0.3 km~s$^{-1}$ for components A and B, respectively.
The two mean velocities differ by less than 2 km~s$^{-1}$, and the velocities 
of both components are similar to the average velocity of \citet{getal99b}.

The 50 {\it Hipparcos} observations from \citet{petal97} did not result in a 
definitive determination of whether the star was constant or variable.  
In contrast, our 45 T12 survey observations indicate clear variability with a 
standard deviation of 0.0038 mag (Table~2).  We acquired approximately 150 
additional observations with the T3 APT that revealed a single period of 
0.46570 days with a $B$ amplitude of 12 mmag (Figs. 34 and 35; Table~5).  
We note that the one-day alias of the 0.46570-day period at a frequency of 
3.1494~day$^{-1}$ is nearly as strong as the 0.46570-day period.  We 
believe we have identified the true period since both the $B$ and $V$ 
data sets converged to that period in an iterative process of removing the 
obvious outliers and redetermining the period.  Figure~35 shows that the 
light curve is sinusoidal; the $B/V$ amplitude ratio is 1.25.  The 
existence of only one period allows the possibility that the light 
variability may be due to the ellipticity effect in a close binary or to 
rotational modulation of the visibility of dark starspots.  The constancy 
of our radial velocities rules out a short-period binary, while the F0 
spectral type and coherency of the low-amplitude photometric variability 
over hundreds of cycles rule out starspots.  Therefore, the variability
must be due to pulsation, and we confirm the brighter F0 component of 
HD~145005 as a new $\gamma$~Doradus star.

\subsection{HD 220091 = HIP 115288}

\citet{l66} provided a spectral type of A9~III for HD~220091 from 
objective prism observations and obtained a mean radial velocity of 
$-$3 km~s$^{-1}$.  Earlier, \citet{y39} had found an average radial 
velocity of $-$19.6 km~s$^{-1}$ from five plates.  Based partially 
on the latter velocity, \citet{metal01a} included the star as a 
candidate for membership in the young IC~2391 supercluster but 
ultimately concluded that it was not a member.  In a survey of 
X-ray emission of A-type stars, \citet{setal95} identified
HD~220091 as having a soft X-ray luminosity level well above that of 
the active Sun.  Because only a small number of A stars have been 
identified as possible X-ray sources, \citet{setal95} urged that a 
search be made for a cooler companion, which would more likely 
be the X-ray emitter.

Like HD~38309, HD 220091 is listed in Part G, the Acceleration 
Solution section, of the Double and Multiple Star Appendix of the 
{\it Hipparcos} catalog \citep{petal97}, indicating that an acceleration 
term was needed to account for a curvature in its stellar trajectory. 
A star with such a solution often turns out to have a companion with 
an orbital period that is too long to be derived from only the 
{\it Hipparcos} data.  \citet{metal01b} used speckle interferometry 
to make follow-up observations of a number of such stars and 
resolved HD~220091 for the first time.  They determined an angular 
separation of 0\farcs114: and estimated an orbital period of 7 yr.  
Thus, HD~220091 does indeed have a binary companion.

Our ten spectrograms show that HD~220091 has a composite spectrum,
consisting of one set of broad lines and a second set of narrow lines,
components A and B, respectively (Figure~5).  We find a spectral class
of F1 for component A and G1: for component B.  The {\it Hipparcos}
parallax \citep{petal97} results in dwarf luminosity classes for both
stars.  Guided by our spectrum addition fits, we estimate $\Delta$$V$
= 2.2 mag.  We also determine {\it v}~sin~{\it i} values of 120: and 3:
for A and B, respectively.  Radial velocities of the components vary
in opposite directions, although the broad-lined star appears to
have additional velocity variations that presumably result from pulsation.
Our observations currently span four yr, during which time the velocity of B
has decreased by nearly 13 km~s$^{-1}$.  A period of 8.5 yr, which is
in approximate agreement with the 7-yr period estimate of \citep{metal01b},
currently provides a reasonable fit to the velocities of the narrow-lined
component.  However, given the modest portion of the velocity orbit covered
so far, the orbital period could certainly be in the 10--25 yr range.
We conclude that the spectroscopic components correspond to those of the 
visual binary.  Our continued observation of HD~220091 should eventually 
allow us to obtain a well-determined orbit for at least the narrow~lined 
component.

The {\it Hipparcos} team \citep{petal97} identified HD~220091 as a 
possible micro-variable with an amplitude less than 0.03 mag.  
\citet{cetal00} included the star in an optical follow-up study of 
EXOSAT serendipitous sources.  Their photometric observations of 
1990 August revealed low-amplitude variability of about 0.02 mag and
a tentative period of 7.5 days.  We acquired 66 observations during the
T12 survey; those data demonstrate photometric variability with a 
standard deviation of 0.0067 mag.  Our analysis of the T3 follow-up 
observations reveals two closely-spaced periods of 0.35364 and 0.36709 days 
with $B$ amplitudes of 19 and 13 mmag, respectively (Figs. 36 and 37; 
Table~5).  The mean $B/V$ amplitude ratio is 1.20, and the light curve is 
sinusoidal when phased with either of the two periods.  Given these 
properties, we confirm that the F1 dwarf component of the HD~220091 
binary system is a $\gamma$~Doradus star.

\subsection{HD 224945 = BU Psc}

\citet{metal04} included this star in a spectroscopic survey of candidate 
and confirmed $\gamma$~Doradus stars.  They obtained just two spectrograms 
and did not comment on any line-profile variability of the star but did 
determine a {\it v}~sin~{\it i} value of 54 km~s$^{-1}$.  Their two radial 
velocities were 5 and 6.5 km~s$^{-1}$.  

From our spectra of HD~224945, we find a spectral class of A9, while it has 
a dwarf luminosity class.  Comparison with HR~1613 [A9~V, \citep{am95}], 
which has solar abundances of iron and calcium \citep{betal02}, suggests 
that HD~224945 is slightly metal poor.  Our {\it v}~sin~{\it i} value of 
55 km~s$^{-1}$ agrees with that of \citet{metal04}.  Our four radial 
velocities have a mean of 7.1 $\pm$ 0.8 km~s$^{-1}$, which is also in 
accord with the results of \citet{metal04}.  We conclude that the star is 
single.

\citet{mmr86} observed HD~224945 as a comparison star for HD~1788 and 
HD~224639 and had trouble characterizing the photometric variations of the 
two program stars.  \citet{mp91} obtained follow-up photometric observations 
of HD~224639 and discovered their comparison star, HD~224945, was also 
variable. \citet{mpz94} analyzed those photometric observations of HD~224945 
and found two periods of 1.494 and 1.072 days. They noted that the variability 
of HD~224945 was similar to that of several other early F stars.  At about 
the same time, \citet{bkc94} extensively analyzed the light variations of 
$\gamma$ Dor and discussed it along with nine other similar variable stars, 
one of which was HD~224945.  They concluded that this group constituted a 
new class of pulsating stars with light variations most likely generated 
by $g$-mode non-radial pulsations.  Several years later, \citet{ks97} gave 
HD~224945 the variable star name BU~Psc.  In 1995 October, HD~224945 was 
the object of a multisite photometric campaign and was extensively monitored 
at five observatories.  As a result, \citet{petal02} found the star to be 
multiperiodic; their Table~2 listed five frequencies corresponding to periods 
of 0.333, 0.353, 0.413, 0.432, and 0.862 days, none of which match the periods 
in \citet{mpz94} given above.  In spite of having multiple sites for this 
observing campaign, \citet{petal02} concluded that, for each detected 
frequency, ``it is not possible to select the true value owing to the numerous 
aliases separated by integer values of $\pm1$~cd$^{-1}$.''

Our T12 photometric survey confirmed that HD~224945 is variable.  The
standard deviation of the 33 observations was 0.0053 mag (Table~2).  Our
analysis of the 300 T3 follow-up observations revealed periods at 0.54318 and
0.77214 days with $B$ amplitudes of 11 and 8 mmag, respectively (Figs. 38
and 39; Table~5).  Our two periods do not match any of the five periods 
in \citet{petal02} and are approximately half the two periods given by 
\citet{mpz94}.  Figure~39 shows that our light curves are sinusoidal when 
phased with each of the two periods.  The mean of the two $B/V$ amplitude 
ratios from Table~5 is $1.27\pm0.14$.  Although the identity of the true 
periods in HD~224945 remains uncertain, this A9 dwarf certainly has multiple 
photometric periods in the range of other $\gamma$~Doradus variables.  We 
agree with the conclusion of \citet{mpz94} and \citet{petal02} that HD~224945 
is a $\gamma$~Doradus variable.

\section{DISCUSSION}

In his review paper presented at the Sixth Vienna Workshop in Astrophysics,
\citet{b00} combined the results of several unbiased variability surveys 
of late-A to early-F field stars searching for pulsation in the lower 
instability strip.  Breger plotted the constant and variable stars in 
the H-R diagram (his Figure~4) and placed the borders of the instability 
strip to contain all the known $\delta$~Scuti variables in his sample.  
There were 222 constant and 56 variable stars (with amplitudes greater 
than a few mmag) inside Breger's instability strip, corresponding to an 
incidence of pulsation within the strip of 25\%.  Breger noted, however, 
that ``the large number of variables with amplitudes near the detection 
limit indicates that many of the so-called constant stars are also 
variable.''

In another study of the incidence of pulsation in the lower instability 
strip, \citet{petal03} searched a proposed {\it CoRoT} mission
field of view before launch to find suitable pulsating targets for $CoRoT$'s
telescope.  They found that 23\% of their sample in the lower instability
strip displayed low-amplitude, multi-periodic variability, a result in 
close agreement with Breger's findings from surveys of more widely-scattered 
field stars.  \citet{petal03} found a similar incidence of pulsation 
($\sim$20\%) in the open cluster NGC~6633, within the same $CoRoT$ 
field of view.

Consequently, we have good ground-based determinations of the incidence of 
pulsation for stars in the solar neighborhood that lie within the 
lower instability strip.  $\delta$~Scuti stars are the primary residents 
of this region of the H-R diagram, generally referred to as the 
$\delta$~Scuti instability strip.

Similar determinations of the incidence of $\gamma$~Doradus pulsations 
within the $\gamma$~Doradus instability strip have yet to be 
done.  Most known $\gamma$~Doradus variables have been found via 
photometric studies of individual candidate field stars \citep[][and 
references therein]{hfh07}.  $\gamma$~Doradus candidates are also being 
found in galactic open star clusters such as NGC~2506 \citep{adg+07}, 
NGC~2516 \citep{zetal98}, NGC~6755 \citep{cetal07}, NGC~6866 \citep{mzetal09}, 
and the Pleiades \citep{mretal08}, but none of these or similar studies 
have defined a complete statistical sample, precluding a good estimate of 
$\gamma$~Doradus incidence.

Our volume-limited sample of 114 $\gamma$~Doradus candidates in this paper 
{\it does} constitute a complete, unbiased, statistical sample of nearby 
field stars.  Our initial photometric survey of this sample with the T12 APT 
found a total of 37 variable stars with intrinsic variability of 0.002 mag 
or more (Table~2; Figure~1).  Higher-cadence observations with the T3 APT 
described in \S3 (above) and spectroscopic observations from KPNO described 
in \S4 (above) allowed us to classify all but three of the 37 variables and 
to describe their properties.  The final results of the survey are listed 
in column~(12) of Table~2.  The results are also plotted in the H-R diagram 
of Figure~40 with separate symbols to designate the 77 constant stars, 
24 $\gamma$~Doradus variables, nine $\delta$~Scuti variables, one hybrid 
pulsator, and the three variables without clear periodicity.  One of the 
24 $\gamma$~Doradus stars is the SB2 binary HD~114447, which is composed 
of two F0 dwarfs (Table~6, Figs. 28 \& 29, \S6.12).  Both components of 
this binary are plotted with open star symbols since we do not know which 
component is pulsating.  Including the hybrid variable HD~8801 among the 
other $\gamma$~Doradus stars gives an observed incidence of $\frac{25}{114}$ 
or 22\% for $\gamma$~Doradus variables among nearby candidate field stars.  
This is similar to the incidence of $\delta$~Scuti variables within the 
$\delta$~Scuti instability strip found in the earlier studies referenced 
above.  For our volume-limited sample, we can also compute the corresponding 
space density of $\gamma$~Doradus stars to be 0.094 stars per $10^3~pc^3$ or 
94 per $10^6~pc^3$.

In addition to the results of our 114 star survey, we also provide an 
updated list of 86 confirmed $\gamma$~Doradus variables in Table~8.  The 
15 new $\gamma$~Doradus variables found in our survey make up the bulk of 
the new stars.  Two more variables (HD~74504 and HD~147787) were discovered 
by \citet{cetal09} and \citet{dcetal09}, respectively. Four new 
$\gamma$~Doradus / $\delta$~Scuti hybrids have been found;  HD~44195 and 
HD~49434 were discovered in ground-based preparatory observations for the 
$CoRoT$ mission by \citet{petal05} and \citet{uetal08}, respectively.  Two 
more hybrids (HD~114839 and BD~+18~4914) were found in $MOST$ spaced-based 
photometry by \citet{kmr+06} and \citet{rmc+06}, respectively.  The 
properties of the five hybrids are listed in bold font in Table~8.  We have 
removed the ``hybrid'' HD~209295 from the list because (1) its 
$\gamma$~Doradus pulsations are excited by tidal effects from the presence 
of a neutron star or white dwarf companion \citep{hetal02}, and (2) it 
lies significantly outside of the $\gamma$~Doradus instability strip 
\citep[][their Figure~25]{hfh07}.  

Table~8 does not include variables in galactic open clusters, such as the 
$\gamma$~Doradus stars found in the clusters NGC~2506, NGC~2516, NGC~6755, 
NGC~6866, and the Pleiades mentioned above.  Cluster variables, as well 
as future discoveries from the $MOST$, $CoRoT$, and $Kepler$ missions, will 
generally be several magnitudes fainter than the stars in Table~8.  So 
we restrict our list in Table~8 to bright, confirmed $\gamma$~Doradus field 
stars.  These bright field stars will continue to be the most useful targets 
for multisite spectroscopic and photometric campaigns that are required to 
identify pulsation modes \citep[e.g.,][]{hetal02,aetal04,metal04}.

Among the 86 $\gamma$~Doradus stars in Table~8 are 49 single stars, eight 
single-lined spectroscopic binaries, 22 double-lined binaries, and 15 
visual double or multiple systems.  Five of the visual double systems 
(HD~7169, HD~23874, HD~38309, HD~147787, and HD~220091) have one component 
that is a spectroscopic binary, so these five stars were counted in both 
the visual double and spectroscopic binary classifications.  In most of 
the cases involving duplicity, it is clear that the primary component is 
the $\gamma$~Doradus variable; in those cases, we have appended an ``A'' 
to the HD number in column~(1) to designate the primary component.  HD~211699, 
however, is a double-lined system for which \citet{hfh07} have shown that 
the {\it secondary} is probably the $\gamma$~Doradus star; we have appended 
a ``B'' to its HD number in Table~8.  Both components of the three 
double-lined binaries HD~86371 \citep{hfh07}, HD~113867 \citep{hf03}, and 
HD~114447 (this paper) have $\gamma$~Doradus-type properties, so it is not 
clear which component is the $\gamma$~Doradus variable.  It is quite 
possible that both components of one or more of these SB2 binaries are 
$\gamma$~Doradus stars.  In those three cases, we have listed both the 
primary and secondary components in Table~8 with designations of ``A:'' 
and ``B:'', indicating uncertainty in identifying the $\gamma$~Doradus 
component.  Therefore, Table~8 actually contains 89 entries.

For the single stars, the wide visual doubles, and the single-lined binaries 
in Table~8, the $V$ magnitudes and $B-V$ colors in columns (4) and (5) are 
from the {\it Hipparcos} catalog \citep{petal97}.  For all of the visual 
double stars and the double-lined spectroscopic binaries, the $V$ magnitudes 
and $B-V$ colors refer to the individual component designated in column~(1); 
\citet{hf03} provide details on the determination of those values.  The 
stellar properties listed in columns (6), (7), and (8) have all been 
determined from the $V$ magnitudes, $B-V$ colors, and parallaxes by the 
method outlined in \citet{hetal01}.  Most of these stars have multiple 
photometric periods; the period given in column~(9) is the one with the 
largest amplitude.  The final column gives the literature reference that 
confirms each star as a $\gamma$~Doradus variable.

We plot all 86 $\gamma$~Doradus stars from Table~8 in Figure~41.  The 15 new 
$\gamma$~Doradus variables discovered in our survey fall within the 
observed limits of the $\gamma$~Doradus instability strip as plotted by 
\citet{fwk03}; those limits remain unchanged with the addition of dozens 
of new $\gamma$~Doradus stars by \citet{hfh05}, \citet{hfh07}, and the 
current paper.   The five known hybrid variables all lie within the 
overlap region between the $\delta$~Scuti and $\gamma$~Doradus instability 
strips.  As mentioned in \S3.3 above, none of the new $\gamma$~Doradus 
stars show evidence of hybrid pulsation.

We note in Figure~40 that there are no $\gamma$~Doradus candidates in the 
upper right corner of the observed instability strip redward of $B-V = 0.38$. 
In hindsight, we might have extended the $B-V$ color index limit of our 
survey to about 0.42.  However, when all 86 known $\gamma$~Doradus stars 
are plotted in Figure~41, we see that $\gamma$~Doradus variability is 
relatively rare but can occur in this corner of the instability strip.
This may cause our incidence and space density values of $\gamma$~Doradus 
variability to be slightly low.

Finally, we anticipate upcoming contributions from the $MOST$, $CoRoT$ and 
$Kepler$ space missions.  The discovery of two new $\gamma$~Doradus / 
$\delta$~Scuti hybrid pulsators from $MOST$ photometry has been mentioned 
above.  A recent search of 131 days of $CoRoT$ observations by 
\citet{hetal10} yielded 418 $\gamma$~Doradus and 274 hybrid candidates 
in the LRa01 field.  From the first few weeks of the $Kepler$ mission, 
\citet{getal10} found 34 new $\gamma$~Doradus stars and 25 hybrids with 
the likelihood of many more to come.  They also made the surprising 
discovery that ``there are practically no pure $\delta$~Scuti or 
$\gamma$~Doradus pulsators.''  They contrast this finding with ground-based 
results that show a clear separation between the two types of pulsation, 
except for a few hybrid pulsators.  \citet{getal10} note that ground-based 
photometric observations are probably sensitive only to pulsations of low 
spherical degree ($\ell = 1,2$).  When candidate pulsators are examined with 
the photometric precision, cadence, and time base enabled by $Kepler$, many 
more modes of higher degree (and lower photometric amplitude) become visible.

\acknowledgements

We thank Lou Boyd for his continuing efforts in support of our automatic
telescopes at Fairborn Observatory.  Astronomy at Tennessee State 
University is supported by NASA, NSF, Tennessee State University, and 
the State of Tennessee through its Centers of Excellence program.  This 
research has made use of the SIMBAD database, operated at CDS, Strasbourg, 
France.

\clearpage
\begin{deluxetable}{cc}
\tabletypesize{\small}
\tablenum{1}
\tablecolumns{2}
\tablewidth{0pt}
\tablecaption{SELECTION CRITERIA FOR THE VOLUME-LIMITED SAMPLE}
\tablehead{
\colhead{Parameter} & \colhead{Range or Value}
}
\startdata
{\it Hipparcos} Catalog Number  &  1 -- 118,218 \\
$V$ (mag)  &  5.5 -- 8.0 \\
$B-V$ (mag)  &  0.25 -- 0.38 \\
Parallax (mas)  &  15.0 -- 20.0 \\
Distance (pc)  &  50.0 -- 66.7 \\
Declination (deg)  &  $-$10 -- $+$65 \\
M$_V$ (mag)  &  1.6 -- 3.7 \\
{\it Hipparcos} Multiplicity Status   &   Single 
\enddata
\end{deluxetable}

\clearpage
\begin{deluxetable}{cccccccccccc}
\rotate
\tabletypesize{\tiny}
\tablenum{2}
\tablewidth{0pt}
\tablecaption{VOLUME-LIMITED SAMPLE OF 114 $\gamma$~DORADUS CANDIDATES}
\tablehead{

\colhead{} & \colhead{} & \colhead{} & \colhead{$V$\tablenotemark{a}} 
& \colhead{{$B-V$}\tablenotemark{a}} & \colhead{$\pi$\tablenotemark{a}} 
& \colhead{$M_V$} & \colhead{{\it Hipparcos}} & \colhead{} 
& \colhead{Date Range} & \colhead{$\sigma_{star}$} & \colhead{Final}  \\

\colhead{HD} & \colhead{HIP} & \colhead{Other Names} 
& \colhead{(mag)} & \colhead{(mag)} & \colhead{(mas)} 
& \colhead{(mag)} & \colhead{Variable Type\tablenotemark{b}} 
& \colhead{$N_{obs}$} & \colhead{(HJD $-$ 2,400,000)} 
& \colhead{(mag)} & \colhead{Status}  \\

\colhead{(1)} & \colhead{(2)} & \colhead{(3)} & \colhead{(4)} 
& \colhead{(5)} & \colhead{(6)} & \colhead{(7)} & \colhead{(8)} 
& \colhead{(9)} & \colhead{(10)} & \colhead{(11)} & \colhead{(12)}
}
\startdata
  1917 &   1892 & \nodata                    & 6.57 & 0.380 &  18.64$\pm$0.69 & 2.92 &   &  39 & 52,172.86--52,287.63 & 0.0000 & Constant \\
  4818 &   3965 & V526 Cas, HR 238           & 6.38 & 0.307 &  15.00$\pm$0.72 & 2.26 & C &  43 & 52,168.82--52,312.62 & 0.0018 & Constant \\
  6288 &   4979 & 26 Cet, HR 301, ADS 875A   & 6.06 & 0.271 &  17.66$\pm$0.72 & 2.30 & M &  30 & 52,161.88--52,288.62 & 0.0000 & Constant \\
{\bf 6568} & {\bf 5209} & \nodata & {\bf 6.93} & {\bf 0.301} & {\bf 15.31$\pm$0.93} & {\bf 2.86} & & {\bf 271} & {\bf 52,162.83--52,991.75} & {\bf 0.0066} & {\bf $\gamma$~Dor} \\
  8801 &   6794 & HR 418, ADS 1151A          & 6.42 & 0.311 &  17.91$\pm$0.75 & 2.68 & U & 532 & 52,161.89--52,980.83 & 0.0054 & $\gamma$~Dor~+~$\delta$~Sct \\
  9780 &   7447 & HR 457                     & 5.95 & 0.259 &  16.75$\pm$0.82 & 2.07 &   &  45 & 52,161.83--52,313.63 & 0.0023 & Variable \\
 12354 &   9452 & \nodata                    & 6.61 & 0.339 &  15.39$\pm$0.77 & 2.55 & C &  55 & 52,162.86--52,317.64 & 0.0000 & Constant \\
 13683 &  10356 & \nodata                    & 6.51 & 0.356 &  15.02$\pm$1.10 & 2.39 & C &  51 & 52,162.85--52,314.63 & 0.0000 & Constant \\
\enddata
\tablenotetext{a}{From the {\it Hipparcos} catalog.}
\tablenotetext{b}{A ``blank'' in this column indicates the star could not 
be classified as variable or constant, C = no variability detected 
(``constant''), D = duplicity induced variability, M = possible micro-variable 
(amplitude $<$0.03 mag), P = periodic variable, U = unresolved variable.}
\tablenotetext{c}{T12 APT survey observations failed because of a
bright, nearby companion.  HD 162898 was added to the T3 follow-up list 
and found to be constant.}
\end{deluxetable}

\clearpage
\begin{deluxetable}{ccccc}
\tabletypesize{\small}
\tablenum{3}
\tablewidth{0pt}
\tablecaption{COMPARISON \& CHECK STARS FOR NEW $\gamma$~DORADUS STARS}
\tablehead{

\colhead{$\gamma$~Doradus} & \colhead{Comparison} & \colhead{Check}
& \colhead{$\sigma_{(V-C)}$\tablenotemark{a}}
& \colhead{$\sigma_{(K-C)}$\tablenotemark{a}} \\

\colhead{Star} & \colhead{Star} & \colhead{Star}
& \colhead{(mag)} & \colhead{(mag)}
}
\startdata
  HD 6568 &   HD 6120 &   HD 7446 & 0.0092 & 0.0046 \\
 HD 17163 &  HD 16647 &  HD 18258 & 0.0055 & 0.0051 \\
 HD 25906 &  HD 25877 &  HD 26755 & 0.0058 & 0.0047 \\
 HD 31550 &  HD 30617 &  HD 30708 & 0.0043 & 0.0033 \\
 HD 38309 &  HD 40259 &  HD 38145 & 0.0050 & 0.0049 \\
 HD 45638 &  HD 45089 &  HD 44853 & 0.0059 & 0.0042 \\
 HD 62196 &  HD 64106 &  HD 60063 & 0.0066 & 0.0046 \\
 HD 63436 &  HD 62724 &  HD 60490 & 0.0277 & 0.0058 \\
 HD 65526 &  HD 66351 &  HD 68145 & 0.0234 & 0.0050 \\
 HD 69682 &  HD 67991 &  HD 71952 & 0.0061 & 0.0044 \\
 HD 99267 &  HD 99207 & HD 101091 & 0.0167 & 0.0039 \\
HD 114447 & HD 114905 & HD 117261 & 0.0099 & 0.0047 \\
HD 138936 & HD 139044 & HD 138562 & 0.0053 & 0.0060 \\
HD 139478 & HD 139778 & HD 139357 & 0.0135 & 0.0048 \\
HD 145005 & HD 146173 & HD 145894 & 0.0067 & 0.0071 \\
HD 220091 & HD 218079 & HD 220044 & 0.0085 & 0.0045 \\
HD 224945 &      HD 6 &    HD 406 & 0.0065 & 0.0061 \\
\enddata
\tablenotetext{a}{In the Johnson $B$ photometric band.}
\end{deluxetable}

\clearpage
\begin{deluxetable}{cccccc}
\tabletypesize{\small}
\tablenum{4}
\tablecolumns{6}
\tablewidth{0pt}
\tablecaption{PHOTOMETRIC OBSERVATIONS OF NEW $\gamma$~DORADUS STARS}
\tablehead{

\colhead{} & \colhead{Date} & \colhead{Var $B$} & \colhead{Var $V$}
& \colhead{Chk $B$} & \colhead{Chk $V$} \\

\colhead{HD} &\colhead{(HJD $-$ 2,400,000)} & \colhead{(mag)} 
& \colhead{(mag)} & \colhead{(mag)} & \colhead{(mag)}

}
\startdata
  6568 & 53,255.7637 & $-$0.088 & $-$0.133 & $-$0.225 & $-$1.033 \\
  6568 & 53,255.8545 & $-$0.095 & $-$0.135 & $-$0.236 & $-$1.037 \\
  6568 & 53,258.7573 & $-$0.099 & $-$0.141 & $-$0.228 & $-$1.033 \\
  6568 & 53,258.8455 & $-$0.093 & $-$0.136 & $-$0.232 & $-$1.035 \\
  6568 & 53,258.9367 & $-$0.092 & $-$0.136 & $-$0.232 & $-$1.040 \\
  6568 & 53,259.7536 & $-$0.098 & $-$0.142 &   99.999 &   99.999 \\
\enddata
\tablecomments{Table 4 is presented in its entirety in
the electronic edition of the Astronomical Journal.  A portion
is shown here for guidance regarding the form and
content.}
\end{deluxetable}

\begin{deluxetable}{ccccccccc}
\rotate
\tabletypesize{\footnotesize}
\tablenum{5}
\tablewidth{0pt}
\tablecaption{PHOTOMETRIC RESULTS FOR NEW $\gamma$~DORADUS STARS}
\tablehead{

\colhead{} & \colhead{} & \colhead{} & \colhead{} & \colhead{} & \colhead{}
& \colhead{Peak-to-Peak} & \colhead{$B/V$} & \colhead{} \\

\colhead{} & \colhead{} & \colhead{Date Range} & \colhead{}
& \colhead{Frequency} & \colhead{Period} & \colhead{Amplitude}
& \colhead{Amplitude} & \colhead{$T_{min}$} \\

\colhead{HD} & \colhead{Photometric Band} & \colhead{(HJD $-$ 2,400,000)}
& \colhead{$N_{obs}$} & \colhead{(day$^{-1}$)} & \colhead{(days)}
& \colhead{(mmag)} & \colhead{Ratio} & \colhead{(HJD $-$ 2,400,000)} \\

\colhead{(1)} & \colhead{(2)} & \colhead{(3)} & \colhead{(4)}
& \colhead{(5)} & \colhead{(6)} & \colhead{(7)} & \colhead{(8)}
& \colhead{(9)}
}
\startdata
  6568 & $B$ & 53,255.76--53,405.59 & 223 & 1.3620 $\pm$ 0.0004 & 0.73421 $\pm$ 0.00022 & 17.6 $\pm$ 1.3 & 1.45 $\pm$ 0.17 & 53,300.612 $\pm$ 0.009 \\
       &     &                      &     & 1.3416 $\pm$ 0.0004 & 0.74538 $\pm$ 0.00022 & 15.2 $\pm$ 1.5 & 1.23 $\pm$ 0.20 & 53,300.055 $\pm$ 0.011 \\
       & $V$ & 53,255.76--53,405.59 & 211 & 1.3629 $\pm$ 0.0004 & 0.73373 $\pm$ 0.00019 & 12.1 $\pm$ 1.2 &                 & 53,300.630 $\pm$ 0.012 \\
       &     &                      &     & 1.3425 $\pm$ 0.0005 & 0.74488 $\pm$ 0.00028 & 12.4 $\pm$ 1.2 &                 & 53,300.036 $\pm$ 0.012 \\
 17163 & $B$ & 53,630.92--53,793.60 & 296 & 2.3612 $\pm$ 0.0003 & 0.42351 $\pm$ 0.00005 &  8.1 $\pm$ 0.8 & 1.16 $\pm$ 0.20 & 53,700.059 $\pm$ 0.007 \\
       & $V$ & 53,630.92--53,793.60 & 279 & 2.3609 $\pm$ 0.0003 & 0.42357 $\pm$ 0.00005 &  7.0 $\pm$ 0.7 &                 & 53,700.047 $\pm$ 0.007 \\
 25906 & $B$ & 53,255.86--53,460.62 & 295 & 1.2632 $\pm$ 0.0003 & 0.79164 $\pm$ 0.00019 &  8.6 $\pm$ 0.9 & 1.28 $\pm$ 0.21 & 53,350.785 $\pm$ 0.012 \\
       &     &                      &     & 1.2248 $\pm$ 0.0002 & 0.81646 $\pm$ 0.00013 &  5.5 $\pm$ 0.9 & 0.91 $\pm$ 0.28 & 53,350.593 $\pm$ 0.022 \\
\enddata
\tablecomments{The individual photometric observations are given in Table 4.}
\tablenotetext{a}{First frequency has slightly asymmetrical light curve.}
\tablenotetext{b}{Frequency identification uncertain due to aliasing.}
\end{deluxetable}

\clearpage
\begin{deluxetable}{ccccccccc}
\rotate
\tabletypesize{\footnotesize}
\tablenum{6}
\tablewidth{0pt}
\tablecaption{BASIC PROPERTIES OF NEW $\gamma$~DORADUS STARS}
\tablehead{

\colhead{} & \colhead{} & \colhead{} & \colhead{$V$\tablenotemark{a}} 
& \colhead{{\it B$-$V}\tablenotemark{a}} & \colhead{Spectral} 
& \colhead{Luminosity} & \colhead{{\it v}~sin~{\it i}\tablenotemark{b}} 
& \colhead{Velocity\tablenotemark{b,c}}  \\

\colhead{HD} & \colhead{Other Names} & \colhead{Component} 
& \colhead{(mag)} & \colhead{(mag)} & \colhead{Class\tablenotemark{b}} 
& \colhead{Class\tablenotemark{b}} & \colhead{(km~s$^{-1}$)} 
& \colhead{(km~s$^{-1}$)}  \\

\colhead{(1)} & \colhead{(2)} & \colhead{(3)} & \colhead{(4)} 
& \colhead{(5)} & \colhead{(6)} & \colhead{(7)} & \colhead{(8)} 
& \colhead{(9)} 
}
\startdata
6568   & \nodata & ... & 6.93 & 0.301 & F1 & Dwarf & 55 & Binary \\
17163  & HR 816  & ... & 6.04 & 0.312 & F1: & Dwarf & 105 & 26.2 $\pm$ 0.3 \\
25906  & \nodata & ... & 7.08 & 0.345 & F1 & Dwarf & 64 & Binary? \\
31550  & \nodata & ... & 6.74 & 0.335 & F0 & Dwarf & 32  & 17.6 $\pm$ 1.8 \\
38309  & HR 1978 & A   & 6.09 & 0.320 &  &  &  & \\
       &         & Aa  & 6.24 & 0.32  & F1 & Dwarf & 95: & Binary \\
       &         & Ab  & 8.34 & 0.61  & G1: & Dwarf & 5: & Binary \\
45638  & HR 2351 & ... & 6.59 & 0.293 & A9 & Dwarf & 38 & 42.4 $\pm$ 0.4 \\
\enddata
\tablenotetext{a}{Single star values and combined values of binary components
are from the {\it Hipparcos} catalog.  For the individual binary components
see the text.}
\tablenotetext{b}{Values from this paper. A colon indicates
greater uncertainty than usual.}
\tablenotetext{c}{Velocity noted as ``Binary'' indicates velocity variability
produced by orbital motion.}
\end{deluxetable}

\clearpage
\begin{deluxetable}{cccl}
\tablenum{7}
\tablewidth{0pt}
\tablecaption{INDIVIDUAL RADIAL VELOCITIES OF NEW $\gamma$~DORADUS STARS}
\tablehead{

\colhead{} & \colhead{Date} & \colhead{Radial Velocity} & \colhead{} \\

\colhead{HD} & \colhead{(HJD $-$ 2,400,000)} & \colhead{(km s$^{-1}$)}
& \colhead{Comments} \\

\colhead{(1)} & \colhead{(2)} & \colhead{(3)} & \colhead{(4)}

}
\startdata
   6568  & 52,940.820  &  $-$13.8  &   \\
         & 52,942.777  &  $-$13.8  &   \\
         & 53,273.902  &   $-$4.1  &   \\
         & 53,534.978  &   $-$6.4  &   \\
         & 53,634.837  &      8.4  &  Asymmetric lines  \\
         & 53,635.907  &      4.2  &  Asymmetric lines  \\
         & 53,637.870  &      6.3  &  Very asymmetric lines \\
         & 54,001.866  &      9.5  &   
\enddata
\tablecomments{Table 7 is published in its entirety in the electronic 
edition of the Astronomical Journal.  A portion is shown here for 
guidance regarding its form and content.}
\end{deluxetable}

\begin{deluxetable}{cccccccccl}
\rotate
\tabletypesize{\tiny}
\tablenum{8}
\tablewidth{0pt}
\tablecaption{PROPERTIES OF $\gamma$ DORADUS FIELD STARS}
\tablehead{

\colhead{} & \colhead{} & \colhead{} & \colhead{$V$}
& \colhead{$(\bv)$} & \colhead{$M_{V}$} & \colhead{$L$}
& \colhead{$R$} & \colhead{Period} & \colhead{} \\

\colhead{HD\tablenotemark{a,b}} & \colhead{Other Names}
& \colhead{Duplicity\tablenotemark{c}} & \colhead{(mag)}
& \colhead{(mag)} & \colhead{(mag)} & \colhead{($L_{\sun}$)}
& \colhead{($R_{\sun}$)} & \colhead{(days)} & \colhead{Reference} \\

\colhead{(1)} & \colhead{(2)} & \colhead{(3)} & \colhead{(4)}
& \colhead{(5)} & \colhead{(6)} & \colhead{(7)} & \colhead{(8)}
& \colhead{(9)} & \colhead{(10)}

}
\startdata
   277    & \nodata                              & single & 8.37  & 0.379 & 3.31 &  3.7 & 1.4 & 0.9005  & \citet{hetal01} \\
  2842    & \nodata                              & single & 7.99  & 0.325 & 2.83 &  5.7 & 1.6 & 0.65070 & \citet{hfh05} \\
  6568A   & \nodata                              &  SB1   & 6.93  & 0.301 & 2.86 &  5.6 & 1.5 & 0.7342  & This paper    \\
  7169A   & HDS 160                              & VB,SB2 & 7.42  & 0.329 & 2.98 &  5.0 & 1.5 & 0.5486  & \citet{hf03} \\
\bf{8801A} & \bf{HR 418, ADS 1151A} & \bf{VB} & \bf{6.42} & \bf{0.311} & \bf{2.68} & \bf{6.5} & \bf{1.7} & \bf{0.40331} & \bf{\citet{hf05}} \\
  9365A   & \nodata                              &   VB   & 8.17  & 0.361 & 2.76 &  6.1 & 1.7 & 0.62582 & \citet{hfh07} \\
 12901    & \nodata                              & single & 6.74  & 0.311 & 2.35 &  8.9 & 1.9 & 0.82270 & \citet{ea00} \\
 17163    & HR 816                               & single & 6.04  & 0.312 & 2.49 &  7.8 & 1.8 & 0.42351 & This paper   \\
\enddata
\tablenotetext{a}{Colon indicates component A and/or B could be the 
$\gamma$~Doradus star.}
\tablenotetext{b}{Bold indicates $\gamma$~Doradus / $\delta$~Scuti hybrid}
\tablenotetext{c}{VB = visual binary or double star. SB = spectroscopic 
binary.}
\tablenotetext{d}{Also has period of 10.959 days with a slightly larger 
amplitude.}
\end{deluxetable}

\clearpage
\begin{figure}
\figurenum{1}
\epsscale{0.8}
\plotone{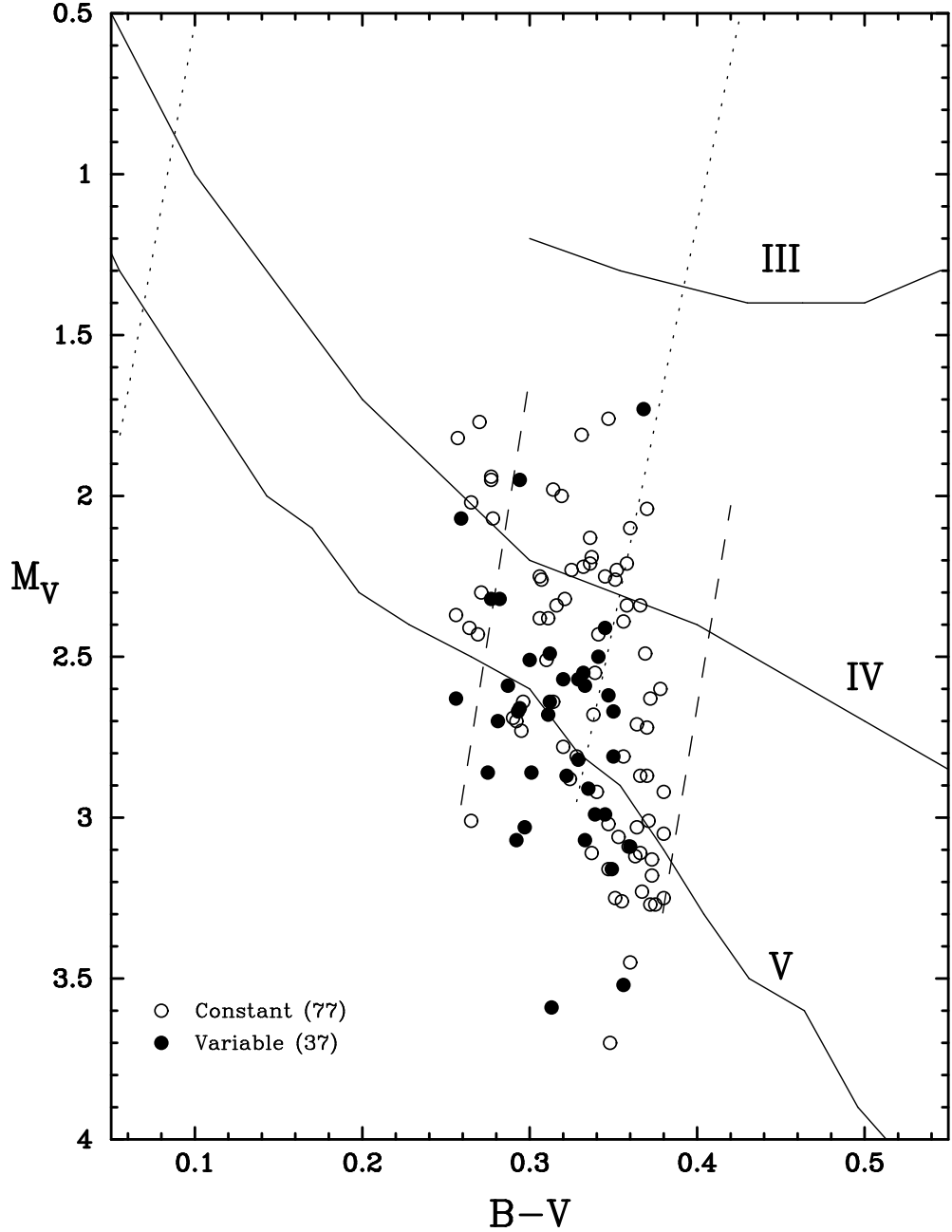}
\figcaption{H-R diagram of the complete volume-limited sample of 114 
$\gamma$~Doradus candidates listed in Table~2.  Solid lines indicate 
observed average locations of normal main-sequence (V), sub-giant (IV), and 
giant (III) stars.  The dotted lines indicate the boundaries of the 
$\delta$~Scuti instability strip, converted from those of \citet{b00}.  The 
dashed lines show the observed domain of the $\gamma$~Doradus pulsators, 
adopted from \citet{fwk03}.  The one-year photometric survey with the 
T12 0.8~m APT found 37 variable stars with $\sigma_{star}~\geq~0.002$ mag 
(filled circles) and 77 constant stars with $\sigma_{star}~<~0.002$ mag 
(open circles).}
\end{figure}

\clearpage
\begin{figure}
\figurenum{2}
\epsscale{0.8}
\plotone{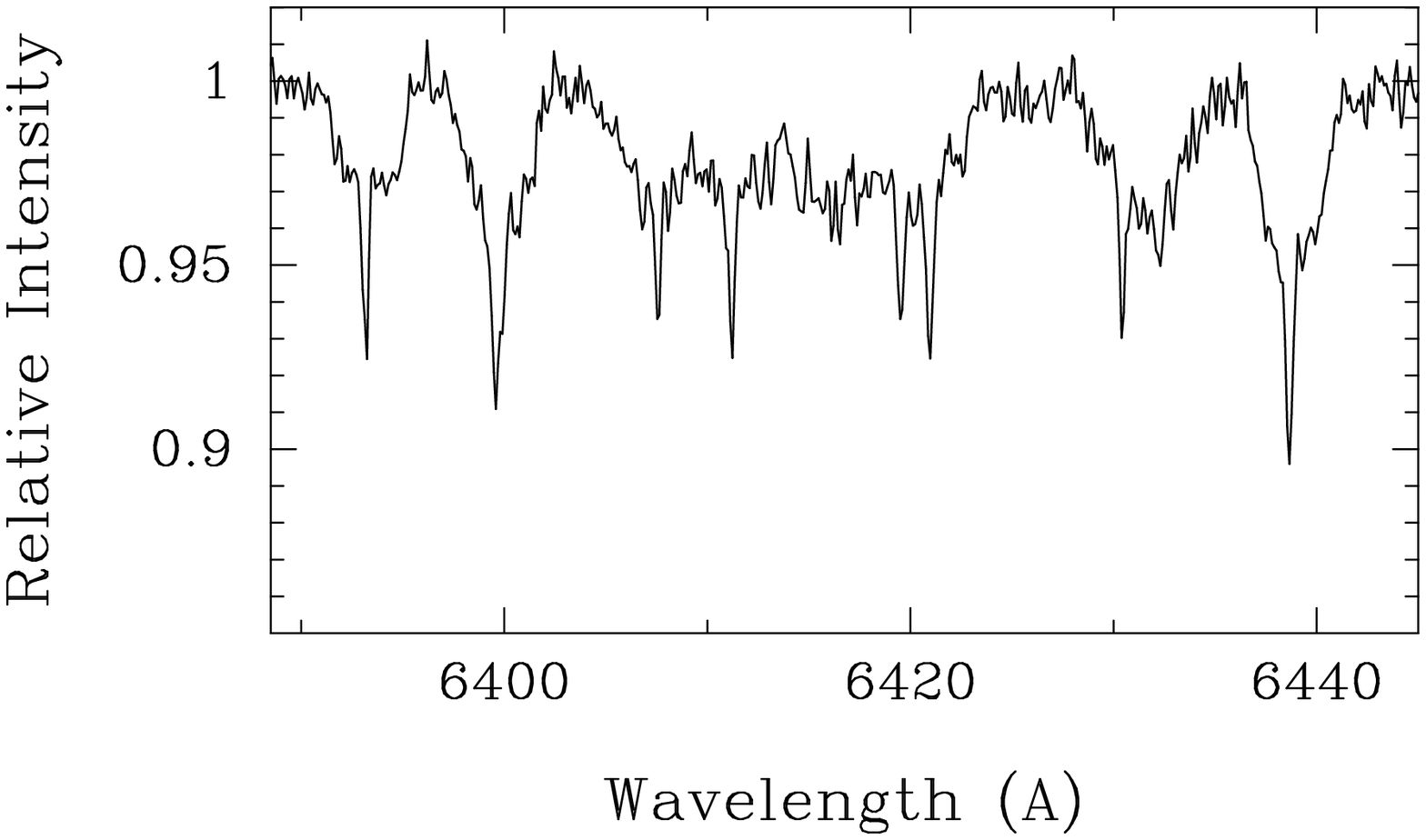}
\figcaption{Spectrum of HD~38309 in the 6430~\AA\ region, which shows
the composite profiles of the lines.  Component A is the broad-lined
star, and component B is the narrow-lined star.}
\end{figure}

\clearpage
\begin{figure}
\figurenum{3}
\epsscale{0.8}
\plotone{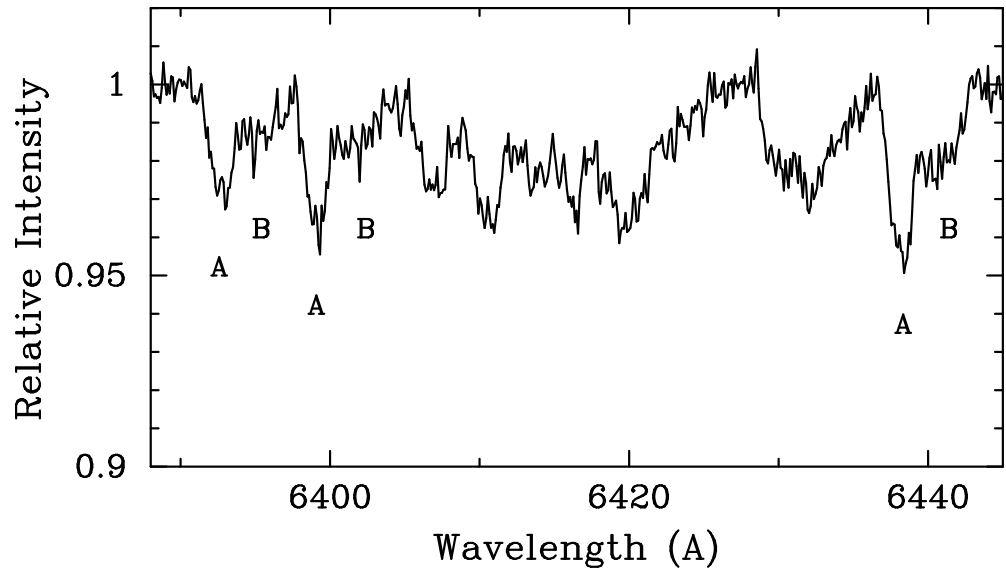}
\figcaption{Spectrum of HD~114447 in the 6430~\AA\ region, which shows
the blended lines of the two components.  Component A has the deeper
blue shifted lines, while component B has the shallower red shifted 
lines.}
\end{figure}

\clearpage
\begin{figure}
\figurenum{4}
\epsscale{0.8}
\plotone{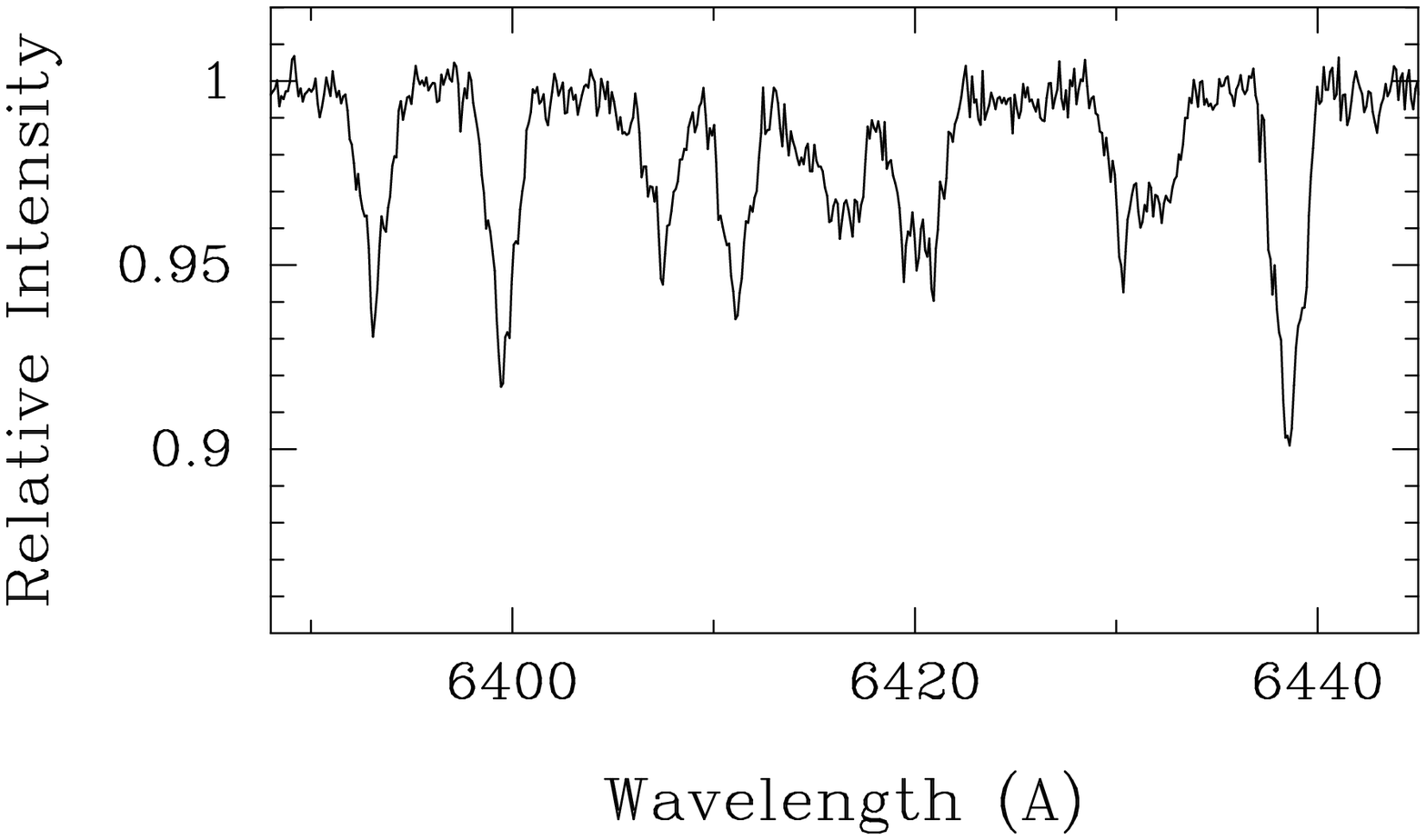}
\figcaption{Spectrum of HD~145005 in the 6430~\AA\ region, which shows
the composite profiles of the lines.  Component A is the broad-lined
star, and component B is the narrow-lined star.}
\end{figure}

\clearpage
\begin{figure}
\figurenum{5}
\epsscale{0.8}
\plotone{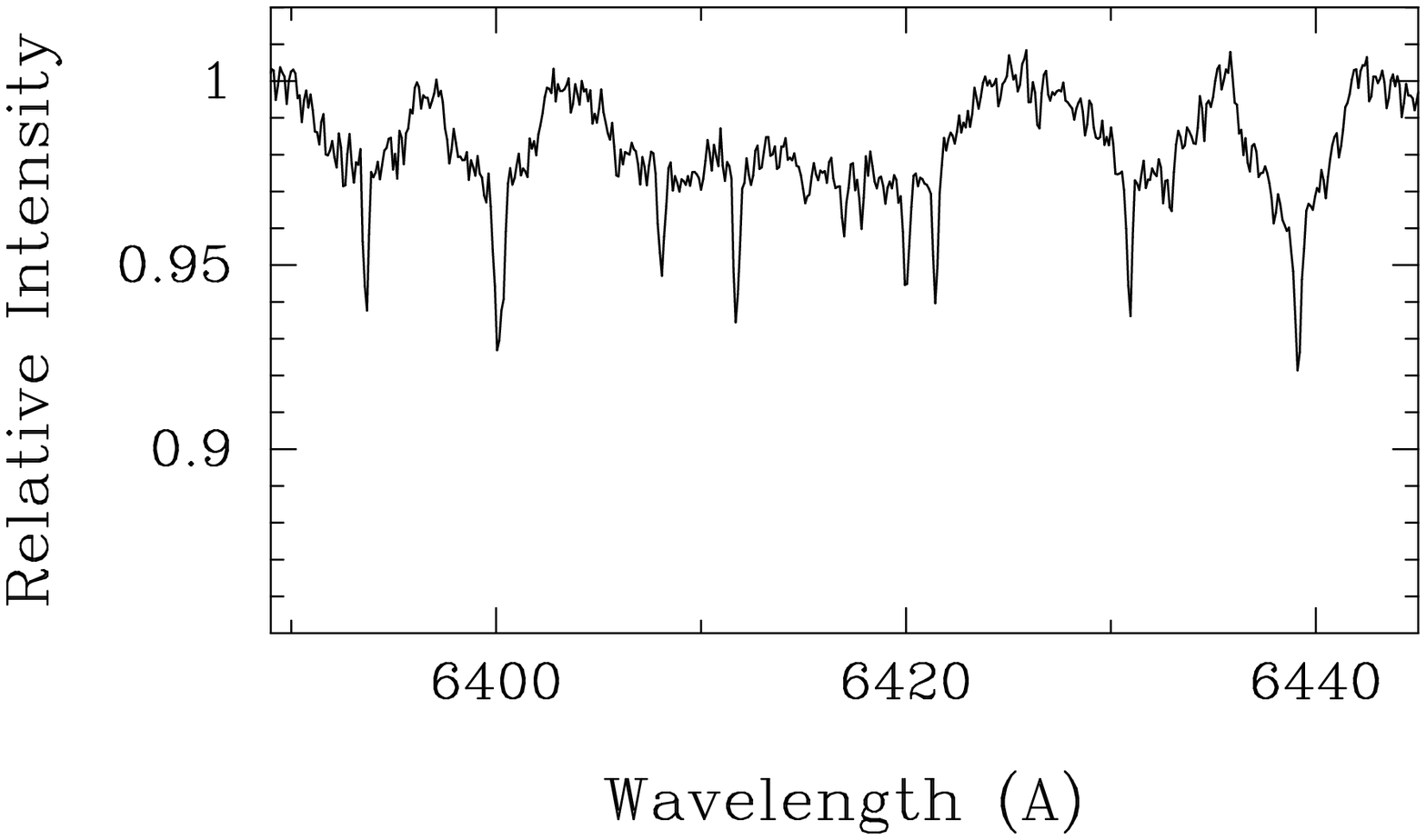}
\figcaption{Spectrum of HD~220091 in the 6430~\AA\ region, which shows
the composite profiles of the lines.  Component A is the broad-lined
star, and component B is the narrow-lined star.}
\end{figure}

\clearpage
\begin{figure}
\figurenum{6}
\epsscale{0.8}
\plotone{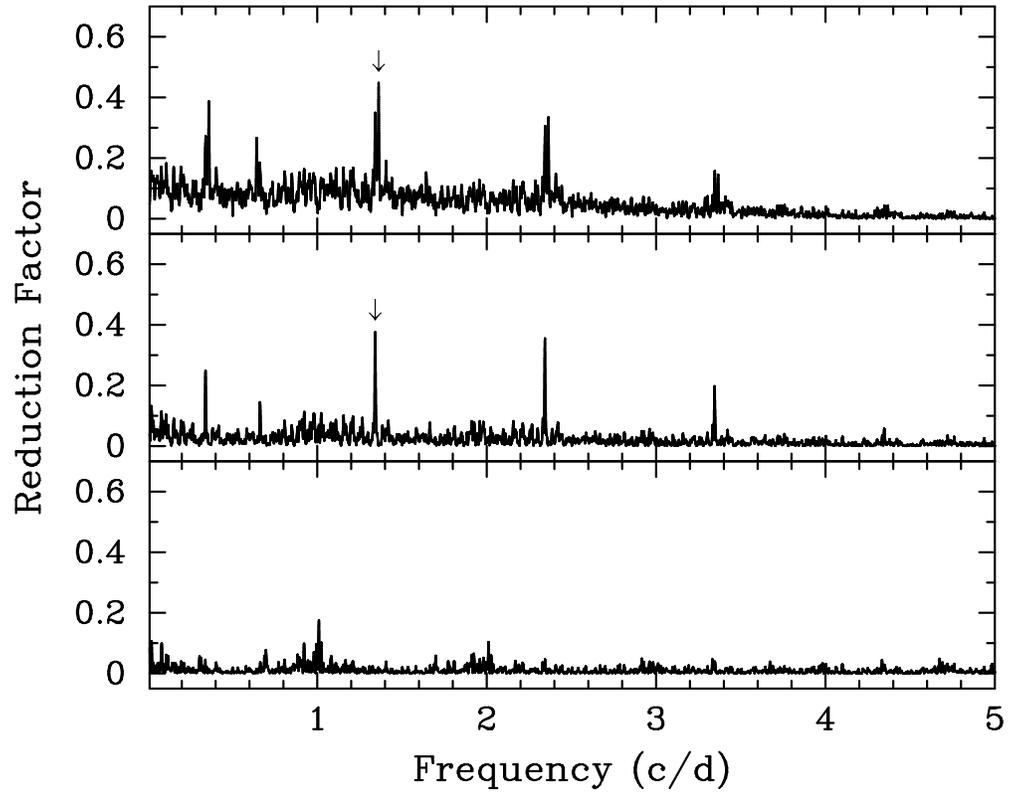}
\figcaption{Least-squares frequency spectra of the HD~6568 Johnson $B$ data 
set, showing the results of progressively fixing the two detected frequencies.
The arrows indicate the two frequencies at 1.3620 day$^{-1}$ ({\it top}) and 
1.3416 day$^{-1}$ ({\it middle}). The bottom panel shows the frequency 
spectrum with these two frequencies fixed.  Both frequencies were confirmed 
in the Johnson $V$ data set.}
\end{figure}

\clearpage
\begin{figure}
\figurenum{7}
\epsscale{0.8}
\plotone{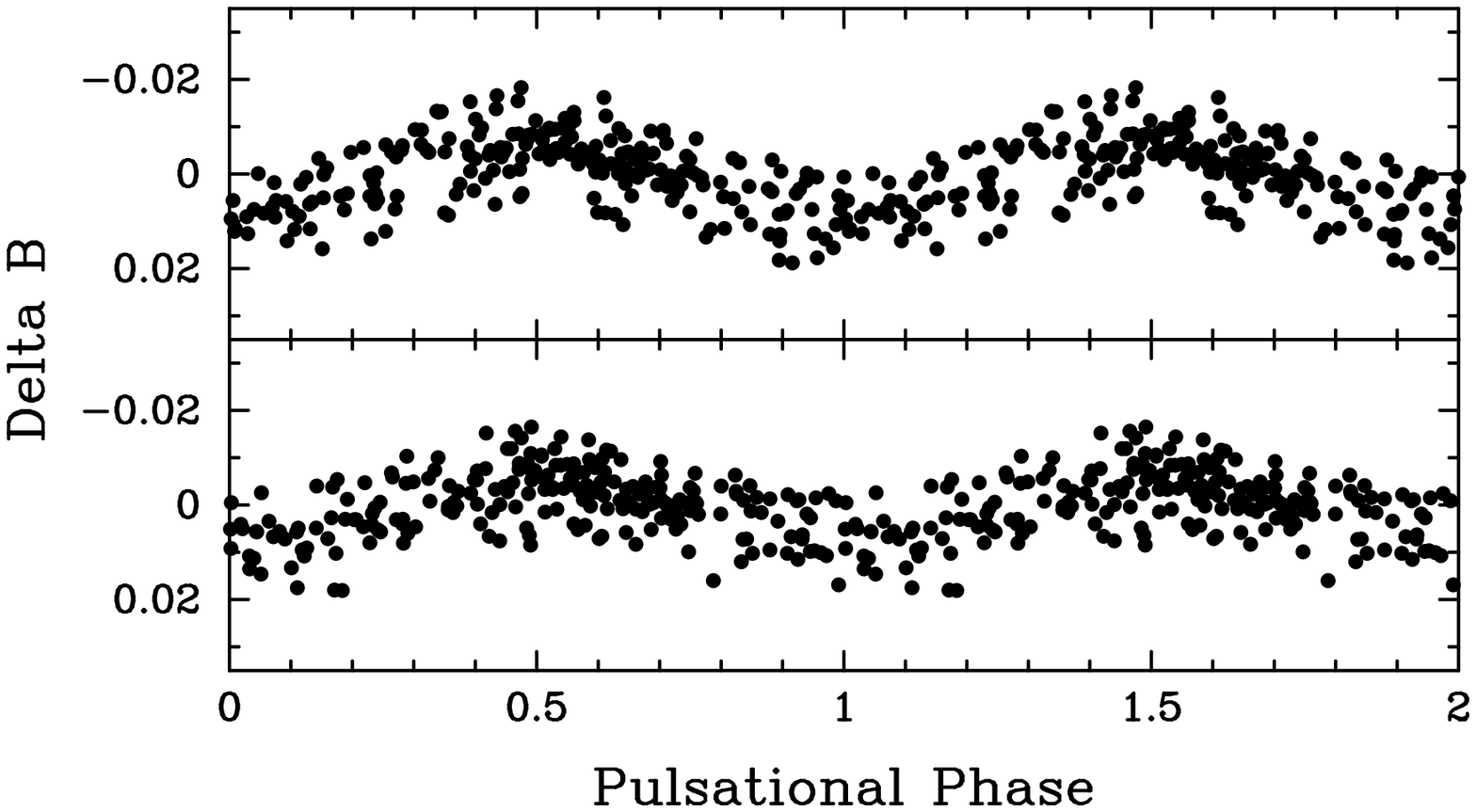}
\figcaption{The Johnson $B$ photometric data for HD~6568, phased with
the two frequencies and times of minimum from Table~5.  The two frequencies
are 1.3620 day$^{-1}$ ({\it top}) and 1.3416 day$^{-1}$ ({\it bottom}).
For each panel, the data set has been prewhitened to remove the other
known frequency.}
\end{figure}

\clearpage
\begin{figure}
\figurenum{8}
\epsscale{0.8}
\plotone{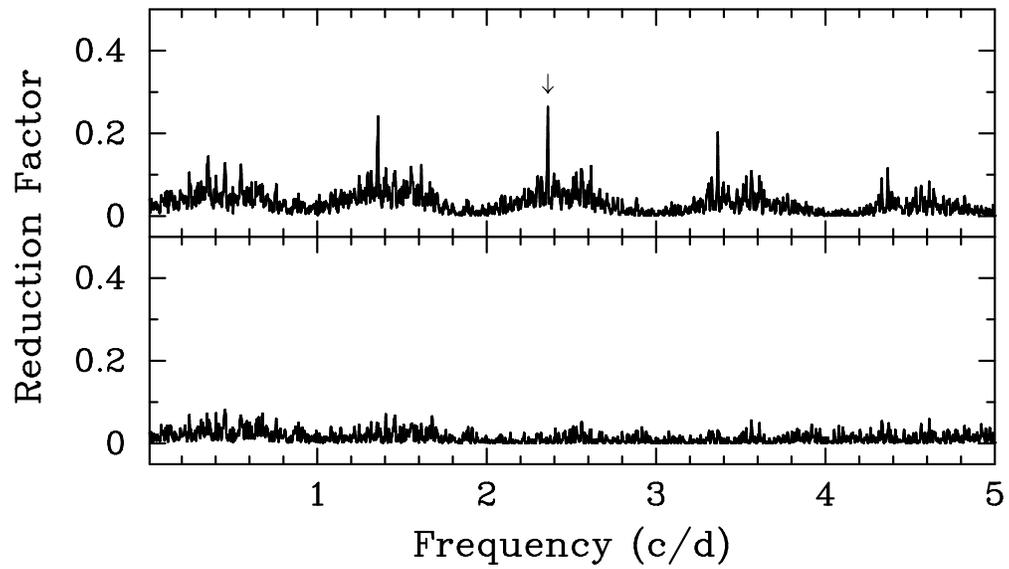}
\figcaption{Least-squares frequency spectra of the HD~17163 Johnson $B$ 
data set. The arrow in the top panel indicates the single detected frequency 
at 2.3612 day$^{-1}$.  The bottom panel shows the frequency spectrum with 
the 2.3612 day$^{-1}$ frequency fixed.  The same frequency was confirmed 
in the Johnson $V$ data set.}
\end{figure}

\clearpage
\begin{figure}
\figurenum{9}
\epsscale{0.8}
\plotone{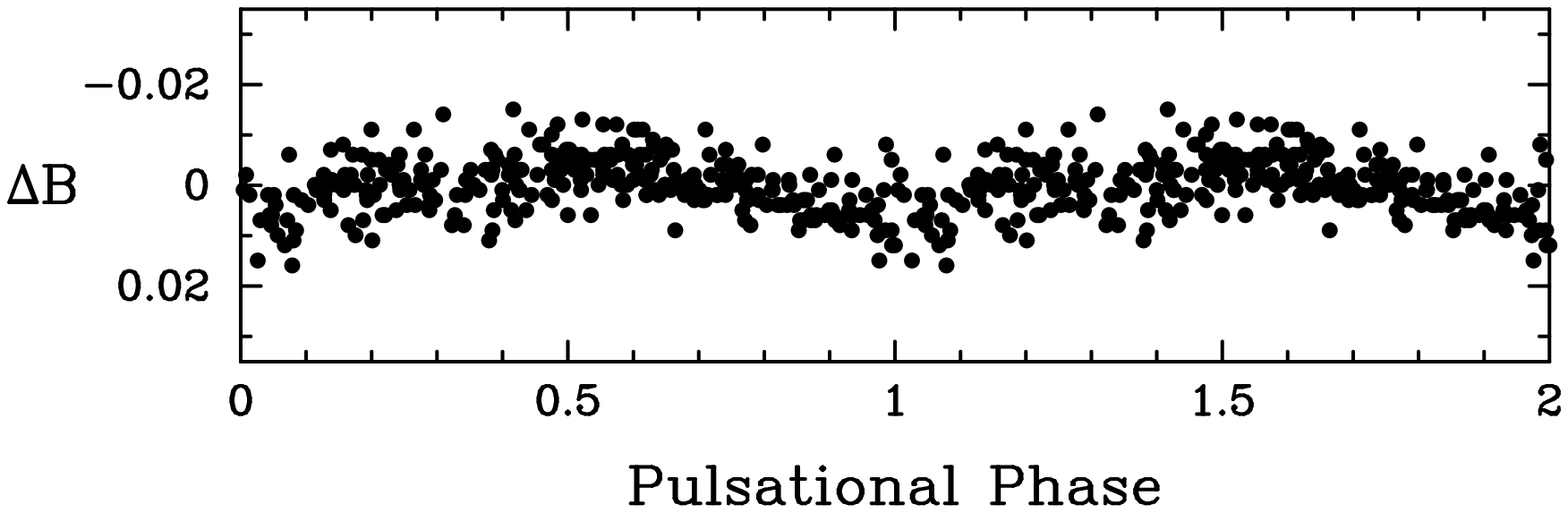}
\figcaption{The Johnson $B$ photometric data for HD~17163 phased with
the single frequency of 2.3612 day$^{-1}$ and the time of minimum from 
Table~5.}
\end{figure}

\clearpage
\begin{figure}
\figurenum{10}
\epsscale{0.8}
\plotone{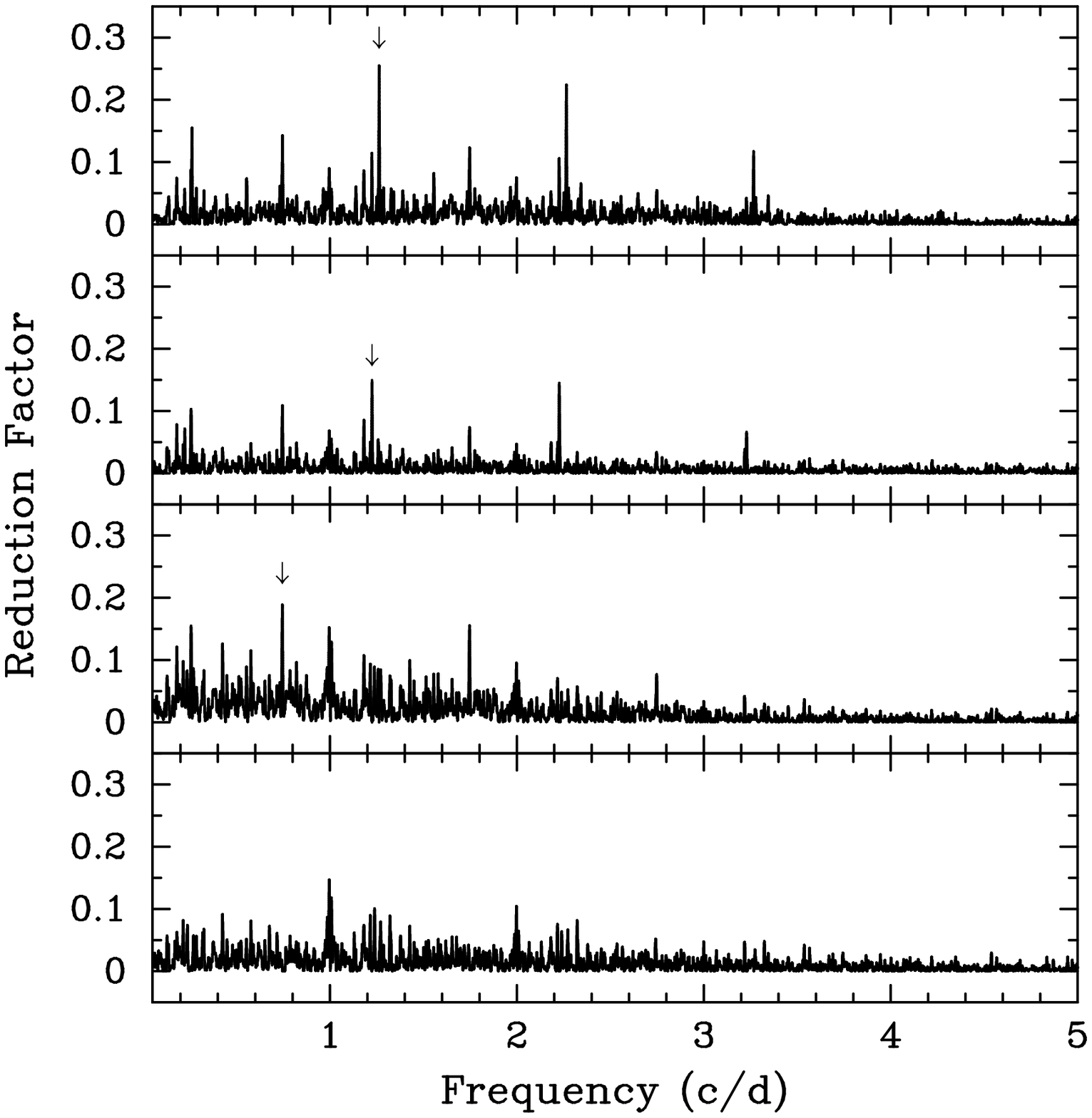}
\figcaption{Least-squares frequency spectra of the HD~25906 Johnson $B$ data 
set, showing the results of progressively fixing the three detected 
frequencies.  The arrows indicate the three frequencies ({\it top to bottom}) 
1.2632, 1.2248, and 0.7452 day$^{-1}$. All three frequencies were confirmed 
in the Johnson $V$ data set.}
\end{figure}

\clearpage
\begin{figure}
\figurenum{11}
\epsscale{0.8}
\plotone{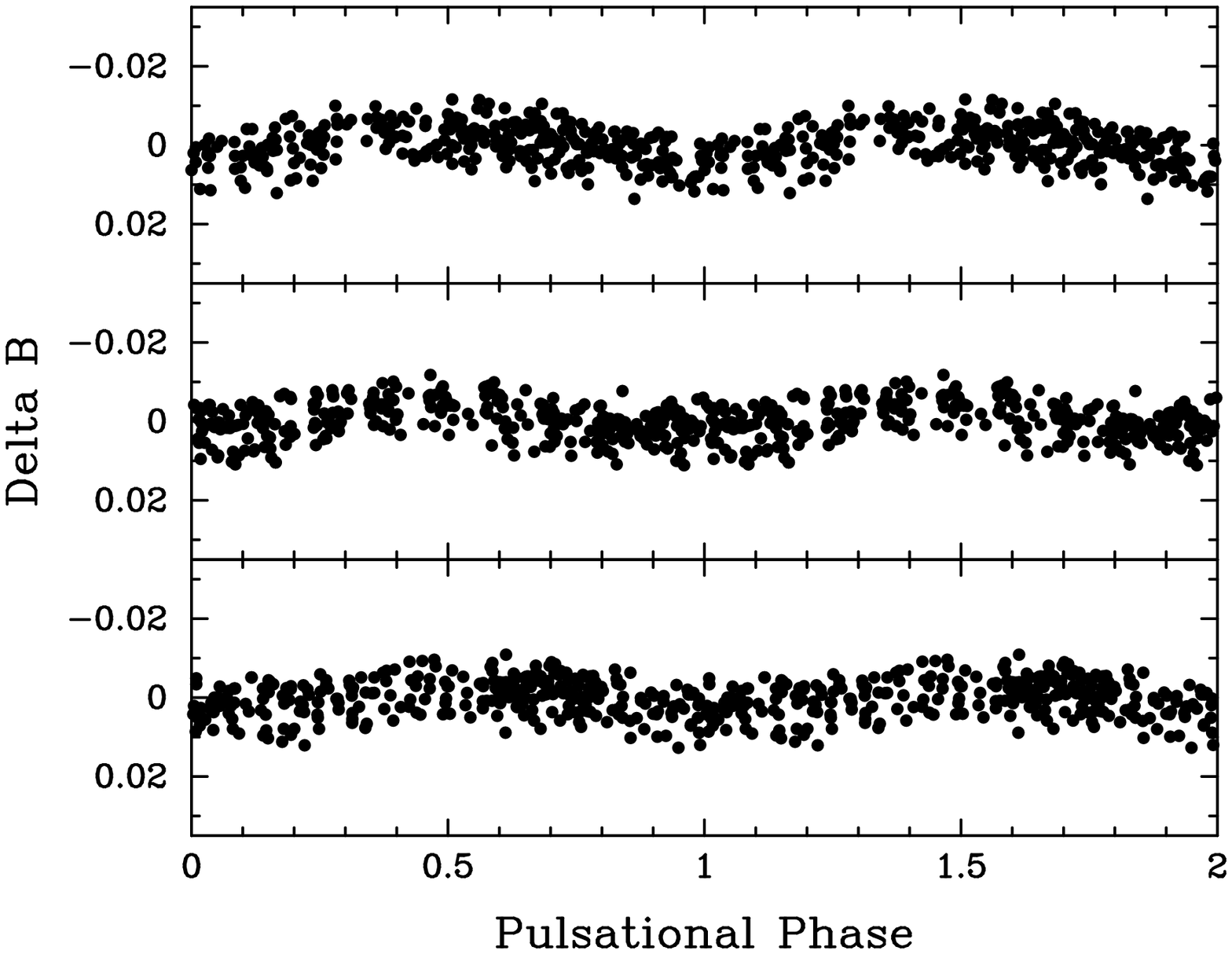}
\figcaption{The Johnson $B$ photometric data for HD~25906, phased with
the three frequencies and times of minimum from Table~5.  The three 
frequencies are ({\it top to bottom}) 1.2632, 1.2248, and 0.7452 day$^{-1}$. 
For each panel, the data set has been prewhitened to remove the other
two known frequencies.}
\end{figure}

\clearpage
\begin{figure}
\figurenum{12}
\epsscale{0.8}
\plotone{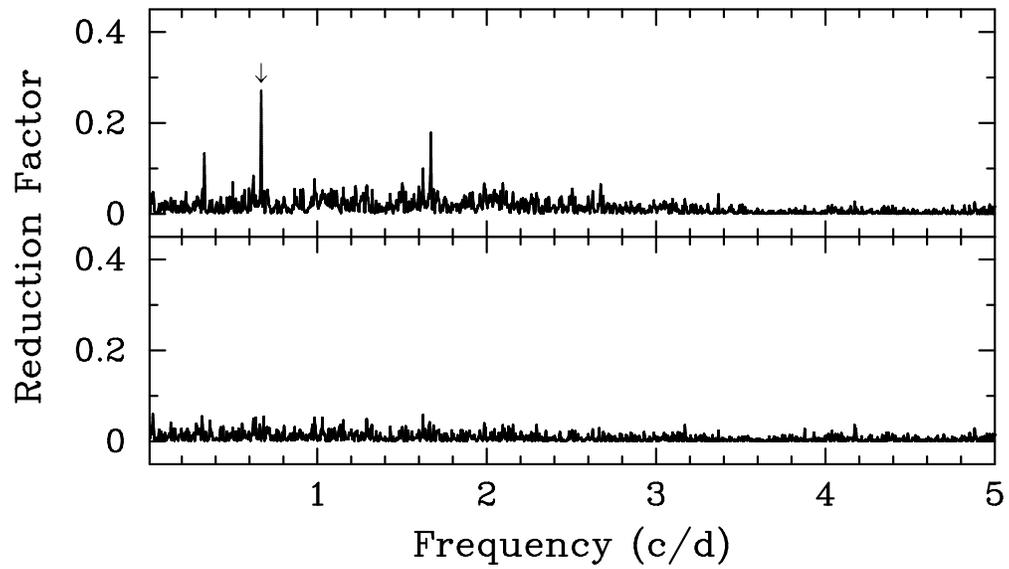}
\figcaption{Least-squares frequency spectra of the HD~31550 Johnson $B$ data 
set.  The arrow in the top panel indicates the single detected frequency of 
0.6686 day$^{-1}$.  The bottom panel is the frequency spectrum resulting when 
the 0.6686 day$^{-1}$ frequency is fixed.  The same frequency was found in 
the Johnson $V$ data set.}
\end{figure}

\clearpage
\begin{figure}
\figurenum{13}
\epsscale{0.8}
\plotone{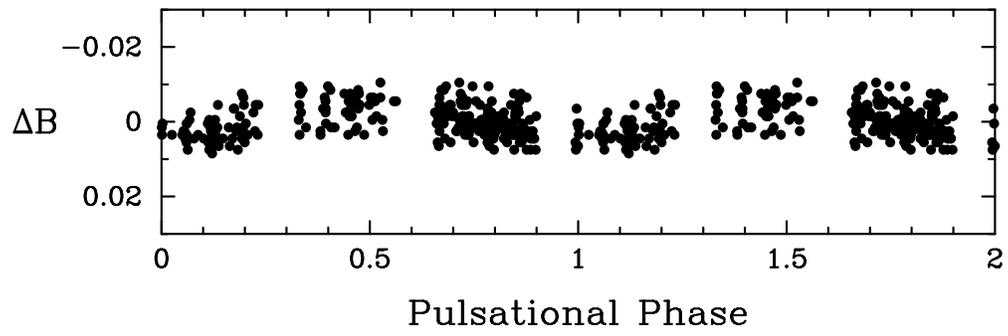}
\figcaption{The Johnson $B$ photometric data for HD~31550, phased with
the 0.6686 day$^{-1}$ frequency and time of minimum from Table~5.}
\end{figure}

\clearpage
\begin{figure}
\figurenum{14}
\epsscale{0.8}
\plotone{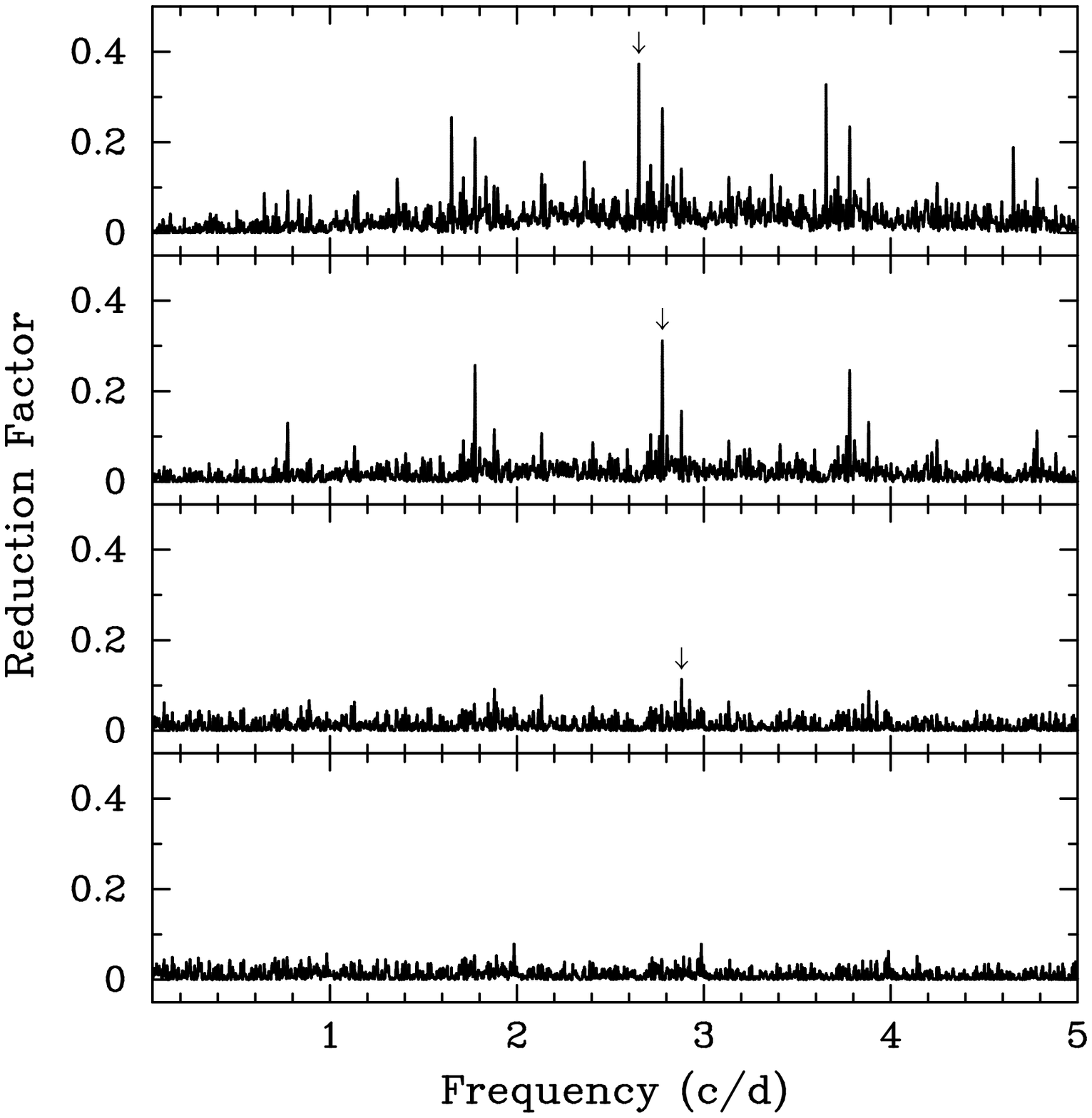}
\figcaption{Least-squares frequency spectra of the HD~38309 Johnson $B$ data 
set, showing the results of progressively fixing the three detected 
frequencies.  The arrows indicate the three frequencies ({\it top to bottom}) 
2.6523, 2.7783, and 2.8808 day$^{-1}$. All three frequencies were confirmed 
in the Johnson $V$ data set.}
\end{figure}

\clearpage
\begin{figure}
\figurenum{15}
\epsscale{0.8}
\plotone{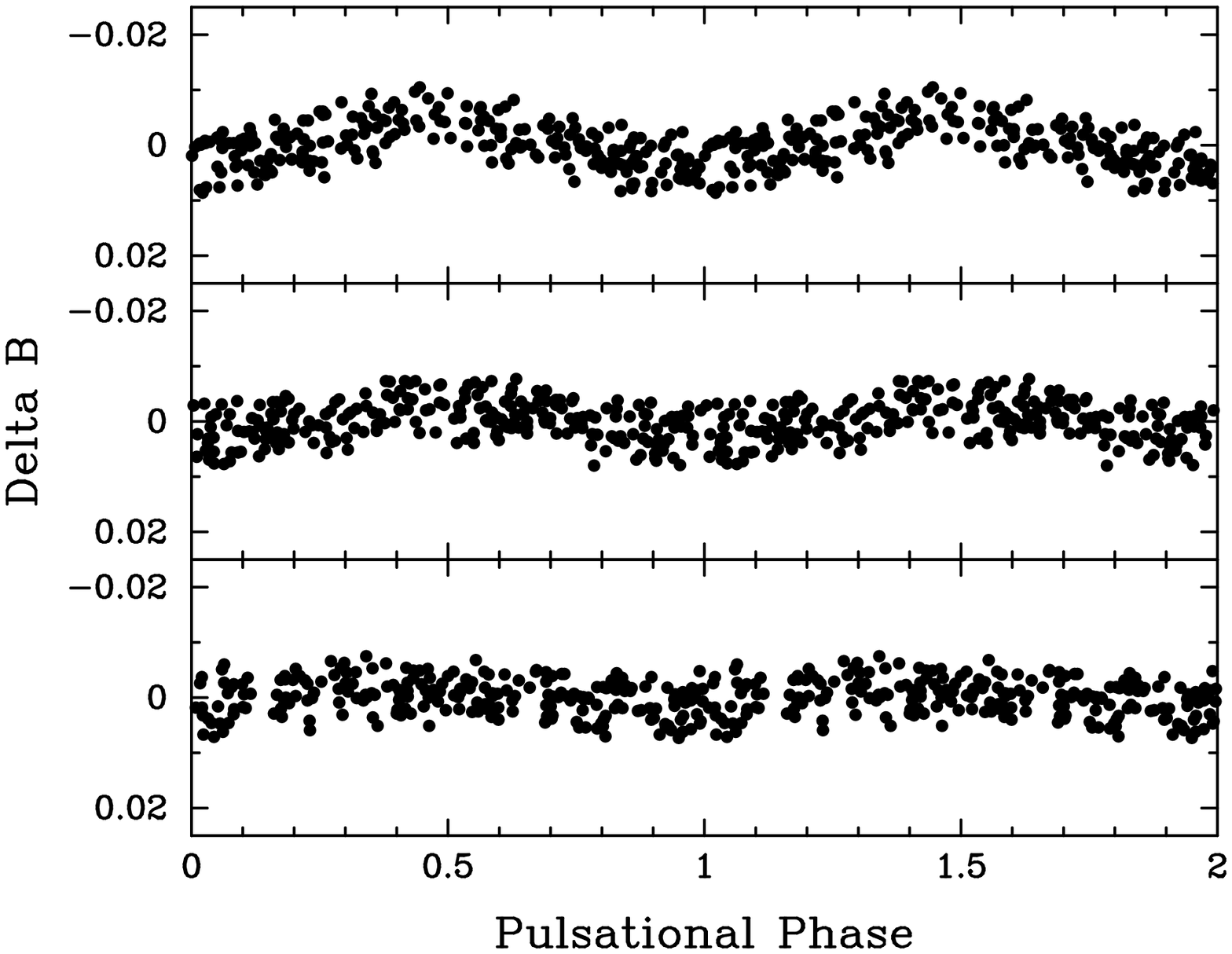}
\figcaption{The Johnson $B$ photometric data for HD~38309, phased with the 
three frequencies and times of minimum from Table~5.  The three frequencies 
are ({\it top to bottom}) 2.6523, 2.7783, and 2.8808 day$^{-1}$.  For each 
panel, the data set has been prewhitened to remove the other known 
frequencies.}
\end{figure}

\clearpage
\begin{figure}
\figurenum{16}
\epsscale{0.8}
\plotone{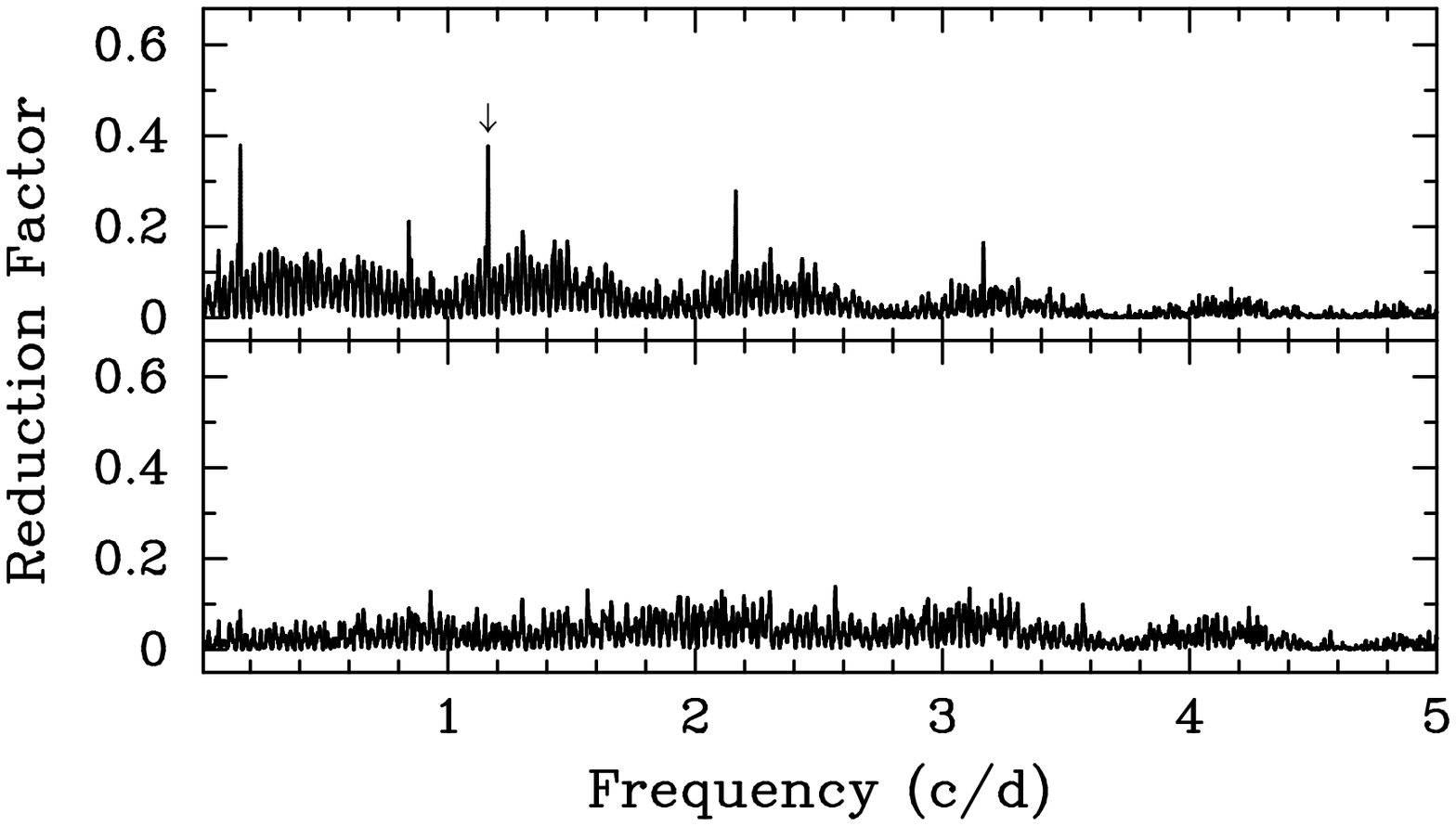}
\figcaption{Least-squares frequency spectra of the HD~45638 Johnson $B$ data 
set.  The arrow in the top panel indicates the single detected frequency at
1.1622 day$^{-1}$.  The bottom panel shows the frequency spectrum with the 
1.1622 day$^{-1}$ frequency fixed.  The same frequency was confirmed in the 
Johnson $V$ data set.}
\end{figure}

\clearpage
\begin{figure}
\figurenum{17}
\epsscale{0.8}
\plotone{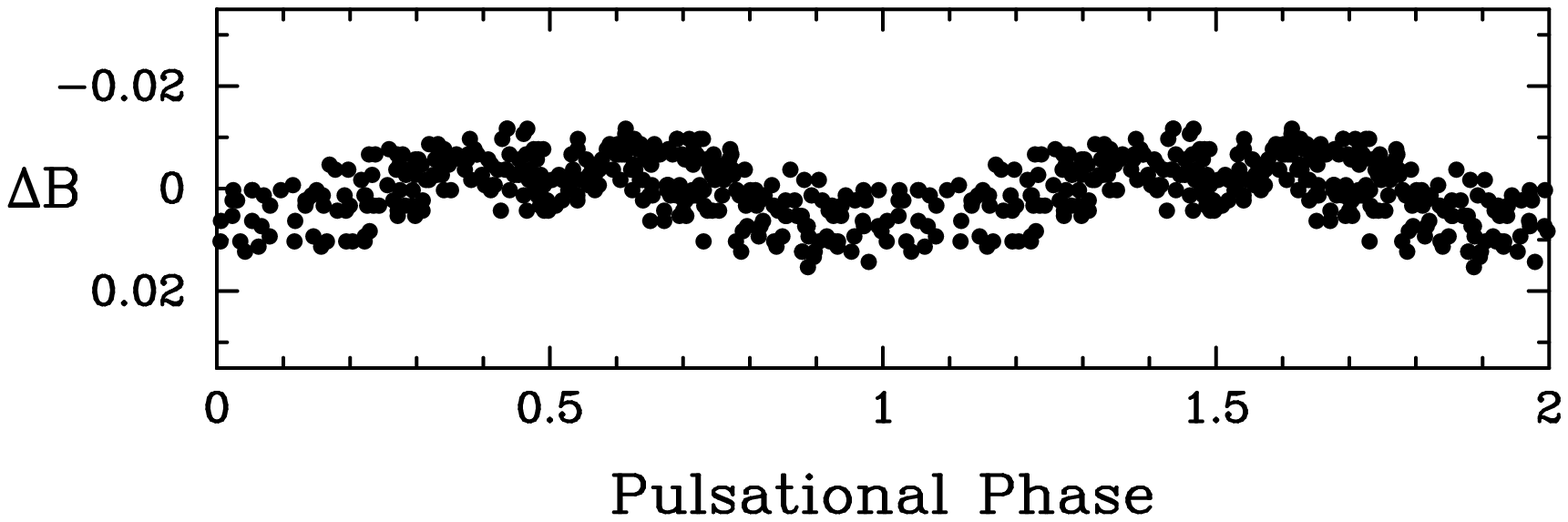}
\figcaption{The Johnson $B$ photometric data for HD~45638 phased with
the single frequency of 1.1622 day$^{-1}$ and the time of minimum from
Table~5.}
\end{figure}

\clearpage
\begin{figure}
\figurenum{18}
\epsscale{0.8}
\plotone{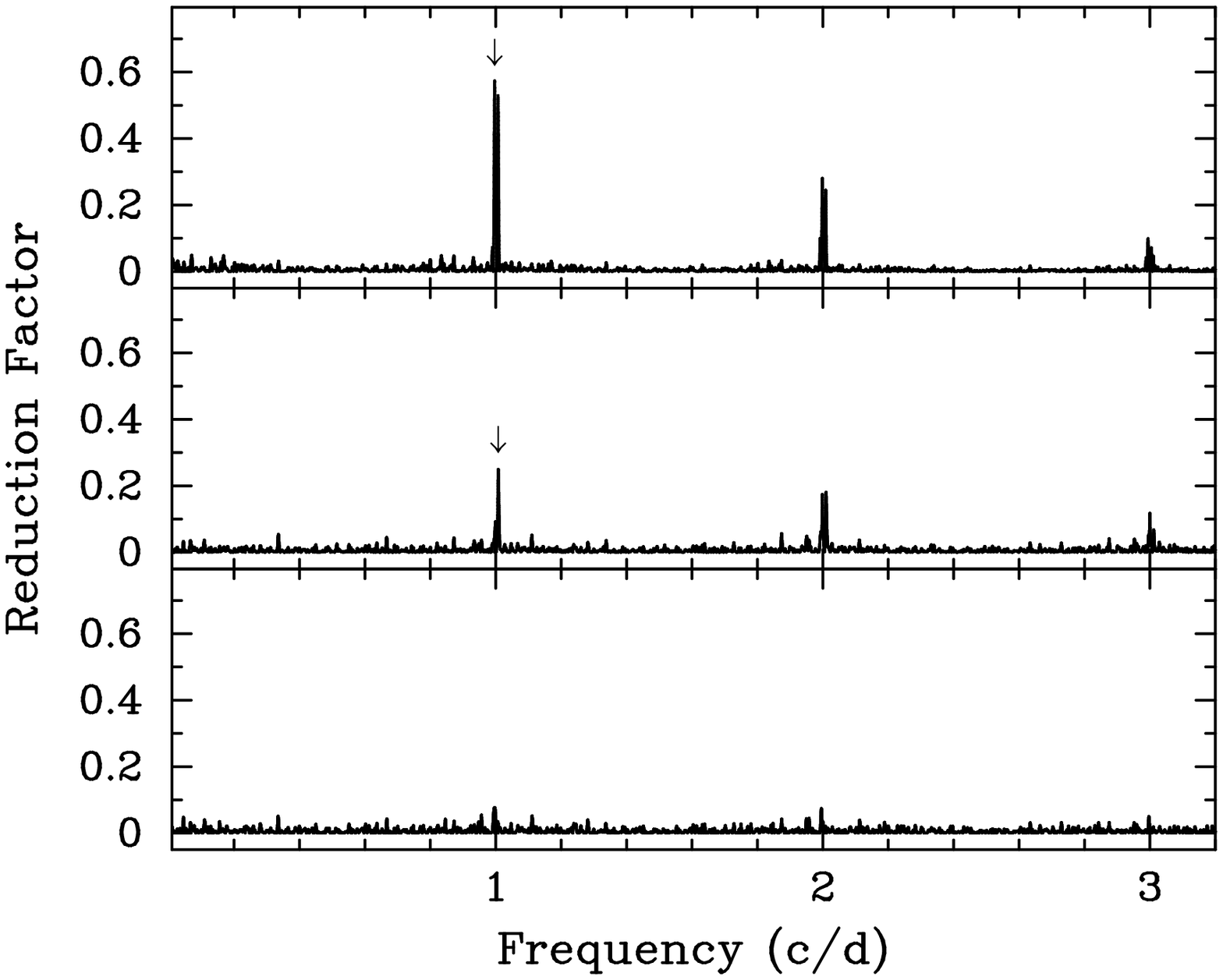}
\figcaption{Least-squares frequency spectra of the HD~62196 Johnson $B$ data 
set, showing the results of progressively fixing the two detected frequencies.
The arrows indicate the two frequencies at 0.9966 day$^{-1}$ ({\it top}) and
1.0077 day$^{-1}$ ({\it middle}). Both frequencies were confirmed in the
Johnson $V$ data set.}
\end{figure}

\clearpage
\begin{figure}
\figurenum{19}
\epsscale{0.8}
\plotone{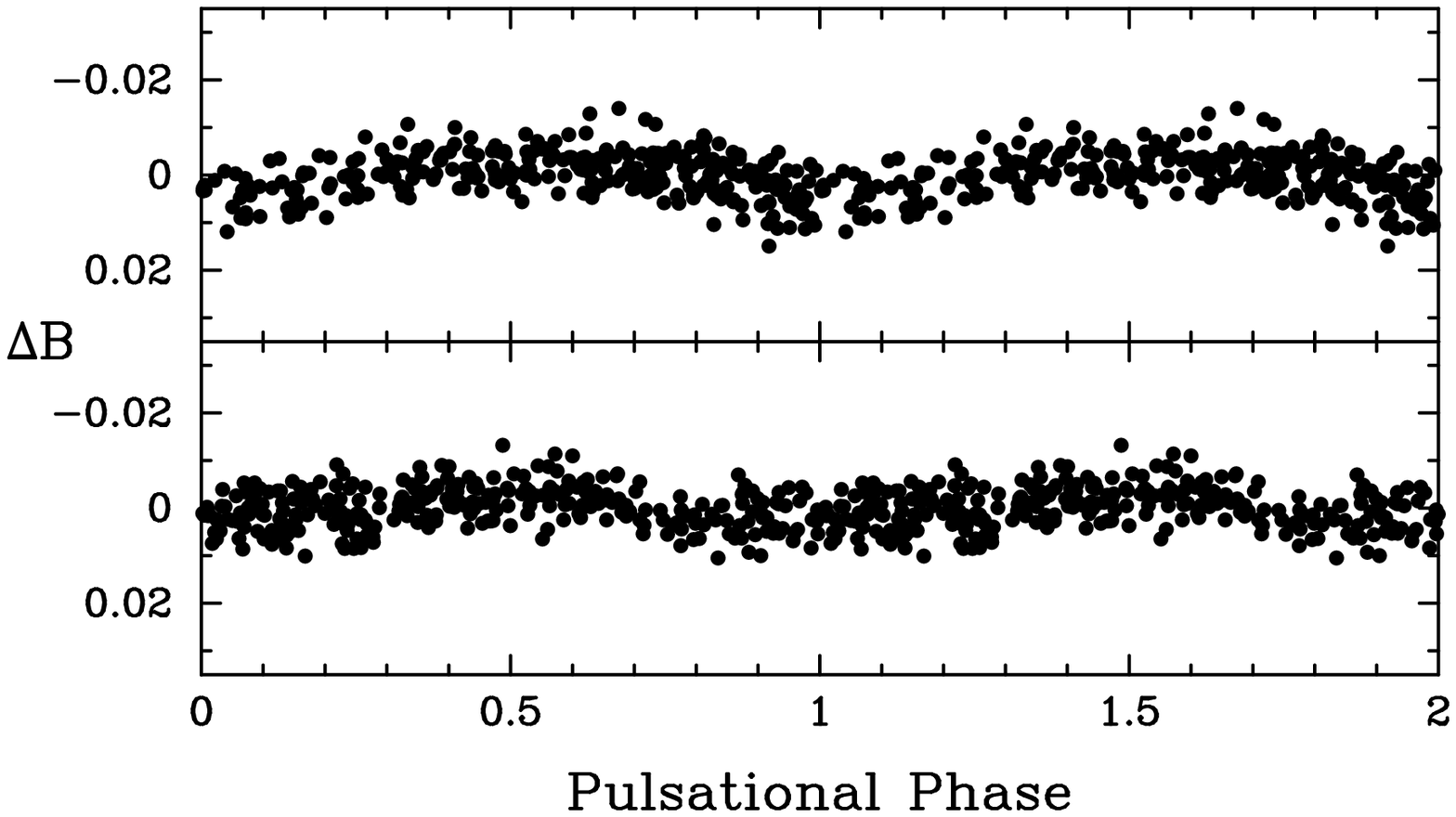}
\figcaption{The Johnson $B$ photometric data for HD~62196, phased with
the two frequencies and times of minimum from Table~5.  The two frequencies
are 0.9966 day$^{-1}$ ({\it top}) and 1.0077 day$^{-1}$ ({\it bottom}).
For each panel, the data set has been prewhitened to remove the other
known frequency.}
\end{figure}

\clearpage
\begin{figure}
\figurenum{20}
\epsscale{0.8}
\plotone{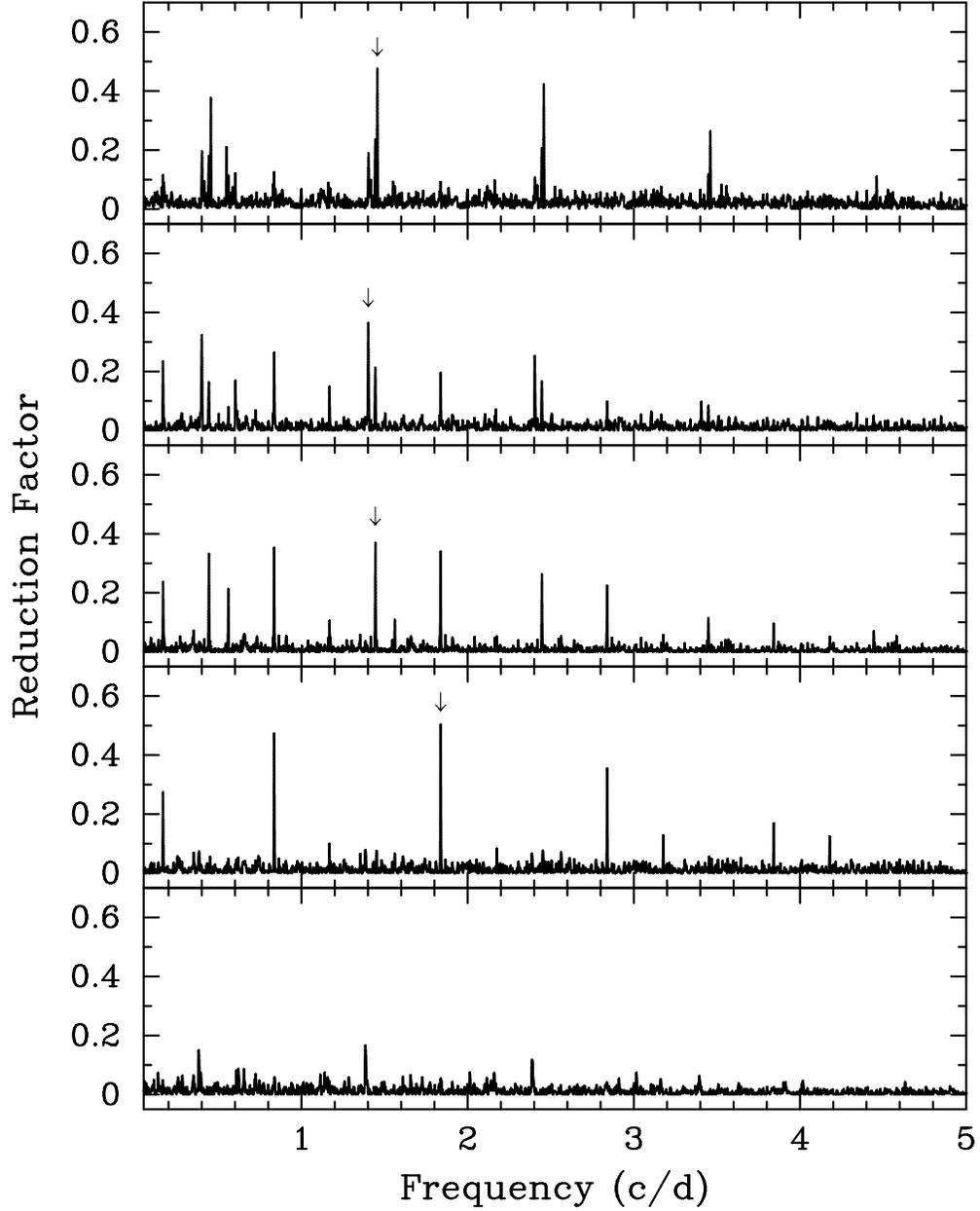}
\figcaption{Least-squares frequency spectra of the HD~63436 Johnson $B$ data 
set, showing the results of progressively fixing the four detected frequencies.
The arrows indicate the four frequencies ({\it top to bottom}) 1.4557,
1.4020, 1.4443, and 1.8372 day$^{-1}$.  All four frequencies were confirmed 
in the Johnson $V$ data set.}
\end{figure}

\clearpage
\begin{figure}
\figurenum{21}
\epsscale{0.8}
\plotone{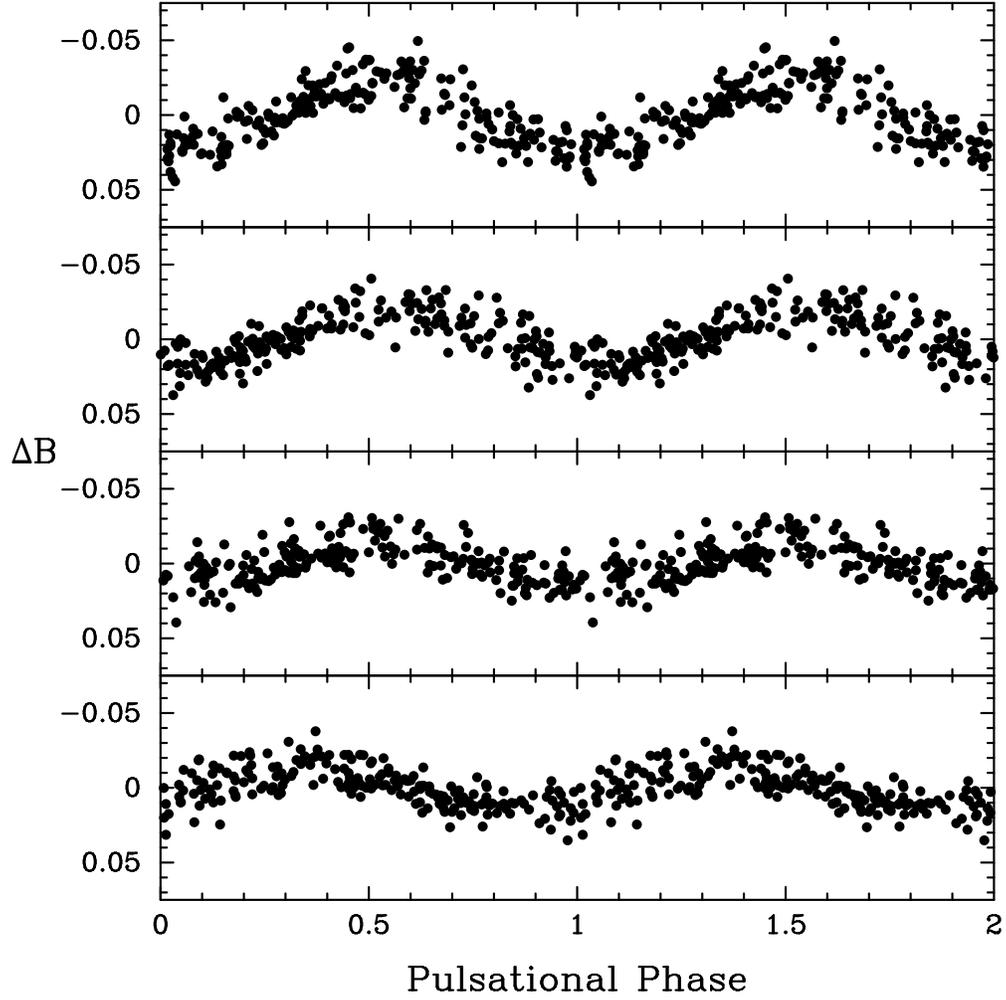}
\figcaption{The Johnson $B$ photometric data for HD~63436, phased with the
four frequencies and times of minimum from Table~5.  The four frequencies
are ({\it top to bottom}) 1.4557, 1.4020, 1.4443, and 1.8372 day$^{-1}$.  
For each panel, the data set has been prewhitened to remove the other known
frequencies.}
\end{figure}

\clearpage
\begin{figure}
\figurenum{22}
\epsscale{0.8}
\plotone{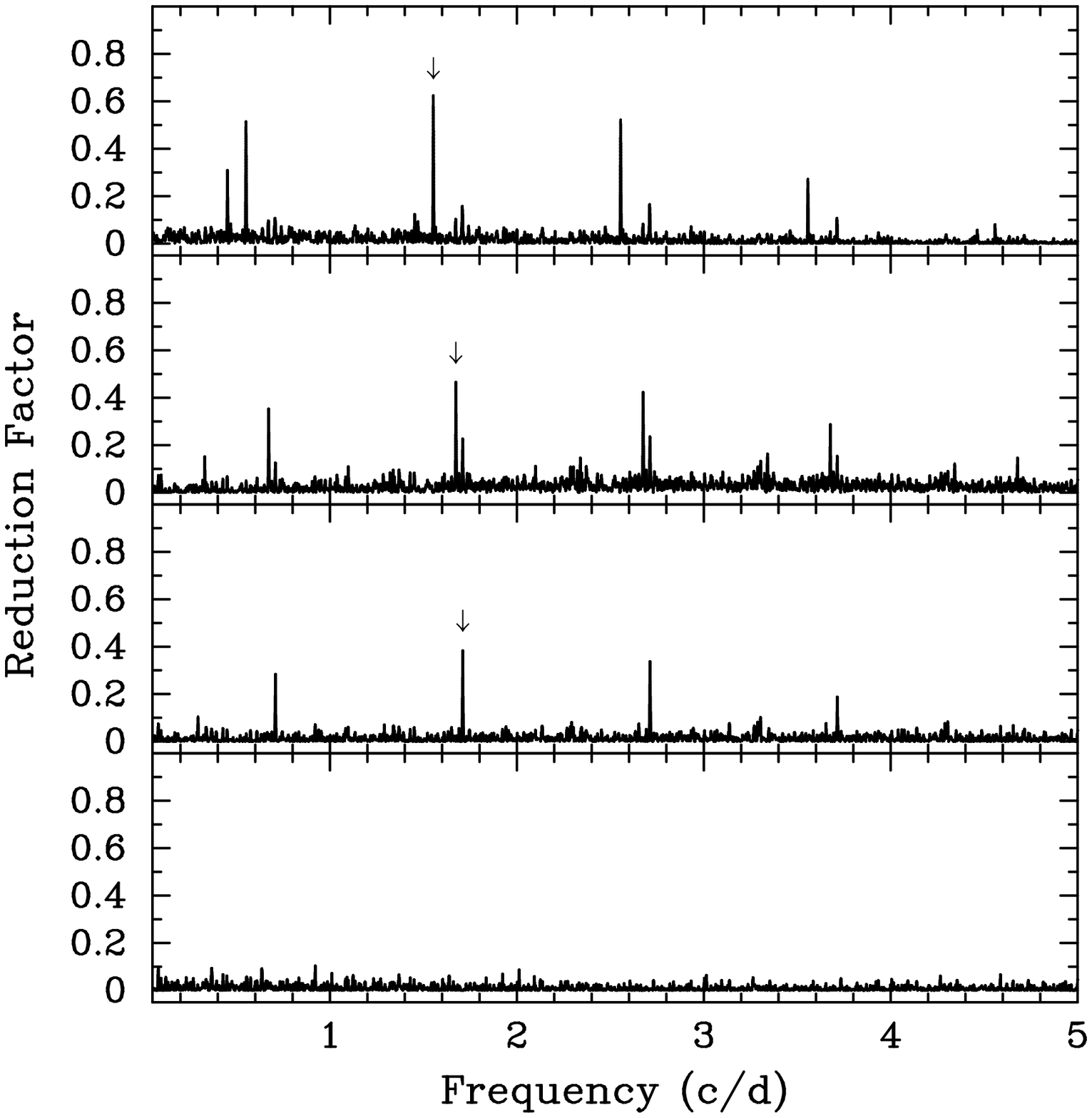}
\figcaption{Least-squares frequency spectra of the HD~65526 Johnson $B$ data 
set, showing the results of progressively fixing the three detected 
frequencies.  The arrows indicate the three frequencies ({\it top to bottom}) 
1.5527, 1.6735, and 1.7101 day$^{-1}$.  All three frequencies were confirmed 
in the Johnson $V$ data set.}
\end{figure}

\clearpage
\begin{figure}
\figurenum{23}
\epsscale{0.8}
\plotone{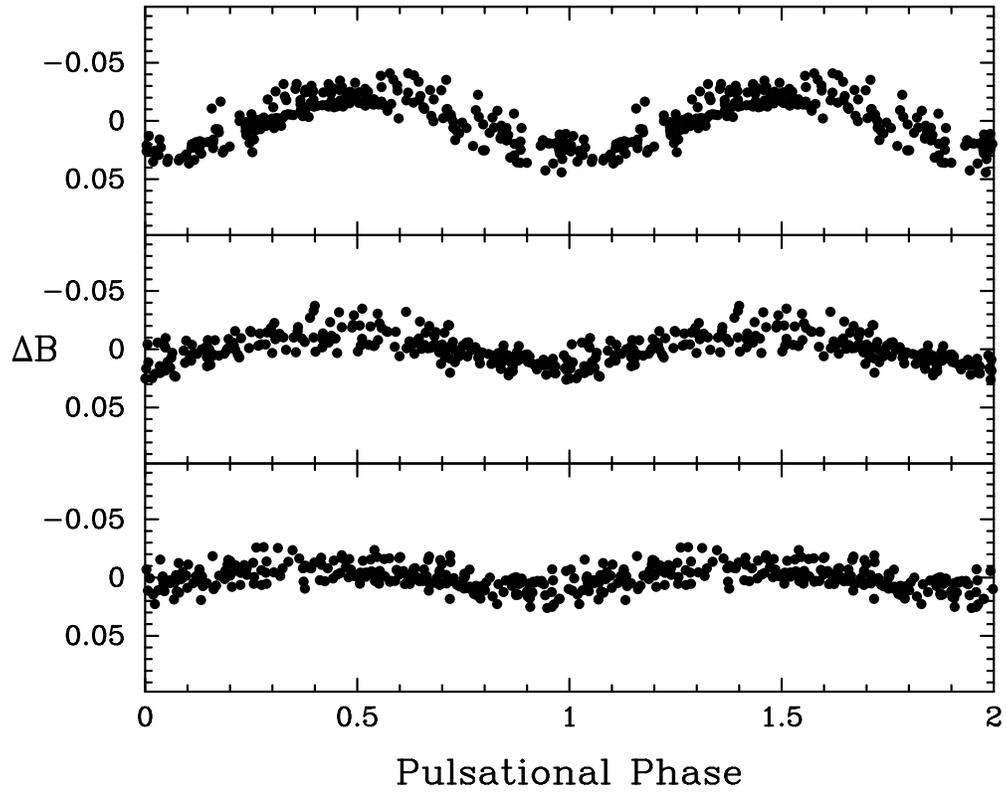}
\figcaption{The Johnson $B$ photometric data for HD~65526, phased with
the three frequencies and times of minimum from Table~5.  The three 
frequencies are ({\it top to bottom}) 1.5527, 1.6735, and 1.7101 day$^{-1}$. 
For each panel, the data set has been prewhitened to remove the other
two known frequencies.}
\end{figure}

\clearpage
\begin{figure}
\figurenum{24}
\epsscale{0.8}
\plotone{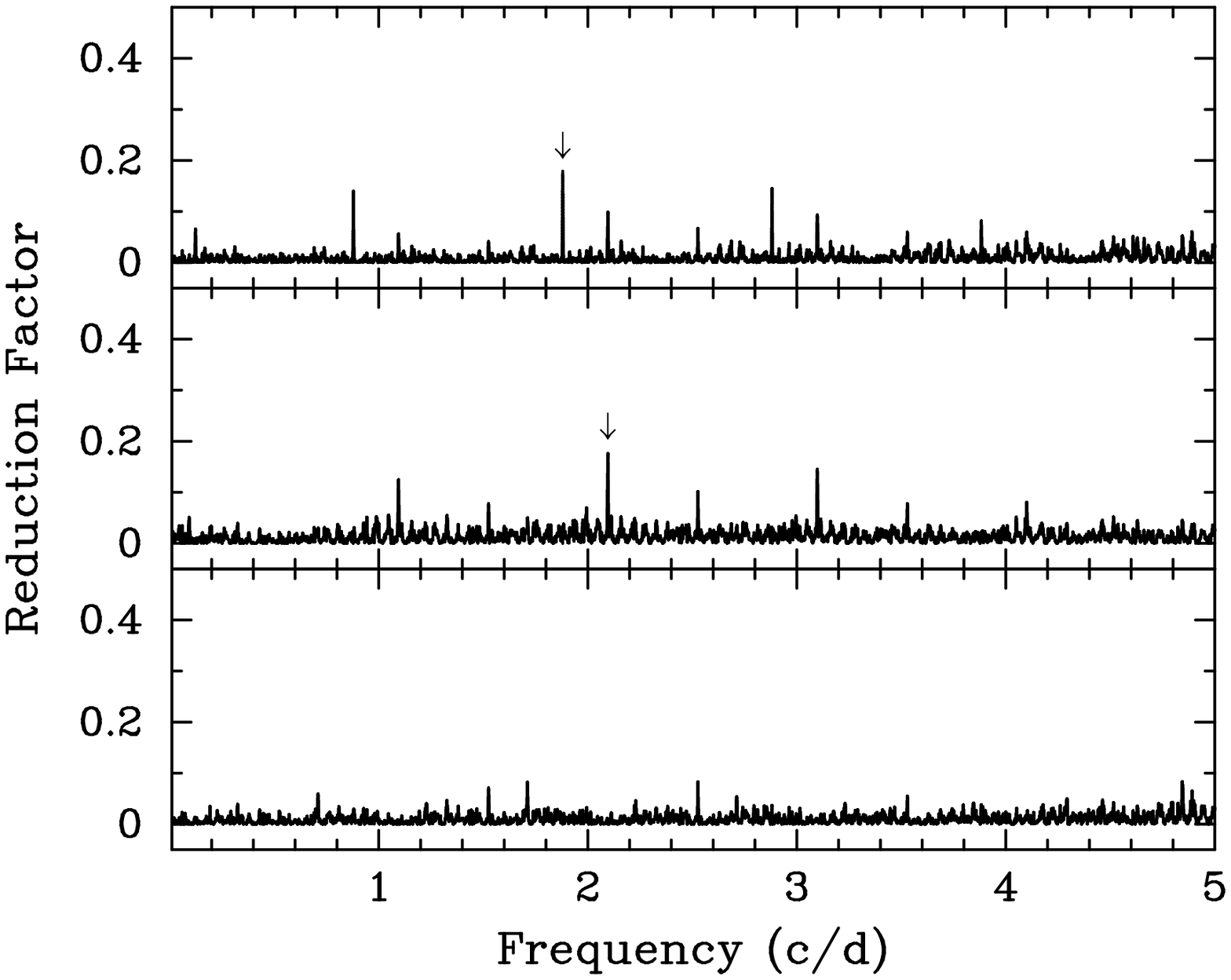}
\figcaption{Least-squares frequency spectra of the HD~69682 Johnson $B$ data 
set, showing the results of progressively fixing the two detected frequencies.
The arrows indicate the two frequencies at 1.8801 day$^{-1}$ ({\it top}) and
2.0963 day$^{-1}$ ({\it middle}). Both frequencies were confirmed in the
Johnson $V$ data set.}
\end{figure}

\clearpage
\begin{figure}
\figurenum{25}
\epsscale{0.8}
\plotone{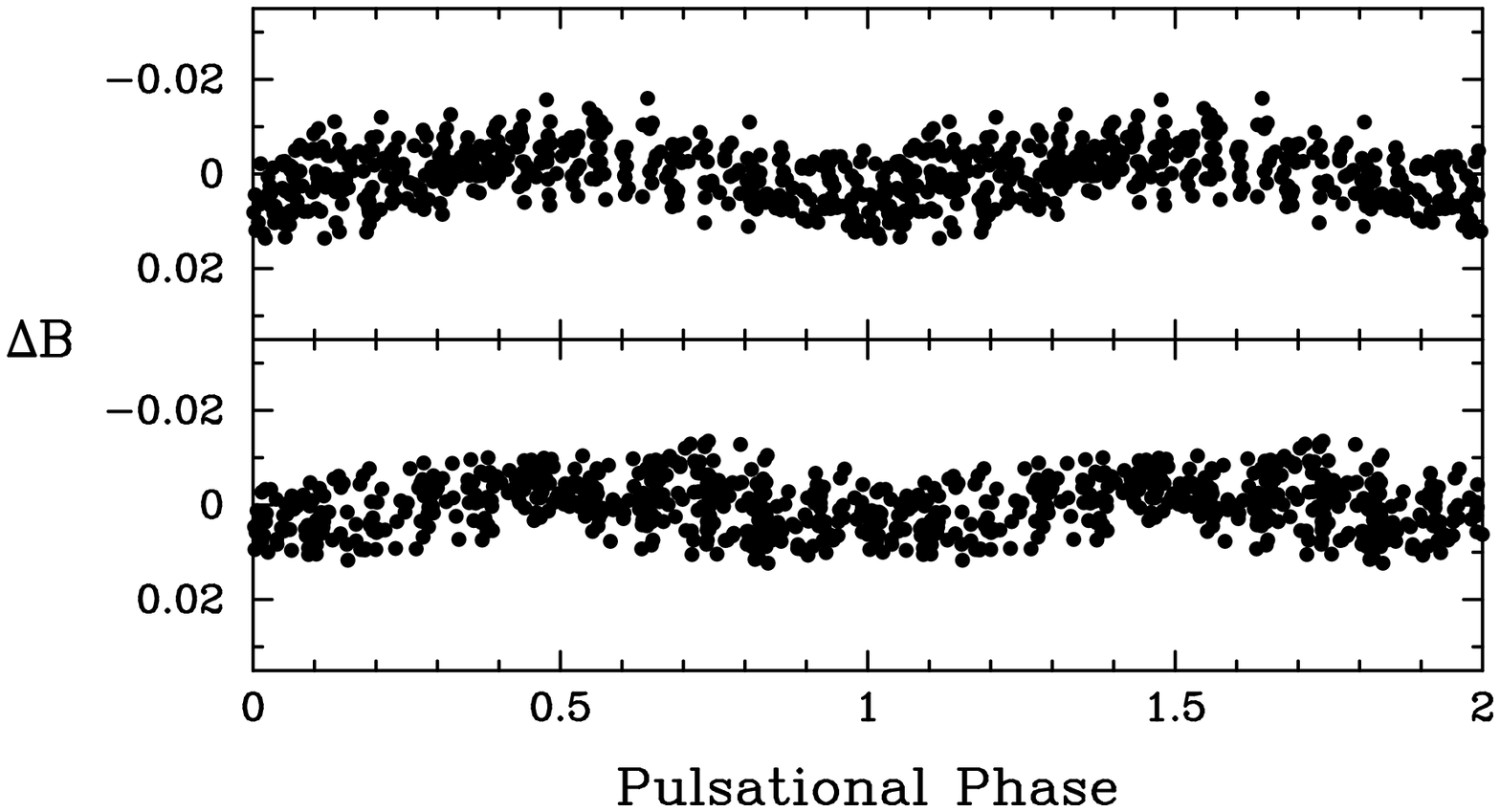}
\figcaption{The Johnson $B$ photometric data for HD~69682, phased with
the two frequencies and times of minimum from Table~5.  The two frequencies
are 1.8801 day$^{-1}$ ({\it top}) and 2.0963 day$^{-1}$ ({\it bottom}).
For each panel, the data set has been prewhitened to remove the other
frequency.}
\end{figure}

\clearpage
\begin{figure}
\figurenum{26}
\epsscale{0.8}
\plotone{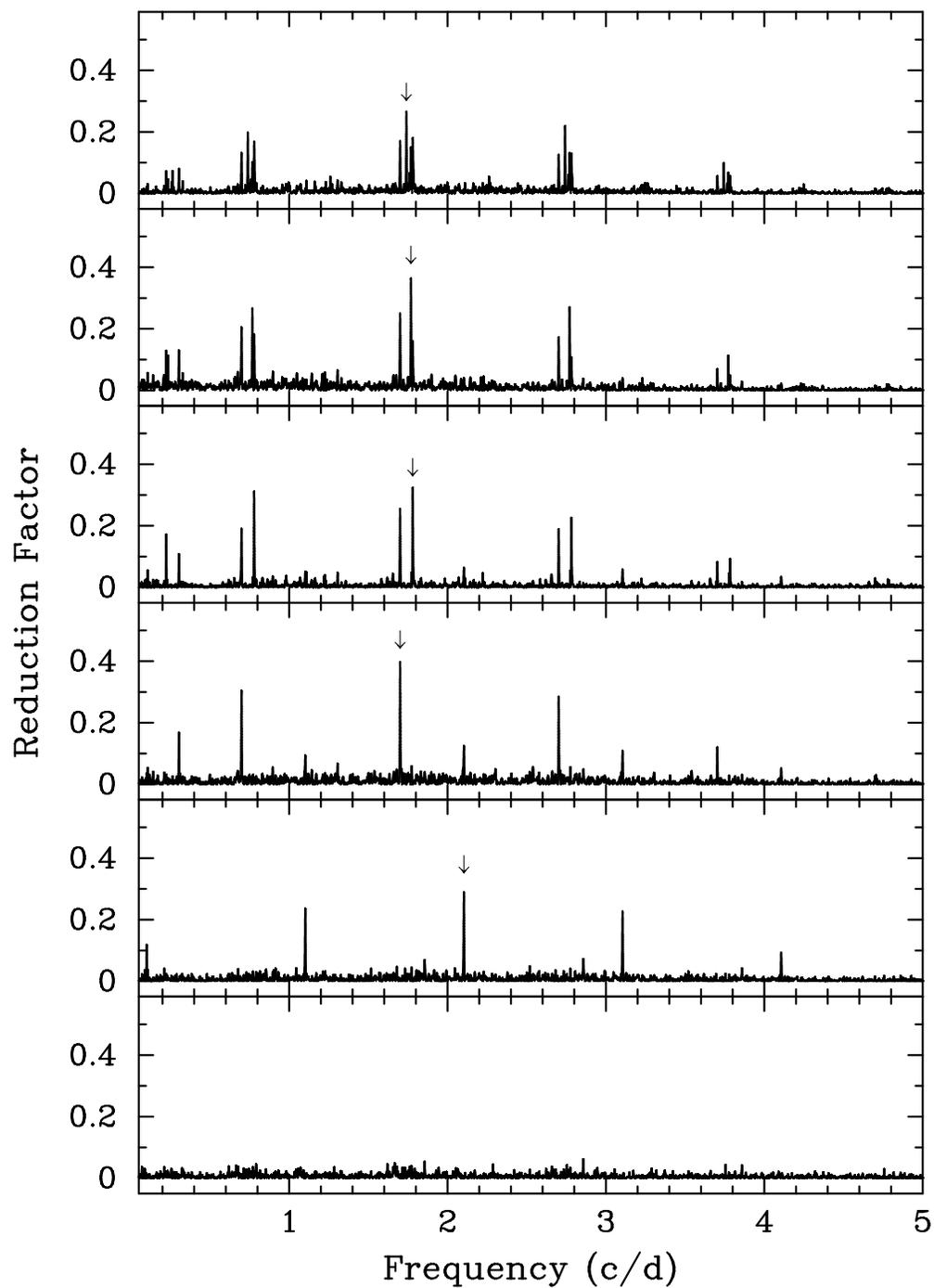}
\figcaption{Least-squares frequency spectra of the HD~99267 Johnson $B$ data 
set, showing the results of progressively fixing the five frequencies.  The 
arrows indicate the five frequencies at ({\it top to bottom}) 1.7402,
1.7690, 1.7802, 1.7001, and 2.1029 day$^{-1}$. All five frequencies were 
confirmed in the $V$ data set.}
\end{figure}

\clearpage
\begin{figure}
\figurenum{27}
\epsscale{0.8}
\plotone{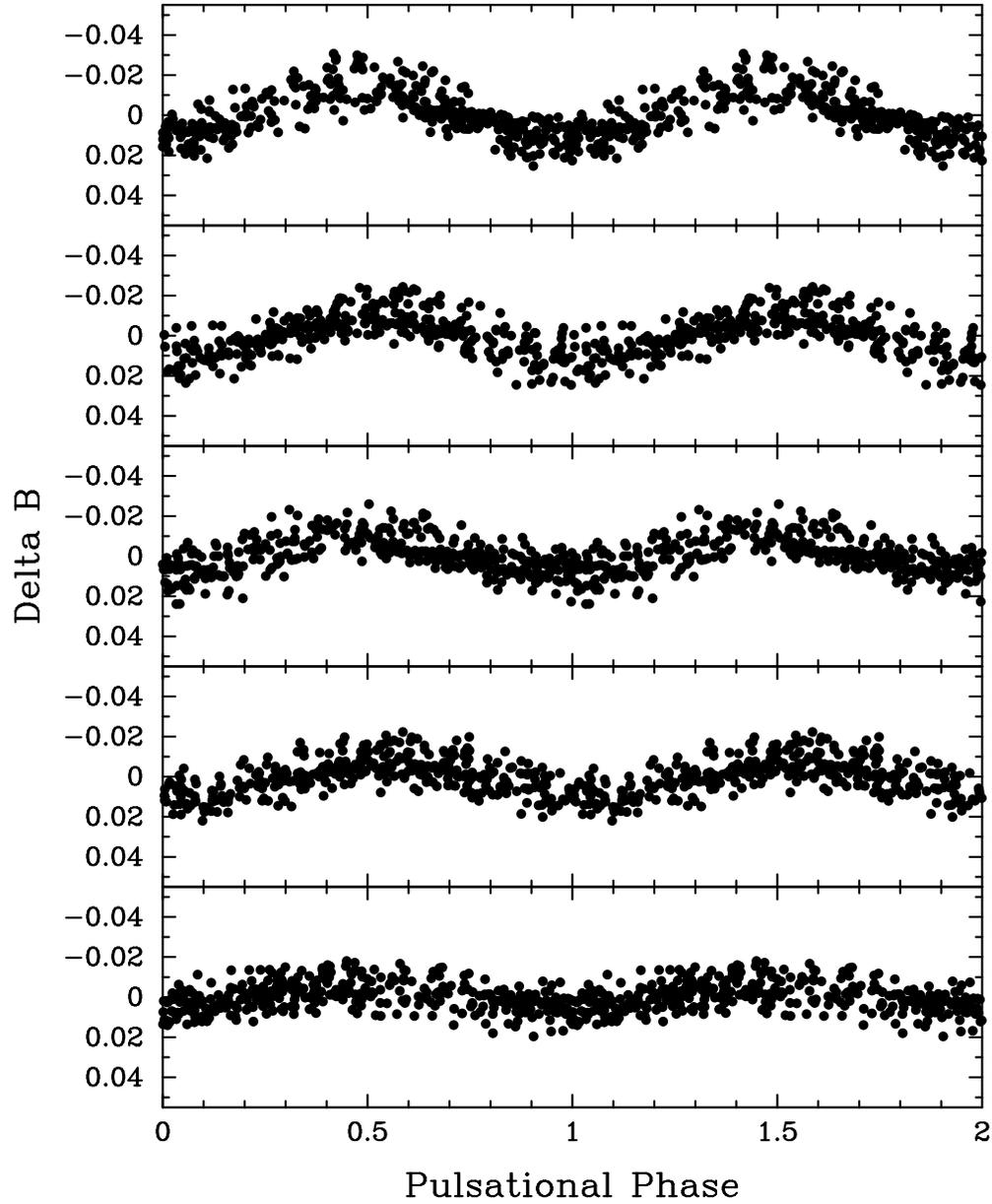}
\figcaption{The Johnson $B$ photometric data for HD~99267, phased with
the five frequencies and times of minimum from Table~5.  The five frequencies
are ({\it top to bottom}) 1.7402, 1.7690, 1.7802, 1.7001, and 
2.1029 day$^{-1}$. For each panel, the data set has been prewhitened to 
remove the other four frequencies.}
\end{figure}

\clearpage
\begin{figure}
\figurenum{28}
\epsscale{0.8}
\plotone{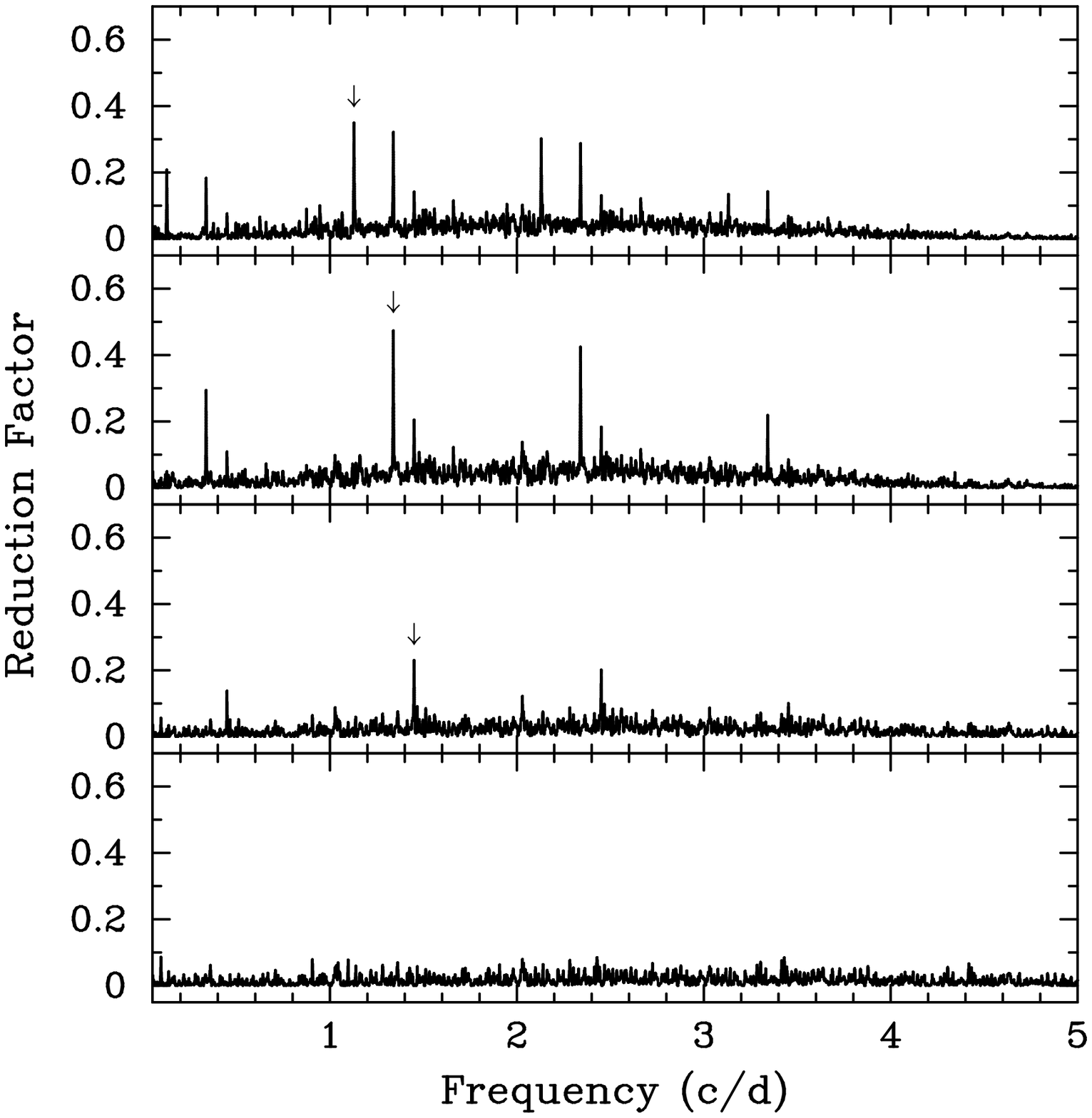}
\figcaption{Least-squares frequency spectra of the HD~114447 Johnson $B$ data 
set, showing the results of progressively fixing the three detected 
frequencies.  The arrows indicate the three frequencies ({\it top to bottom}) 
1.1284, 1.3386, and 1.4499 day$^{-1}$. All three frequencies were confirmed in 
the Johnson $V$ data set.}
\end{figure}

\clearpage
\begin{figure}
\figurenum{29}
\epsscale{0.8}
\plotone{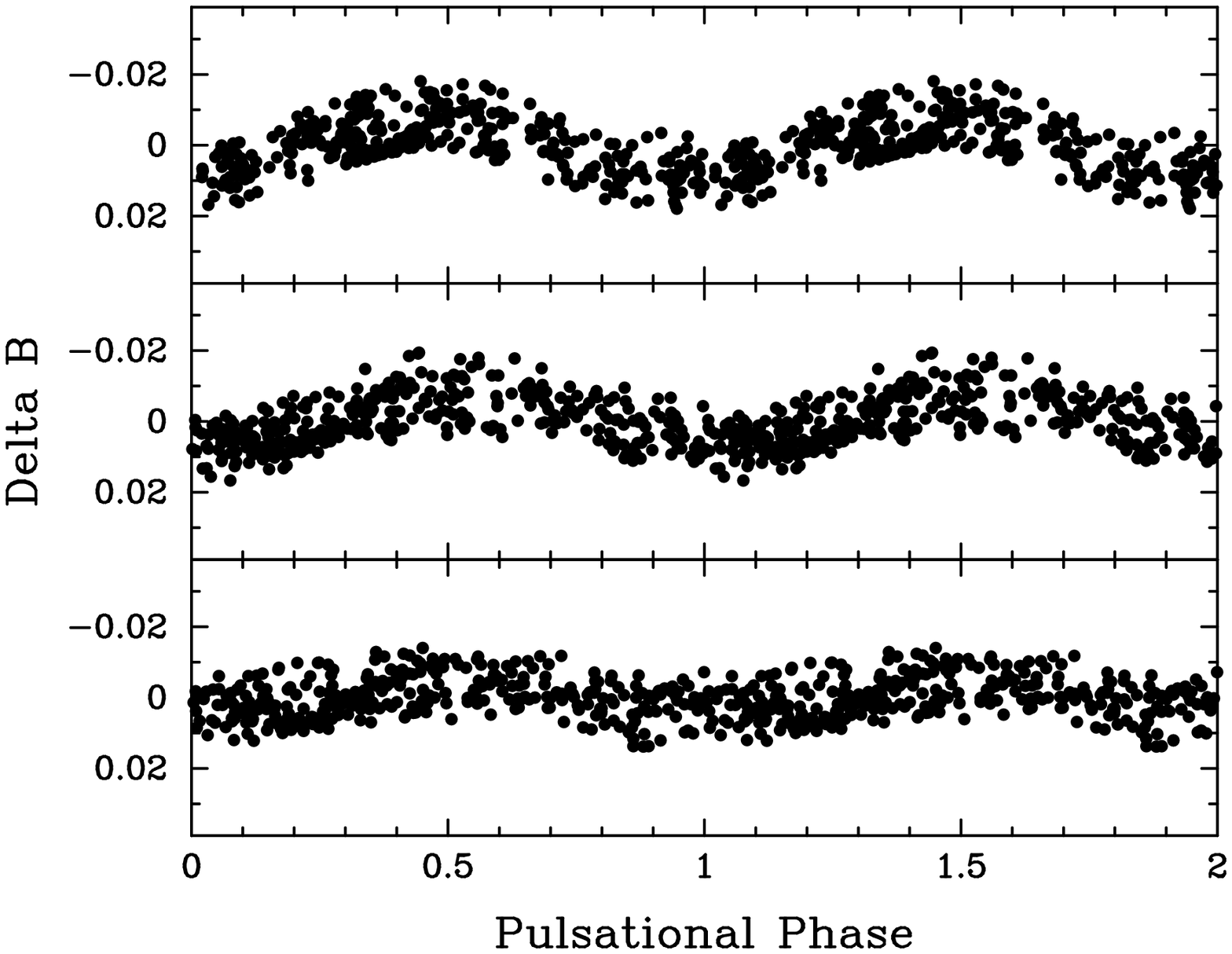}
\figcaption{The Johnson $B$ photometric data for HD~114447, phased with
the three frequencies and times of minimum from Table~5.  The three
frequencies are ({\it top to bottom}) 1.1284, 1.3386, and 1.4499 day$^{-1}$.
For each panel, the data set has been prewhitened to remove the other
two known frequencies.}
\end{figure}

\clearpage
\begin{figure}
\figurenum{30}
\epsscale{0.8}
\plotone{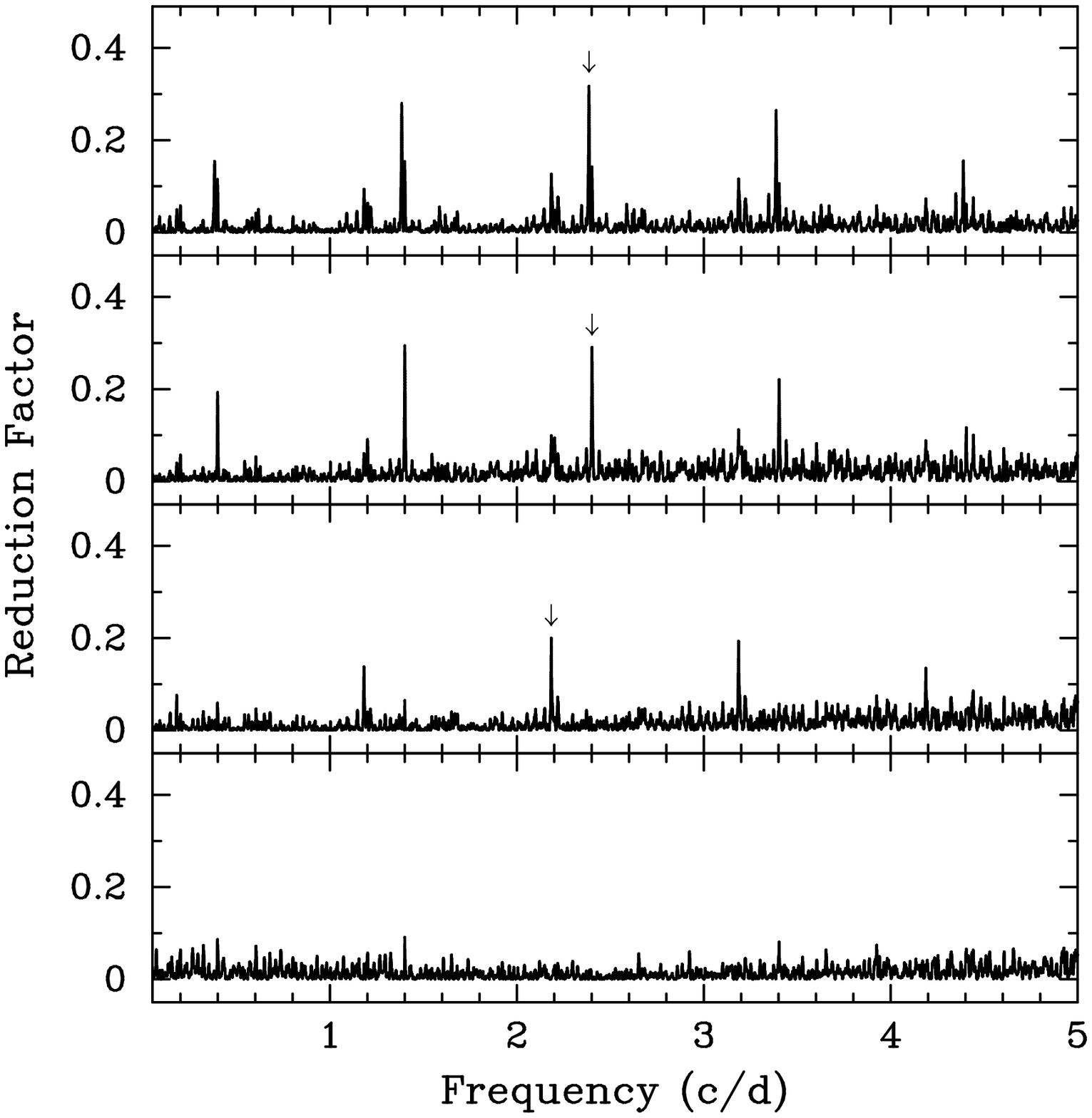}
\figcaption{Least-squares frequency spectra of the HD~138936 Johnson $B$ data 
set, showing the results of progressively fixing the three detected 
frequencies.  The arrows indicate the three frequencies ({\it top to bottom}) 
2.3855, 2.4018, and 2.1839 day$^{-1}$. All three frequencies were confirmed 
in the Johnson $V$ data set.}
\end{figure}

\clearpage
\begin{figure}
\figurenum{31}
\epsscale{0.8}
\plotone{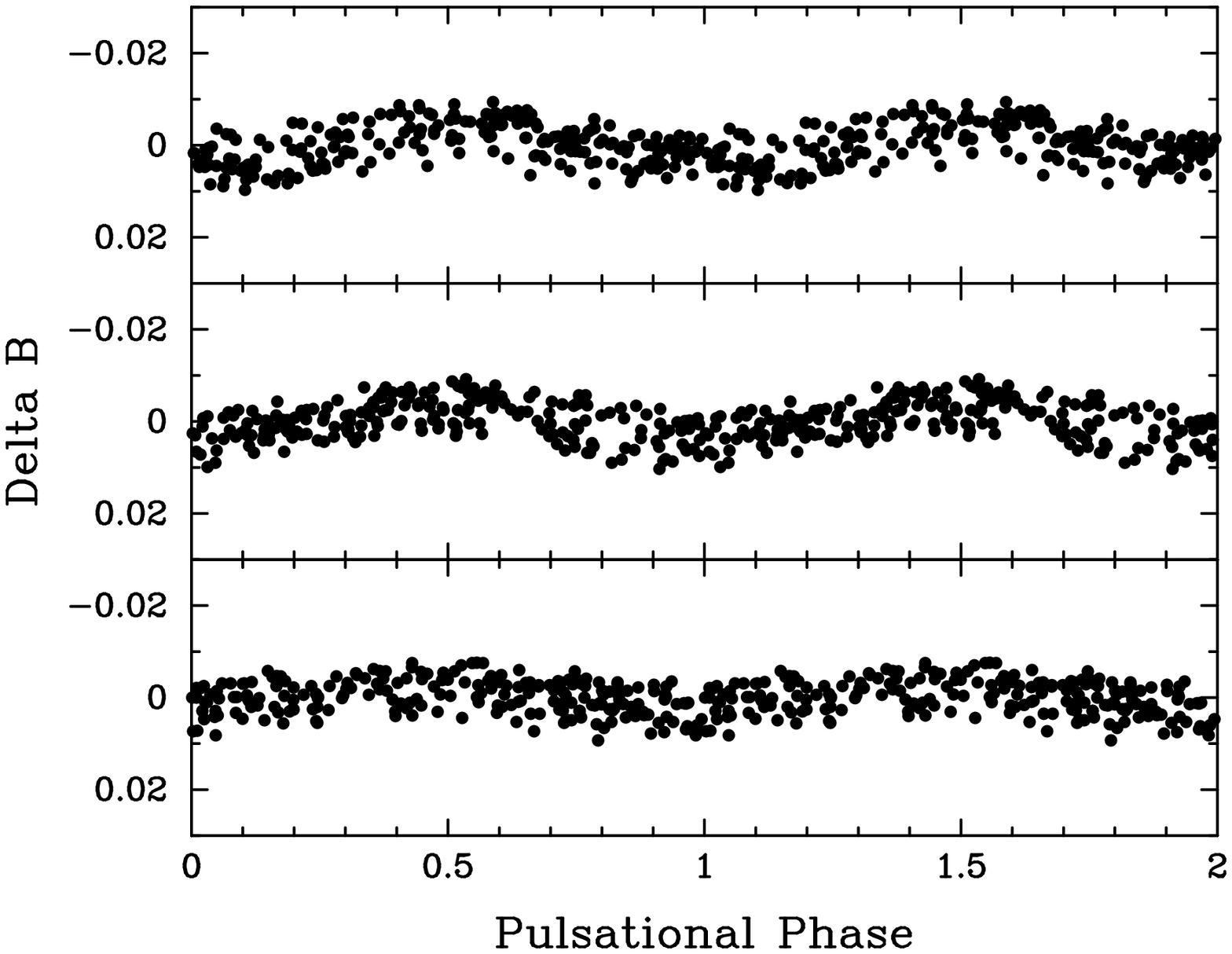}
\figcaption{The Johnson $B$ photometric data for HD~138936, phased with
the three frequencies and times of minimum from Table~5.  The three
frequencies are ({\it top to bottom}) 2.3855, 2.4018, and 2.1839 day$^{-1}$.
For each panel, the data set has been prewhitened to remove the other
two known frequencies.}
\end{figure}

\clearpage
\begin{figure}
\figurenum{32}
\epsscale{0.8}
\plotone{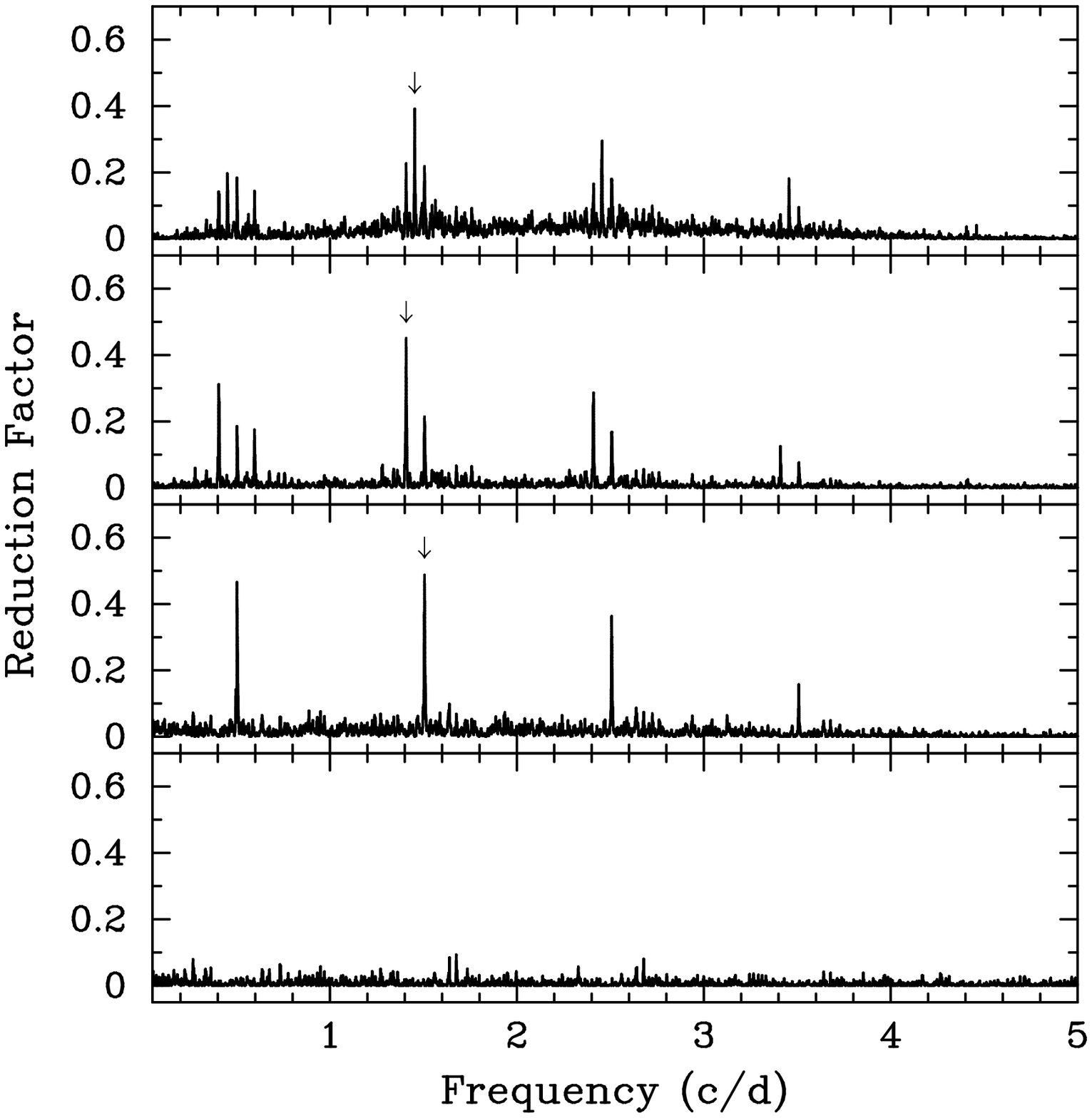}
\figcaption{Least-squares frequency spectra of the HD~139478 Johnson $B$ data 
set, showing the results of progressively fixing the three detected 
frequencies.  The arrows indicate the three frequencies ({\it top to bottom}) 
1.4531, 1.4073, and 1.5056 day$^{-1}$. All three frequencies were confirmed in 
the Johnson $V$ data set.}
\end{figure}

\clearpage
\begin{figure}
\figurenum{33}
\epsscale{0.8}
\plotone{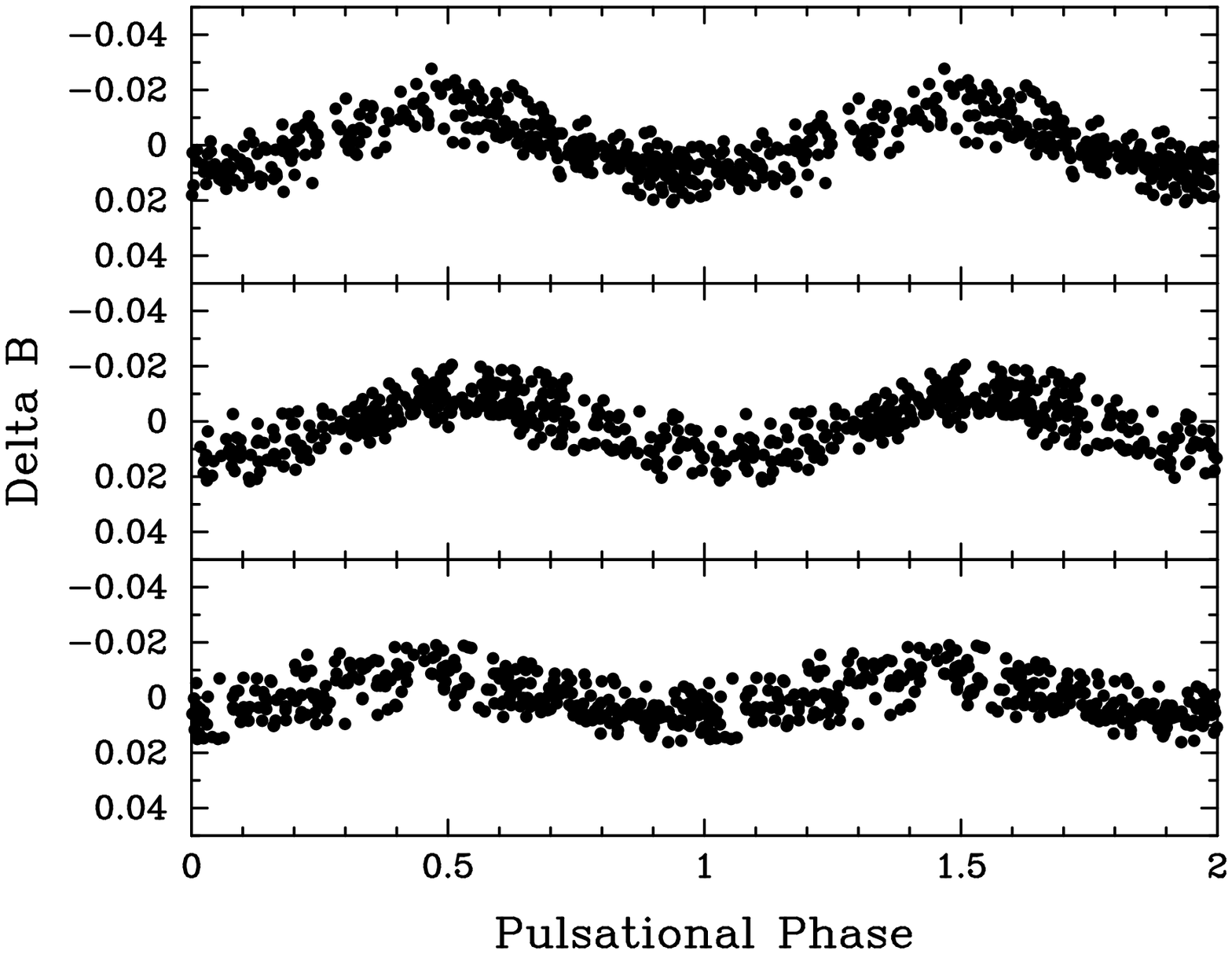}
\figcaption{The Johnson $B$ photometric data for HD~139478, phased with
the three frequencies and times of minimum from Table~5.  The three
frequencies are ({\it top to bottom}) 1.4531, 1.4073, and 1.5056 day$^{-1}$.
For each panel, the data set has been prewhitened to remove the other
two known frequencies.}
\end{figure}

\clearpage
\begin{figure}
\figurenum{34}
\epsscale{0.8}
\plotone{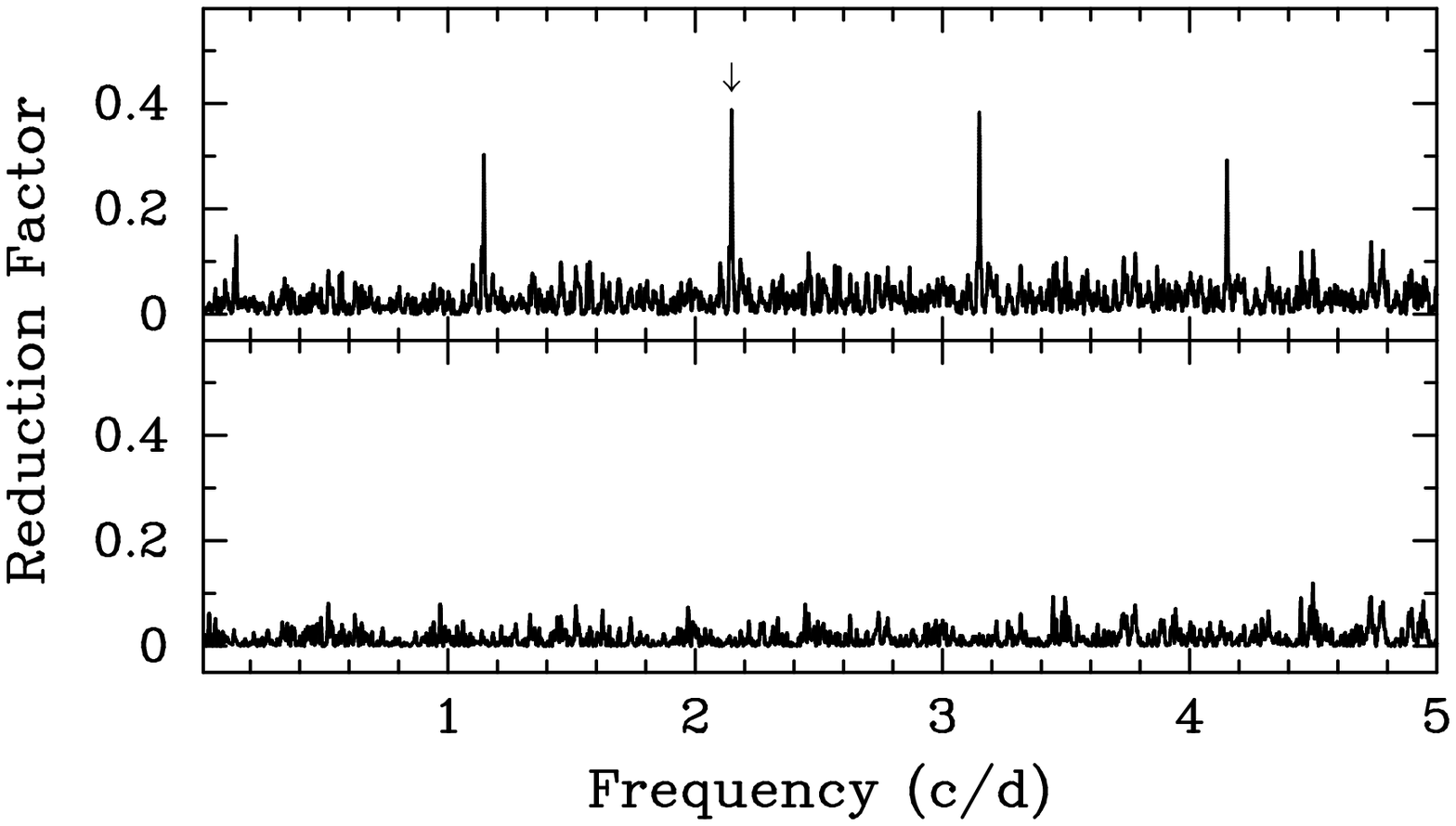}
\figcaption{Least-squares frequency spectra of the HD~145005 Johnson $B$ data 
set.  The arrow in the top panel indicates the single detected frequency of 
2.1473 day$^{-1}$.  The bottom panel is the frequency spectrum resulting when 
the 2.1473 day$^{-1}$ frequency is fixed.  The same frequency was found in 
the Johnson $V$ data set.}
\end{figure}

\clearpage
\begin{figure}
\figurenum{35}
\epsscale{0.8}
\plotone{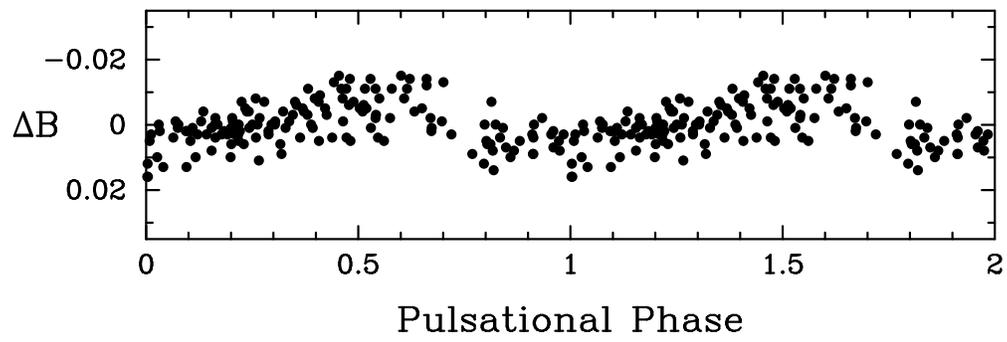}
\figcaption{The Johnson $B$ photometric data for HD~145005, phased with
the 2.1473 day$^{-1}$ frequency and time of minimum from Table~5.}
\end{figure}

\clearpage
\begin{figure}
\figurenum{36}
\epsscale{0.8}
\plotone{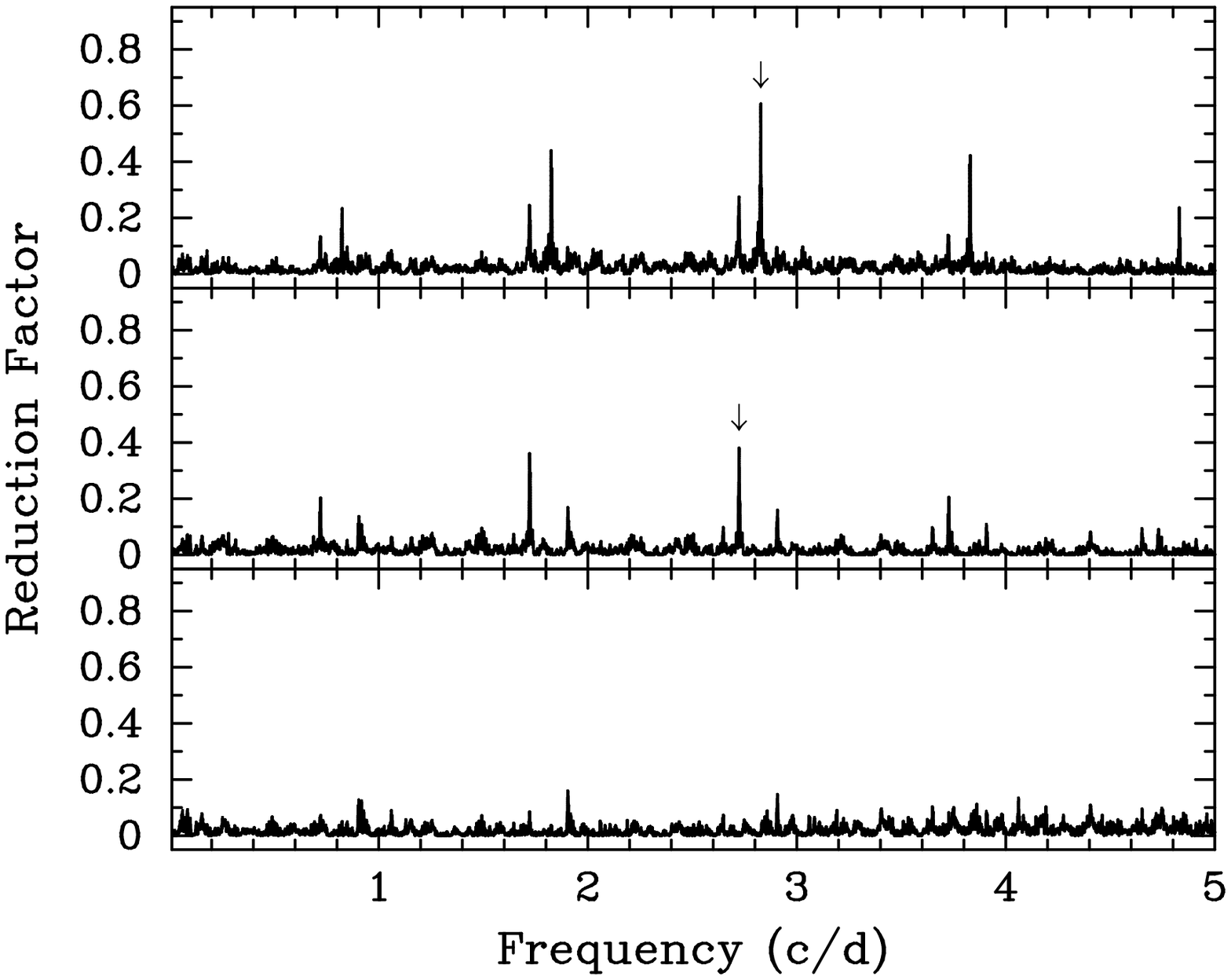}
\figcaption{Least-squares frequency spectra of the HD~220091 Johnson $B$ data 
set, showing the results of progressively fixing the two detected frequencies.
The arrows indicate the two frequencies at 2.8277 day$^{-1}$ ({\it top}) and
2.7241 day$^{-1}$ ({\it middle}). Both frequencies were confirmed in the
Johnson $V$ data set.}
\end{figure}

\clearpage
\begin{figure}
\figurenum{37}
\epsscale{0.8}
\plotone{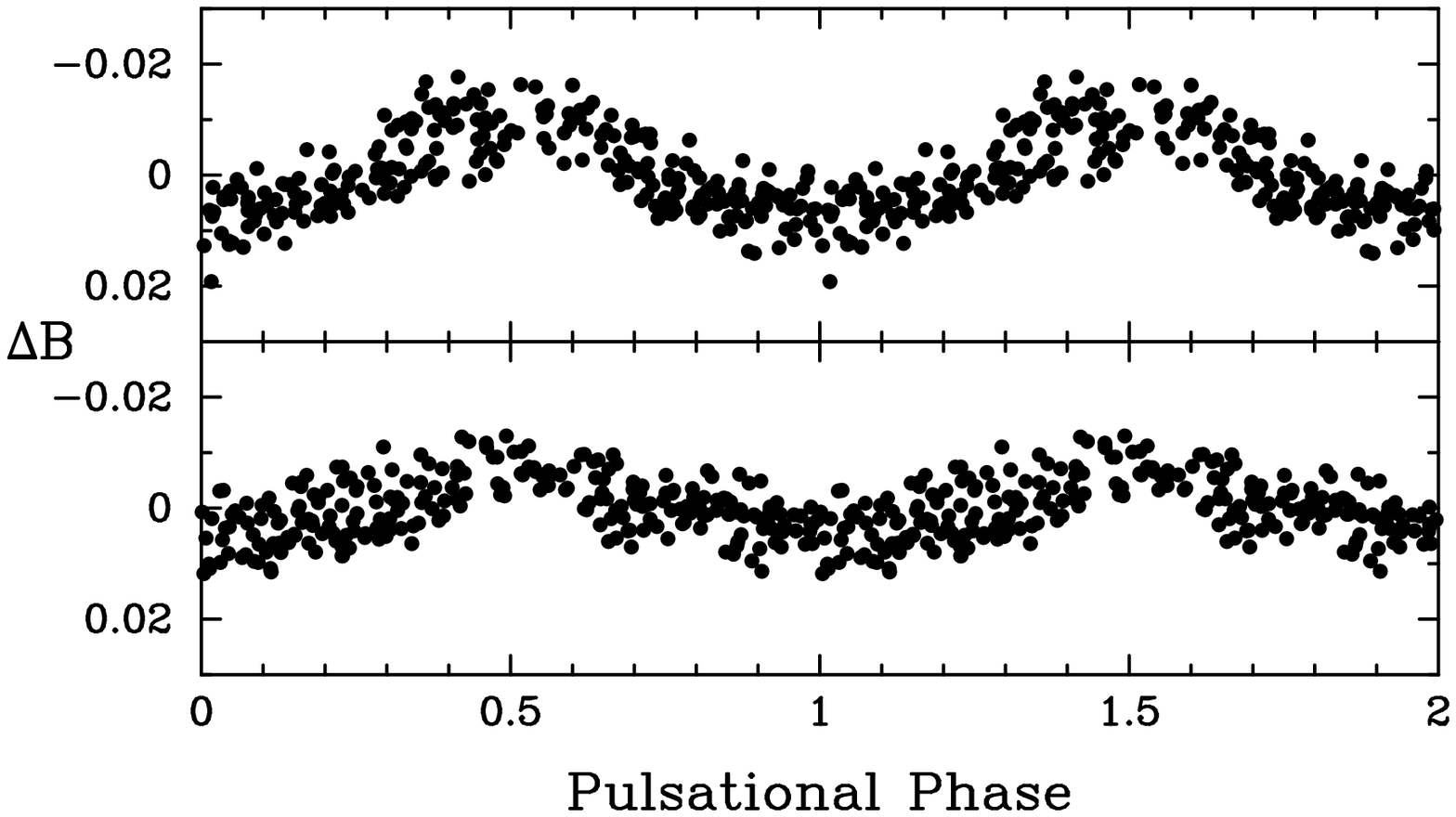}
\figcaption{The Johnson $B$ photometric data for HD~220091, phased with
the two frequencies and times of minimum from Table~5.  The two frequencies
are 2.8277 day$^{-1}$ ({\it top}) and 2.7241 day$^{-1}$ ({\it bottom}).
For each panel, the data set has been prewhitened to remove the other
known frequency.}
\end{figure}

\clearpage
\begin{figure}
\figurenum{38}
\epsscale{0.8}
\plotone{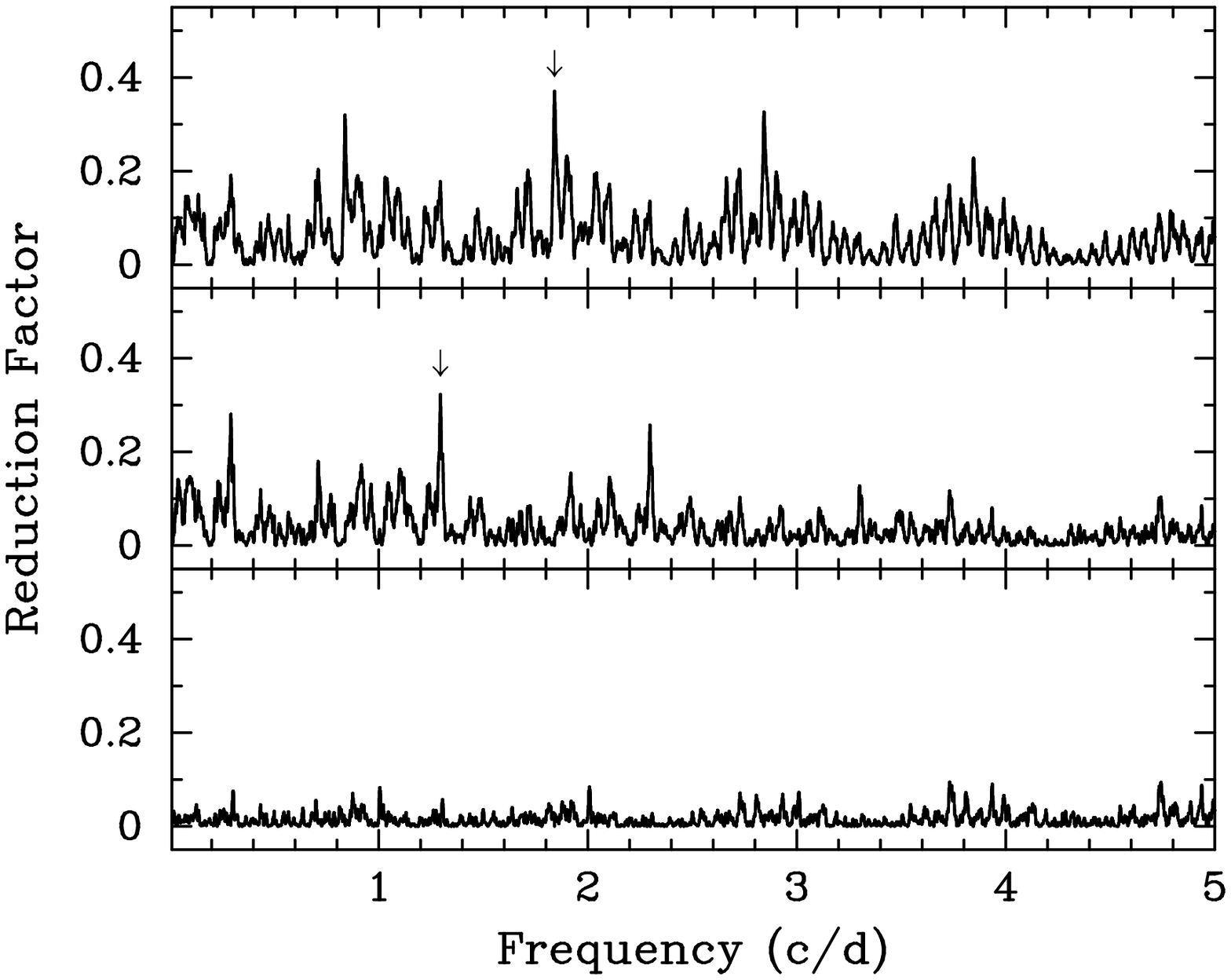}
\figcaption{Least-squares frequency spectra of the HD~224945 Johnson $B$ data 
set, showing the results of progressively fixing the two detected frequencies.
The arrows indicate the two frequencies at 1.8410 day$^{-1}$ ({\it top}) and
1.2951 day$^{-1}$ ({\it middle}). Both frequencies were confirmed in the
Johnson $V$ data set.}
\end{figure}

\clearpage
\begin{figure}
\figurenum{39}
\epsscale{0.8}
\plotone{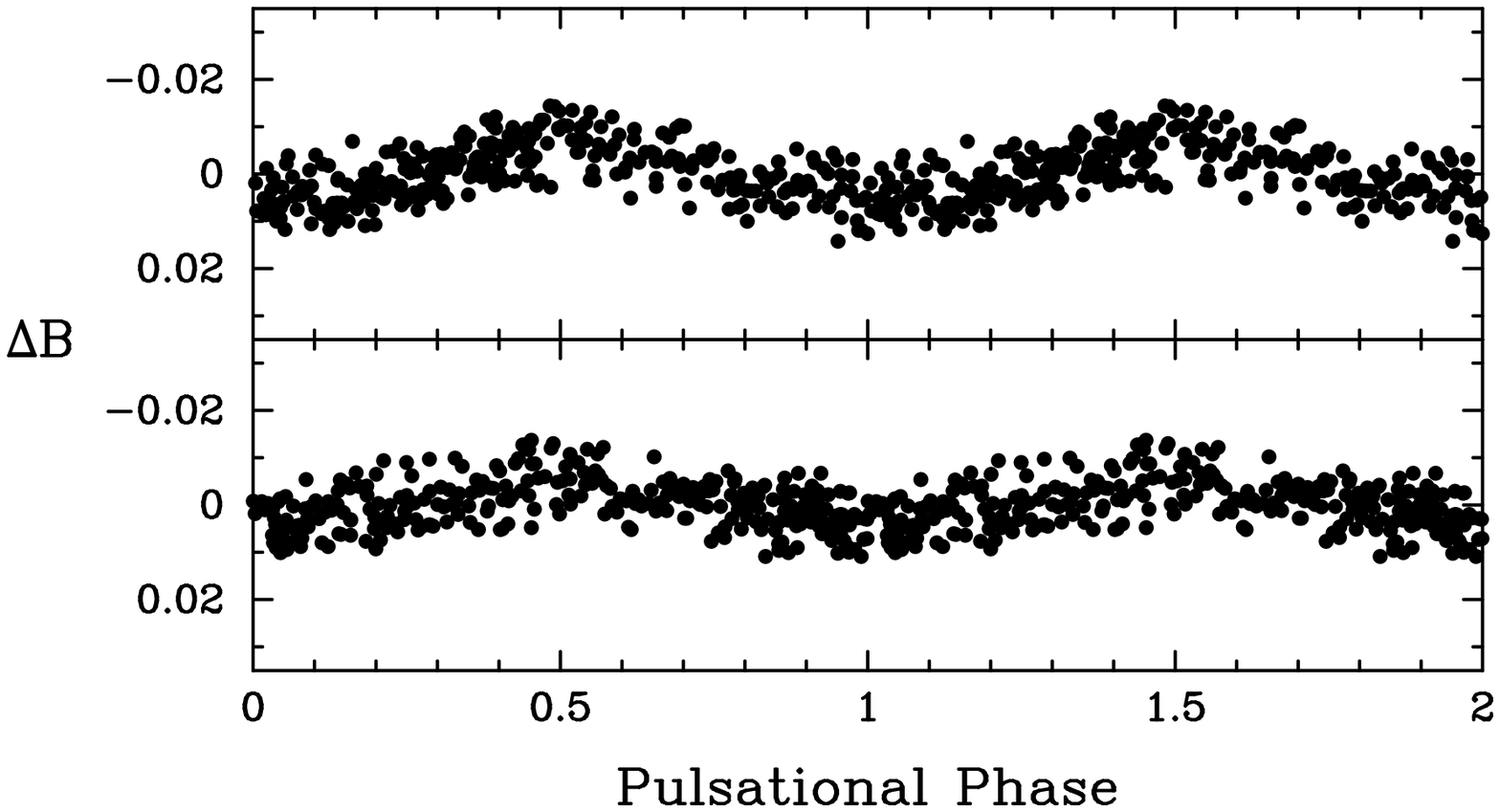}
\figcaption{The Johnson $B$ photometric data for HD~224945, phased with
the two frequencies and times of minimum from Table~5.  The two frequencies
are 1.8410 day$^{-1}$ ({\it top}) and 1.2951 day$^{-1}$ ({\it bottom}).
For each panel, the data set has been prewhitened to remove the other
known frequency.}
\end{figure}

\clearpage
\begin{figure}
\figurenum{40}
\epsscale{0.8}
\plotone{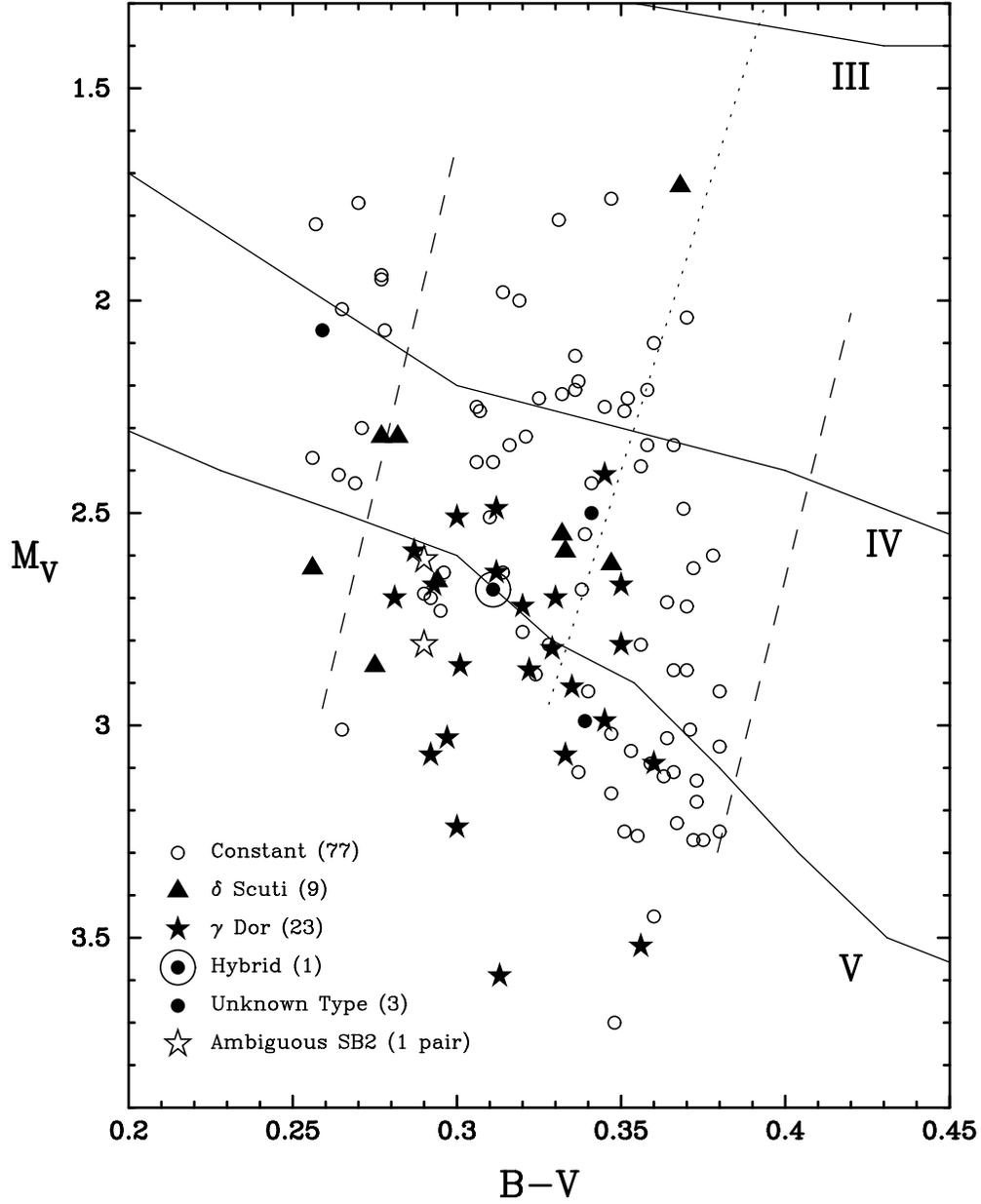}
\figcaption{Results of our follow-up photometric and spectroscopic 
observations of the 37 variable stars found in the T12 survey.  The 
luminosity classes and the $\gamma$~Doradus and $\delta$~Scuti instability 
strips are as described in Figure~1, except here we see only the cool 
edge of the $\delta$~Scuti instability strip (dotted line).  The 77 
constant stars are again plotted as open circles.  Among the 37 variable
stars, we find 15 new and nine previously discovered $\gamma$~Doradus stars 
(filled stars), eight new and one previously discovered $\delta$~Scuti stars 
(filled triangles), three new variables of unknown type (filled circles), and 
HD~8801, previously discovered to exhibit both $\gamma$~Doradus and 
$\delta$~Scuti pulsations (circled point).}
\end{figure}

\clearpage
\begin{figure}
\figurenum{41}
\epsscale{0.8}
\plotone{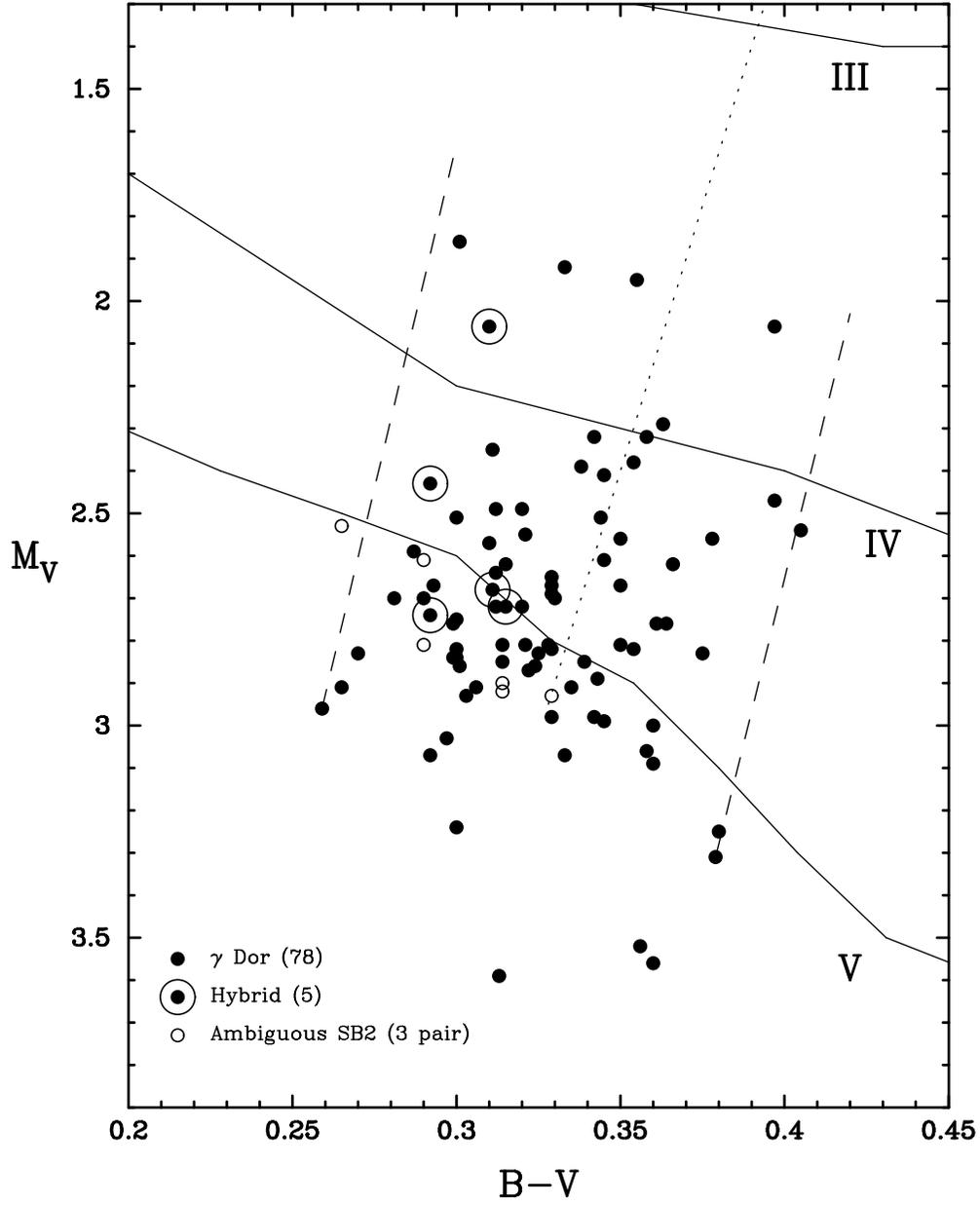}
\figcaption{Location in the H-R diagram of all 86 confirmed $\gamma$~Doradus 
field stars from Table~8. The 86 stars include 78 $\gamma$~Doradus pulsators, 
five self-excited hybrid stars HD~8801, HD~44195, HD~49434, HD~114839, and 
BD +18 4914 (circled points) and the six components of three SB2 binaries for 
which either (or both) components could be the pulsator (open circles).  All 
86 stars lie within the observed $\gamma$~Doradus instability strip defined 
in \citet{fwk03}.  All five hybrid pulsators lie within the overlap region
of the $\gamma$~Doradus and $\delta$~Scuti instability strips.  The star
furthest below the main sequence is the metal-poor star HD~62196, with
[Fe/H] $\sim -$0.5.}
\end{figure}

\end{document}